\documentclass[onefignum,onetabnum]{siamonline220329}

\usepackage{amsfonts}
\usepackage{mathabx}
\usepackage{graphicx}
\usepackage{epstopdf}
\usepackage{algorithmic}
\usepackage{amsmath}
\usepackage{subfig}
\usepackage{booktabs}
\usepackage{hyperref}
\usepackage{pgfplots}
\usepackage{mathtools}
\usepackage{float}
\usepackage{makecell}
\usepackage{placeins} 
\usepackage{amssymb}
\usepackage{array}
\usepackage[x11names]{xcolor}
\pgfplotsset{compat=1.18}

\usepackage{cite} 

\usepackage{tikz}
\usetikzlibrary{positioning}
\usetikzlibrary{shapes.geometric}
\usetikzlibrary{decorations.markings}

\tikzset{
  mycircle/.style={
    circle,
    draw,
    thick,
    fill=black,
    minimum size=1.5mm, 
    inner sep=0pt
  }
}

\ifpdf
  \DeclareGraphicsExtensions{.eps,.pdf,.png,.jpg}
\else
  \DeclareGraphicsExtensions{.eps}
\fi

\newcommand{\pdev}[2]{\frac{\partial{#1}}{\partial{#2}}}

\newcommand{\dev}[2]{\frac{\mathrm{d}{#1}}{\mathrm{d}{#2}}}

\newcommand*\conj[1]{\overline{#1}}

\newcommand{\expkone}{e^{i\boldsymbol{k_1} \cdot \boldsymbol{x}}}

\newcommand{\expktwo}{e^{i\boldsymbol{k_2} \cdot \boldsymbol{x}}}

\newcommand{\expq}{e^{i\boldsymbol{q_1} \cdot \boldsymbol{x}}}

\newcommand{\eone}{\boldsymbol{s_1}}

\newcommand{\etwo}{\boldsymbol{s_2}}

\usepackage{enumitem}
\setlist[enumerate]{leftmargin=.5in}
\setlist[itemize]{leftmargin=.5in}

\newsiamremark{remark}{Remark}
\newsiamremark{hypothesis}{Hypothesis}
\crefname{hypothesis}{Hypothesis}{Hypotheses}
\newsiamthm{claim}{Claim}

\headers{Pattern Formation with Two Length Scales: Spatiotemporal Chaos}{L. Pinkney, A. M. Rucklidge and C. Beaume}

\title{Pattern Formation with Two Length Scales: Spatiotemporal Chaos}

\author{Laura Pinkney\thanks{School of Mathematics, University of Leeds, Leeds, LS2 9JT, UK .}
\and Alastair M. Rucklidge\footnotemark[1] \and C\'{e}dric Beaume\footnotemark[1]}

\usepackage{amsopn}

\ifpdf
\hypersetup{
  pdftitle={Pattern Formation with Two Length Scales: Spatiotemporal Chaos},
  pdfauthor={L. Pinkney, A. M. Rucklidge, and C. Beaume} 
}
\fi

\begin{document}

\maketitle

\begin{abstract}
Three-wave interactions (or resonant triads) are the lowest-order nonlinear interaction in pattern formation and arise between waves with different orientations when the sum of two wavevectors equals a third one. 
When a pattern has only one length scale, stripe patterns are possible but three-wave interactions are responsible for the prevalence of hexagons close to onset.
In problems with two length scales, there is a much wider range of possible three-wave interactions, leading to more complex structures such as superhexagons, stars, quasipatterns and even spatiotemporal chaos. 
We investigate the role that nonlinear wave interactions play in the formation of spatiotemporal chaos in a model partial differential equation~\hbox{(PDE)} in the case that the length scale ratio is~$\sqrt{7}$, relevant to superlattice patterns in the Faraday wave experiment.
The simpler aspects of the dynamics can be represented by a system of ordinary differential equations (ODEs) derived from the PDE using weakly nonlinear theory.
We analyze the equilibrium patterns in these ODEs and evaluate their stability, comparing the results with direct numerical simulations of the model~\hbox{PDE}.
The ODEs predict parameter regimes where there are no stable simple equilibria, which is where we typically find complex behavior in the~\hbox{PDE}.
We have conducted a careful study of the transition from simple patterns (stripes and hexagons) to patterns that include modes beyond the finite-dimensional subspace imposed in the reduction to the ODEs, to time-dependent competition between different triads, ending up with fully developed spatiotemporal chaos.
For our choice of length scale ratio, we show that four-wave interactions also play an important role.
Our analysis is relevant to any pattern-forming system with three-wave interactions involving two length scales, such as the Faraday wave experiment, coupled reaction--diffusion systems, and pattern formation in dryland vegetation. 
\end{abstract}



\section{Introduction}
\label{introduction}
Patterns often arise in non-equilibrium systems, an example of which is Rayleigh--B\'{e}nard convection, where a fluid is confined between two horizontal plates maintained at different temperatures.
When the lower plate is the warmer, the fluid at the bottom is less dense than the fluid at the top and convection may arise, creating convection rolls, which appear as a horizontal pattern of stripes when viewed from above.
Stripes are the simplest patterns on the two-dimensional horizontal $(x,y)$ plane, as they depend only on one spatial coordinate~($x$) and they have only a single length-scale, the spacing from one stripe to the next.
This implies that their horizontal Fourier transform will be dominated by a single Fourier mode~$e^{ikx}$ (and its complex conjugate), with one wavevector $(k,0)$ and its negative~$(-k,0)$.

Other simple patterns, such as squares and hexagons, have also been observed in Rayleigh--B\'{e}nard convection~\cite{BESTEHORN1992,Cross1993,Bajaj1998} and in other fluid experiments, such as the Faraday wave experiment.
Square and hexagon patterns contain two and three wavevectors respectively (and their negatives), all with the same wavenumber.
In the Faraday wave experiment~\cite{Faraday1831,WESTRA2003}, a container with a thin layer of fluid is sinusoidally forced up and down, and if the forcing is strong enough, patterns of standing waves form on the fluid surface, leading to stripes, squares and hexagons~\cite{Fauve1992,Daudet1995,KUDROLLI1996,Arbell2002}.

\begin{figure}
\centering
\begin{tikzpicture}[scale=0.11]
\draw[->] (-18,0)--(18,0) node[right]{$k_x$};
\draw[->] (0,-20.5)--(0,20.5) node[above]{$k_y$};
\draw[ thick] (0,0) circle [radius=16.1];
\draw[ thick] (0,0) circle [radius =6];
\draw [very thick][->](0,0)--(6,0);
\draw[very thick][->](0,0)--(3,16);
\draw[very thick][->](0,0)--(3,-16);
\draw [dashed,very thick][->](0,0)--(-6,0);
\draw[dashed,very thick][->](0,0)--(-3,-16);
\draw[dashed,very thick][->](0,0)--(-3,16);
\node at (8.1,1.3){\small {$\boldsymbol{q_1}$}};
\node at (3.6,18){\small {$\boldsymbol{k_2}$}};
\node at (3.6,-18){\small {$\boldsymbol{k_1}$}};
\draw[black] (0,-9) arc [start angle=270, end angle=280, radius=9 ];
\draw[black] (0,-9) arc [start angle=270, end angle=260, radius=9 ];
\draw[black][-](0.5,-9.5)--(4.7,-12.5);
\node at (6,-13){ {\color{black}$\alpha$}};
\end{tikzpicture}
\caption{Schematic of three-wave interactions between wavevectors on two circles with radius~1 and $q<\frac{1}{2}$ (outer and inner circles respectively).
The wavevectors satisfy $\boldsymbol{k_1} + \boldsymbol{k_2} = \boldsymbol{q_1}$, where $|\boldsymbol{k_1}|=|\boldsymbol{k_2}|=1$ and $|\boldsymbol{q_1}|=q$.
Two of the shorter waves are not long enough to add up to one of the longer waves.
The angle~$\alpha$ is the acute angle between $\boldsymbol{k_1}$ and $\boldsymbol{k_2}$, with $q=2\sin(\alpha/2)$.}
\label{figure1a}
\end{figure}
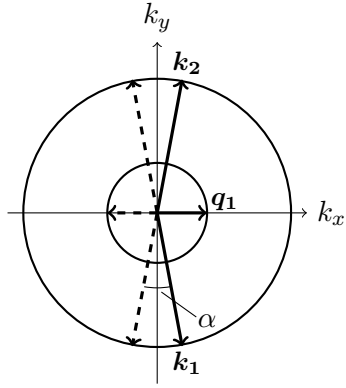

Later Faraday wave experiments introduced two-frequency forcing~\cite{Edwards1993}, which allows the possibility of two (or more) length-scales in the emergent pattern, and so can lead to rhombus patterns, superlattice patterns and quasipatterns~\cite{edwards1994,KUDROLLI1998,Arbell2002}.
These patterns are stabilized by the nonlinear interaction between waves with the two length-scales~\cite{Muller1993,Beyer1995,Silber1999,SILBER2000,PORTER2004,Ding2006,Rucklidge2009,Skeldon2015}, where by \emph{wave}, we mean a horizontal Fourier mode $e^{i\boldsymbol{k}\cdot\boldsymbol{x}}$ with a steady or time-dependent amplitude: these waves are the basic ingredients in the theory of pattern formation~\cite{Hoyle2006}.
Nonlinear three-wave interactions (3WIs, also known as triadic interactions) play an important role: when two waves with the same wavenumber have wavevectors that add up to a third wavevector with a different wavenumber, as in \cref{figure1a}, the presence of the first two waves in a pattern can influence the amplitude of the third.
The stabilization of complex patterns occurs when the nonlinear 3WIs act to reinforce the presence of all three waves in the pattern.
In contrast, when the 3WIs act so that the waves compete with each other, this can lead to time-dependent patterns and possibly spatiotemporal chaos (STC)~\cite{PORTER2004,Rucklidge2012,CASTELINO2020}.

Three-wave interactions can be investigated by considering the ordinary differential equations (ODEs) that govern the evolution of the small-amplitude waves.
The standing wave amplitude equations for the Faraday wave experiment can be computed in principle from the Navier--Stokes equations for free-surface fluid dynamics~\cite{Skeldon2007}.
For the two length-scale case, the ODE coefficients in these amplitude equations were derived in~\cite{PORTER2004} starting from the Zhang--Vi\~{n}als equations~\cite{zhang_vinals_1997,zhang_vinals_1997_2}, which are a set of quasi-potential equations modeling surface waves.

In this paper we investigate in detail how the nonlinear interaction between waves with two different wavenumbers can lead to \hbox{STC}.
We consider the case where the smaller wavenumber is less than half the larger (\cref{figure1a}) since, in this case, the 3WIs can only happen when two of the larger wavenumber waves add up to one of the smaller, which avoids the complications discussed in~\cite{Rucklidge2012}.
We primarily choose wavenumber ratio $q=1/\sqrt{7}\approx0.3780$, as this leads to the simplest of the superlattice patterns and is an example that is readily found in Faraday wave experiments~\cite{edwards1994,Besson1996,KUDROLLI1998,Arbell2002,Epstein2006}.
We write the pattern arising from the three waves (triad) in \cref{figure1a} as
 \begin{equation}
 u(x,y,t) = z_1(t) \expkone + z_2(t) \expktwo +w_1(t) \expq +c.c., \label{u:one_triad}
 \end{equation}
where $u$ represents the pattern (e.g., height of the fluid surface), and $z_1(t)$, $z_2(t)$ and $w_1(t)$ are the complex, time-dependent amplitudes of the three waves, and $c.c.$ denotes the complex conjugates.
The system of three complex ODEs for one triad can describe stripes and rhombs~\cite{PORTER2004}.
When hexagonal 3WIs, defined as 3WIs between waves separated by $120^\circ$ and with the same wavenumber, are included, the equations are extended to nine complex ODEs (six with one wavenumber and three with the other)~\cite{Subramanian2026}.
This allows more complex structures including hexagons, hexa-rolls~\cite{Iooss2022} and superlattice patterns.

Porter and Silber~\cite{PORTER2004} showed that the amplitude equations for a single triad had the possibility of Hopf bifurcations, traveling waves, structurally stable heteroclinic cycles and chaotic dynamics.
The temporal chaos in the amplitude equations still represents spatially ordered patterns, with only three wavevectors and their negatives.
However, temporal chaos within the three-mode ODEs and the availability of modes of all orientations in experiments done in large domains led to the conjecture that this combination could lead to spatiotemporal chaos~\cite{Rucklidge2012}.
Having modes of all orientations allows for \emph{competing triads}, by which we mean two triads that each have two $k=1$ modes and one $k=q$ mode, with one $k=1$ mode in common.
This enables modes outside of those originally considered to play a role in the dynamics.
The quadratic terms in the three-mode ODEs play an important role in the existence of time dependence: in particular, the coefficients of the quadratic terms must have different signs for Hopf bifurcations and chaotic dynamics to be possible~\cite{PORTER2004}.

Here, we test the hypothesis of~\cite{Rucklidge2012}, and show that in fact the situation is more subtle than the original conjecture, though having time-dependent competition between modes with $k=1$ and modes with $k=q$ still plays a central role in the development of spatiotemporal chaos.
We use an extension of the partial differential equation (PDE) model introduced in~\cite{Rucklidge2012} (based on an earlier model from~\cite{Lifshitz_Petrich_1997}).
The model has easily controllable growth rates at two wavenumbers, and we include here a wider range of nonlinear terms. 
We use weakly nonlinear theory to establish the relationship between the PDE parameters and the coefficients in complex amplitude equations for nine waves, six with wavenumber $k=1$ and three with wavenumber $k=q$. 
We compute eigenvalues of the Jacobian matrix to determine the stability of simple patterns in the amplitude equations and so to predict pattern selection within the~\hbox{PDE}. 
The presence of Hopf bifurcations in the amplitude equations indicates where time-dependent dynamics arise in the ODEs and guides our search for \hbox{STC} in the~\hbox{PDE}. 
We find good agreement between regions in the ODE parameter space where there are Hopf bifurcations and no stable simple equilibria, and regions in the PDE parameter space where there is \hbox{STC}.

The exact form of the nine complex amplitude equations (truncated at cubic order) depends on the wavenumber ratio.
The terms in the equations that are present for all wavenumber ratios are called \emph{generic} by~\cite{Subramanian2026} but, for $q=1/\sqrt{7}$, there are four-wave interactions (4WIs) that lead to additional \emph{non-generic} cubic terms.
The additional cubic terms introduce the possibility of further Hopf bifurcations and greatly extend the region in parameter space where we find time dependence in the ODEs and \hbox{STC} in the~\hbox{PDE}. 
To the best of our knowledge, no example of \hbox{STC} has previously been reported in this model PDE when the ratio of wavenumbers is less than~$\frac{1}{2}$.

Spatiotemporal chaos occurs, at least potentially, in other pattern formation problems with two length scales, including the Faraday wave experiment~\cite{KUDROLLI1996} and two-layer reaction--diffusion systems~\cite{Berenstein2004,CASTELINO2020,Fan2025}. 
We anticipate that the link between \hbox{STC} and time dependence in the underlying amplitude equations that we have found for our particular model PDE will extend to PDEs for these other cases, and indeed to other two length scale problems such as vegetation pattern formation~\cite{Bennett2023}.

The paper is organized as follows. 
We introduce the idea of 3WIs with two critical wavenumbers in \cref{amplitude eqs}. 
We consider the ODE systems of amplitude equations governing 3WIs for the cases of a single triad (three complex ODEs), as well as the nine complex ODEs that combine both rhombic and hexagonal 3WIs.
The general form of the ODEs differ in the case $q\neq1/\sqrt{7}$, where only 3WIs are present in the cubic truncation, and $q=1/\sqrt{7}$, where 4WIs lead to
additional cubic terms.
We also give some of the conditions for Hopf bifurcations in both cases.
In \cref{model PDE} we present a model PDE with two linearly unstable wavenumbers, an extension of the one investigated by~\cite{Rucklidge2012,Subramanian2026}, analyzing the linear and weakly nonlinear behavior. 
The full derivation of the weakly nonlinear approximation can be found in \hbox{\cref{app wnlt}}. 
\Cref{patternclassification} introduces the criteria we use to classify steady and time-dependent patterns in the~\hbox{PDE}, with further details in \cref{app:pattern classification}.
In \cref{numerical}, we present three sets of numerical results, investigating the roles of the 3WI Hopf and 4WI Hopf bifurcations on the generation of \hbox{STC}.
We make direct comparisons between the ODE predictions and the fully nonlinear PDE behavior.
Where the ODEs predict a parameter region with stable patterns, we find that the PDE has the same stable patterns for similar parameter values.
Where the ODEs predict a parameter region with no stable simple patterns, we find that the PDE has time-dependent solutions, sometimes involving the same nine modes as in the amplitude equations, but sometimes involving a wider range of modes, including the possibility of full spatiotemporal chaos and intermittent chaos. 
We test the hypothesis of~\cite{Rucklidge2012} most closely in \cref{onset of chaos}, where we show, for parameters close to a Hopf bifurcation, how the transition to chaos occurs due to the growth of modes driven by 3WIs, going beyond the nine modes. 
A summary of our findings and ideas for future work are in \cref{sec:conclusions}.

This paper is closely connected to~\cite{Subramanian2026}, which discusses in more detail the relation between the value of the wavenumber ratio and the selection of which Fourier modes to include in the weakly nonlinear theory, discusses various two-wavenumber PDE models, including the one used here, and addresses the challenge of finding all of the equilibria of the nine complex cubic amplitude equations and their stability.

\begin{figure}
\centering
\begin{tikzpicture}
\begin{axis} [width=0.6\textwidth,height=0.4\textwidth,
  axis lines=center,
  clip mode = individual,  
    xlabel={$k$},
  ylabel={$~~~\sigma$},
  xmin=0, xmax=1.1,
  ymin=-11.1, ymax=2.5,
  xticklabel=\empty,
  yticklabel=\empty,
x label style={at={(axis description cs:1.05,0.81)}, anchor=east},
y label style={at={(axis description cs:-0.03,.5)},rotate=90,anchor=south},]
\addplot [domain=0:1.1, smooth, color = cyan, thick] {    x^2*(0.5*(-x^2*(3-1/7)-2*1/7+4)*(1/7-x^2)^2*1/49+1*(-x^2*(1-3/7)+2/7-4/49)*(1-x^2)^2)/(1/49*(1-1/7)^3)  -6*49*(1-x^2)^2*(1/7-x^2)^2 };   
 \draw (0.3780, 0  ) node[below=1pt]{$q$};
 \draw (1,     0  ) node[below=1pt]{$1$};
 \draw (0.3780, 1  ) node[above=1pt]{$\nu$};
 \draw (1,     0.5) node[above=1pt]{$\mu$};
 \draw (0,      -6) node[left=1pt]{$\sigma_0$};
\end{axis}
\end{tikzpicture}
\caption{Growth rate~$\sigma$ as a function of the wavenumber~$k$. There are maxima at wavenumbers $k=q$ and $k=1$ with corresponding growth rates $\nu$ and $\mu$ respectively. The growth rate at $k=0$ is~$\sigma_0$.}
\label{fig:growthrate}
\end{figure}
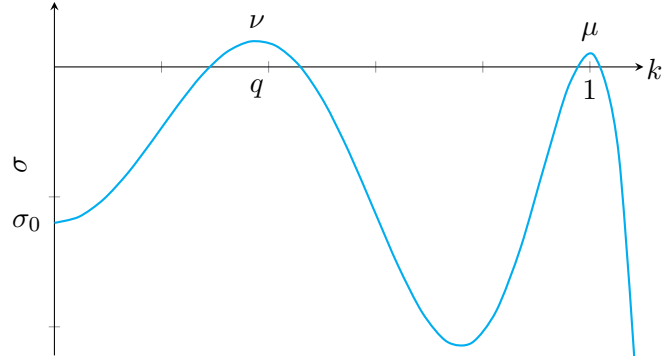

\section{Amplitude Equations} \label{amplitude eqs}
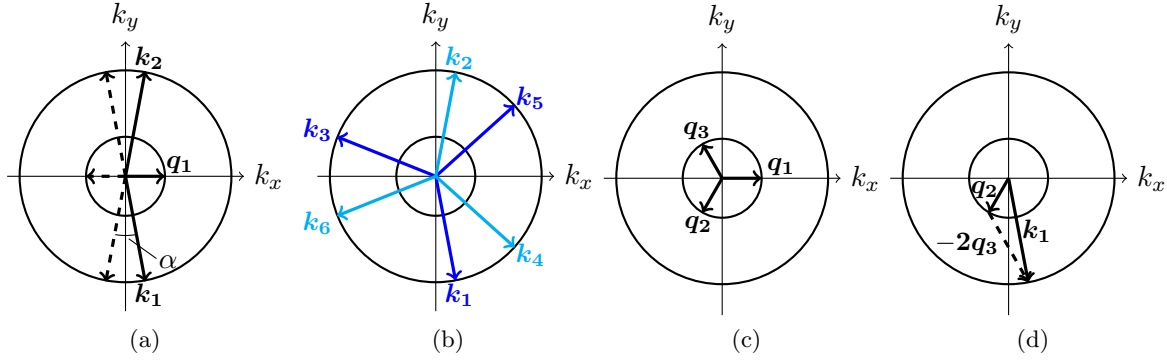
\begin{figure}
\centering
\subfloat[\label{figure:1a}]{
\begin{tikzpicture}[scale=0.087]
\draw[->] (-18,0)--(18,0) node[right]{$k_x$};
\draw[->] (0,-20.5)--(0,20.5) node[above]{$k_y$};
\draw[ thick] (0,0) circle [radius=16.1];
\draw[ thick] (0,0) circle [radius =6];
\draw [very thick][->](0,0)--(6,0);
\draw[very thick][->](0,0)--(3,16);
\draw[very thick][->](0,0)--(3,-16);
\draw [dashed,very thick][->](0,0)--(-6,0);
\draw[dashed,very thick][->](0,0)--(-3,-16);
\draw[dashed,very thick][->](0,0)--(-3,16);
\node at (8.3,1.7){\small {$\boldsymbol{q_1}$}};
\node at (3.6,18){\small {$\boldsymbol{k_2}$}};
\node at (3.6,-18){\small {$\boldsymbol{k_1}$}};
\draw[black] (0,-9) arc [start angle=270, end angle=280, radius=9 ];
\draw[black] (0,-9) arc [start angle=270, end angle=260, radius=9 ];
\draw[black][-](0.5,-9.5)--(4.7,-12.5);
\node at (6.5,-13){ {\color{black}$\alpha$}};
\end{tikzpicture}} 
\subfloat[\label{figure:1b}]{\begin{tikzpicture}[scale=0.087]
\draw[->] (-18,0)--(18,0) node[right]{$k_x$};
\draw[->] (0,-20.5)--(0,20.5) node[above]{$k_y$};
\draw[thick] (0,0) circle [radius=16.1];
\draw[thick] (0,0) circle [radius =6];
\draw[blue, very thick][->](0,0)--(3,-15.8);
\draw[blue, very thick][->](0,0)--(12,10.8);
\draw[blue,very thick][->](0,0)--(-15,6);
\draw[cyan,very thick][->](0,0)--(3,15.8);
\draw[cyan,very thick][->](0,0)--(12,-10.8);
\draw[cyan,very thick][->](0,0)--(-15,-6);
\node at (3.6,-18){\small {\color{blue}$\boldsymbol{k_1}$}};
\node at (14.4,12){\small {\color{blue}$\boldsymbol{k_5}$}};
\node at (-18,7.1){\small {\color{blue}$\boldsymbol{k_3}$}};
\node at (3.6,18){\small {\color{cyan}$\boldsymbol{k_2}$}};
\node at (14.4,-12){\small {\color{cyan}$\boldsymbol{k_4}$}};
\node at (-18,-7.1){\small {\color{cyan}$\boldsymbol{k_6}$}};
\end{tikzpicture}}
\subfloat[\label{figure:1c}]{\begin{tikzpicture}[scale=0.087]
\draw[->] (-18,0)--(18,0) node[right]{$k_x$};
\draw[->] (0,-20.8)--(0,20.5) node[above]{$k_y$};
\draw[thick] (0,0) circle [radius=16.1];
\draw[thick] (0,0) circle [radius =6];
\draw [very thick][->](0,0)--(6,0);
\draw [very thick][->](0,0)--(-3,5.2);
\draw [very thick][->](0,0)--(-3,-5.2);
\node at (8.3,1.7){\small { $\boldsymbol{q_1}$}};
\node at (-4.3,-7){\small { $\boldsymbol{q_2}$}};
\node at (-4.5,7.2){\small { $\boldsymbol{q_3}$}};
\end{tikzpicture}}
\subfloat[\label{figure:1d}]
{\begin{tikzpicture}[scale=0.087]
\draw[->] (-18,0)--(18,0) node[right]{$k_x$};
\draw[->] (0,-20.8)--(0,20.5) node[above]{$k_y$};
\draw[thick] (0,0) circle [radius=16.1];
\draw[thick] (0,0) circle [radius =6];
\draw [dashed, very thick][->](-3,-5.2)--(3,-15.8);
\draw [very thick][->](0,0)--(-3,-5.2);
\draw[very thick][->](0,0)--(3,-15.8);
\node at (-4.3,-1.8){\small { $\boldsymbol{q_2}$}};
\node at (-7,-10){\small { $\boldsymbol{-2q_3}$}};
\node at (4.3,-8){\small {$\boldsymbol{k_1}$}};
\end{tikzpicture}}
\caption{Schematic of nonlinear interactions between wavevectors on two critical circles with radius~1 and $q<\frac{1}{2}$ (outer and inner circles respectively).
(a)~The wavevectors satisfy $\boldsymbol{k_1} + \boldsymbol{k_2} = \boldsymbol{q_1}$. 
(b)~and (c) show the hexagonal 3WIs on the outer and inner circles respectively. 
(d)~Four-wave interactions satisfying $\boldsymbol{k_1} = \boldsymbol{q_2} - 2\boldsymbol{q_3}$ for $q=1/\sqrt{7}$.  }
\label{figure1}
\end{figure}

We consider pattern forming systems of the form
\begin{equation}
    \pdev{u}{t}  = \mathcal{L}u + \mathcal{N}(u), \label{generalPDE}
\end{equation}
where $u(x,y,t)$ represents the pattern, $\mathcal{L}$~is a linear partial differential operator on~$u$ and $\mathcal{N}(u)$~denotes the nonlinear terms. 
We consider a dispersion relation for a growth rate~$\sigma$ as a function of a wavenumber~$k$, so $\mathcal{L}e^{ikx}=\sigma(k)e^{ikx}$.
Since we are interested in the competition between two length scales, we want our dispersion relation to have maxima at two critical wavenumbers. 
Without loss of generality, we assume these wavenumbers to be $k=1$ and $k=q$ where $0<q<1$, as shown in \cref{fig:growthrate}.
The resulting nonlinearly interacting wavevectors thus satisfy the relation $\boldsymbol{q_1}=\boldsymbol{k_1} +\boldsymbol{k_2} $, as shown in  \cref{figure:1a}, where $|\boldsymbol{k_1}|=|\boldsymbol{k_2}|=1$ and $|\boldsymbol{q_1}|=q$. 

We are interested in the patterns formed as a result of 3WIs between waves with these two wavenumbers.
Close to onset, the pattern forming field $u(x,y,t)$ is given by
\begin{equation}
    u(x,y,t) = \sum_j z_j(t) e^{i\boldsymbol{k_j} \cdot \boldsymbol{x}} + \sum_j w_j(t) e^{i\boldsymbol{q_j} \cdot \boldsymbol{x}} + c.c., \label{u:2}
\end{equation}
where $z_j$ and $w_j$ are complex time-dependent amplitudes, $\boldsymbol{x}=(x,y)$, $|\boldsymbol{k_j}|=1$ and $|\boldsymbol{q_j}|=q$. 
The sums are taken over the number of modes that we choose to include.
Considering only one triad, $u$ takes the form of \cref{u:one_triad}, the first sum in \cref{u:2} has two terms and the second one term. 
The nonlinear interactions within this triad may be summarized by a system of ODEs governing the evolution of each amplitude. 
This system is invariant under the following transformations:
\begin{subequations}
\begin{align}
\kappa: & \ (z_1,z_2,w_1) \rightarrow (z_2,z_1,w_1), \label{ksym} \\
T_{\boldsymbol{\phi}}: & \ (z_1,z_2,w_1) \rightarrow (z_1 e^{i \phi_1}, z_2 e^{i \phi_2}, w_1 e^{i (\phi_1+\phi_2)}), \hspace{0.3cm}  \boldsymbol{\phi} = (\phi_1,\phi_2), \ \phi_1, \phi_2 \in [0,2\pi),
\label{Tsym}\\
R_{180} : & \ (z_1,z_2,w_1) \rightarrow \left(\bar{z}_1, \bar{z}_2, \bar{w}_1 \right), \label{rotsym}
\end{align} 
\end{subequations}
which are a reflection, a translation and the $180^\circ$ rotation respectively. 
These symmetries are used to construct the amplitude equations, which are defined (up to cubic order) as
\begin{subequations}
\begin{align}
\dot{z}_1 & = \mu z_1 +Q_{zw} \conj{z}_2w_1 +z_1 \left(A |z_1|^2 +B_{\alpha} |z_2|^2 +C_{90-\alpha/2} |w_1|^2\right), \label{z1dot:1} \\
\dot{z}_2 & = \mu z_2 +Q_{zw} \conj{z}_1w_1 + z_2 \left(B_{\alpha} |z_1|^2 +A |z_2|^2 +C_{90-\alpha/2} |w_1|^2\right), \label{z2dot:1}\\
\dot{w}_1 & = \nu w_1 +Q_{zz} z_1 z_2 + w_1\left(E_{90-\alpha/2} |z_1|^2 +E_{90-\alpha/2} |z_2|^2 +D |w_1|^2\right), \label{wdot:1}
\end{align}
\end{subequations}
where $\mu$ and $\nu$ are the growth rates corresponding to wavenumbers~1 and $q$ respectively, $Q_{zw}$ and $Q_{zz}$ are quadratic coefficients, and $A$, $B_{\alpha}$, $C_{90-\alpha/2}$, $D$ and $E_{90-\alpha/2}$ are cubic coefficients. 
These coefficients are all real due to the $180^{\circ}$ rotation symmetry and can be computed from the PDE using weakly nonlinear theory.
The subscripts of the cubic coefficients correspond to the (acute) angle between pairs of modes. 
For example, the modes for $z_1$ and $z_2$ are separated by an angle of $\alpha$ (see \cref{figure:1a}), where $\alpha$ is selected by the ratio of the critical wavenumbers: $\alpha = 2\sin^{-1}\frac{1}{2}q$. For our choice of $q=1/\sqrt{7}$, $\alpha \approx 22^\circ$.
The reflection symmetry~$\kappa$ allows for the $z_1$ and $z_2$ amplitudes to be interchanged, resulting in the same coefficients in \cref{z1dot:1} and \cref{z2dot:1}.
Since the amplitudes are complex, these equations are complemented with equations for the complex conjugates, bringing the total dimension of the system to six. 
We note that the working dimension of the system can be reduced to four via the introduction of an invariant phase~\cite{PORTER2004} but that this should be done with care: this phase becomes undefined when any of the amplitudes vanish.

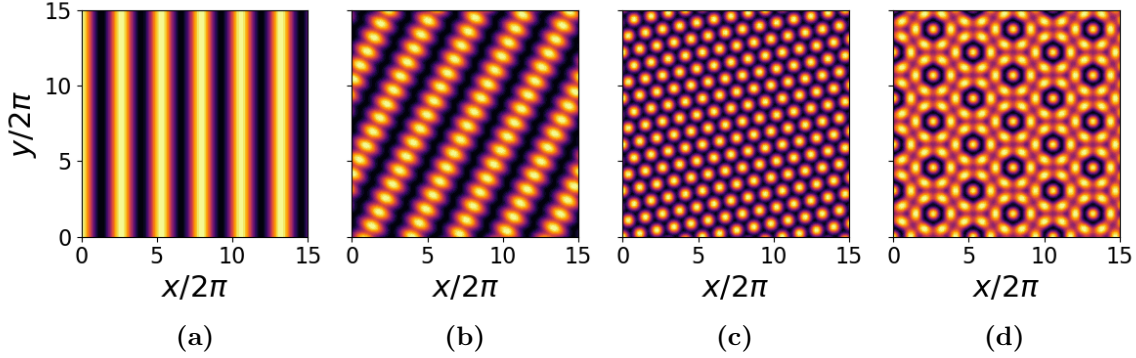
\begin{figure}
\centering
\begin{tikzpicture}
\node[anchor=south west,inner sep=0] (image) at (0,0) {\includegraphics[width=155mm]{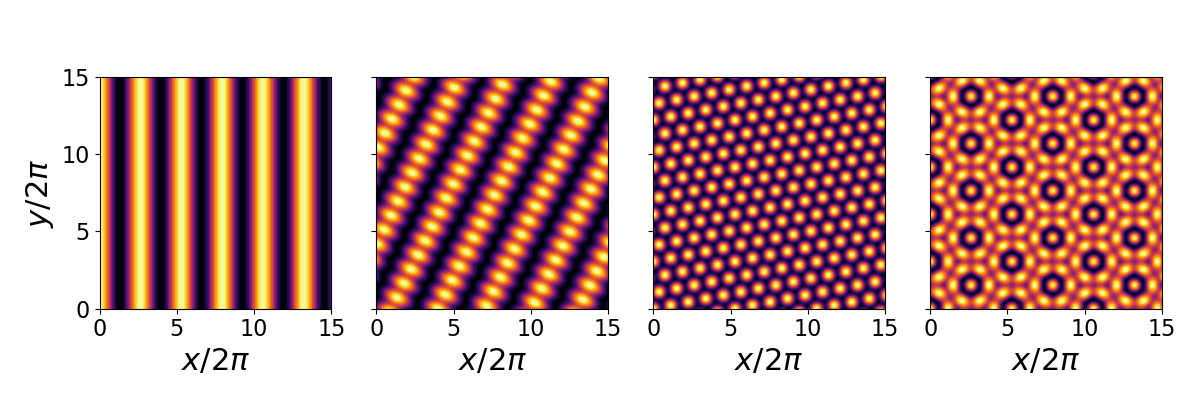}};
\begin{scope}[x={(image.south east)},y={(image.north west)}]
  \node[fill=none] at (0.18,-0.03) {\small \textbf{(a)}};
\end{scope}
\begin{scope}[x={(image.south east)},y={(image.north west)}]
  \node[fill=none] at (0.41,-0.03) {\small \textbf{(b)}};
\end{scope}
\begin{scope}[x={(image.south east)},y={(image.north west)}]
  \node[fill=none] at (0.64,-0.03) {\small \textbf{(c)}};
\end{scope}
\begin{scope}[x={(image.south east)},y={(image.north west)}]
  \node[fill=none] at (0.87,-0.03) {\small \textbf{(d)}};
\end{scope}
\end{tikzpicture}
\caption{Examples of simple patterns governed by 3WIs: (a)~$w$-stripes, (b)~rhombs, (c)~$z$-hexagons, (d)~superhexagons.}
\label{simple patterns} 
\end{figure}

The amplitude equations \cref{z1dot:1}--\cref{wdot:1} have been analyzed in depth by Porter and Silber~\cite{PORTER2004}, who observed that simple patterns such as $z$-stripes ($z_1\neq0$, $z_2=w_1=0$), $w$-stripes ($z_1=z_2=0$, $w_1\neq0$; \cref{simple patterns}{\color{siaminlinkcolor}a}) and rhombs ($|z_1|=|z_2|\neq 0$, $w_1\neq0$; \cref{simple patterns}{\color{siaminlinkcolor}b}) dominate.
Multiple examples of time-periodic solutions were also found, which bifurcate off the equilibrium branches via Hopf bifurcations in addition to heteroclinic cycles between two $w$-stripe solutions separated by a phase shift of~$\pi$. 
Porter and Silber also documented an example of a chaotic attractor. 

Quadratic terms are only present when we have 3WIs, and the sign of the product of their coefficients, $Q_{zz}Q_{zw}$, heavily influences the possible states this system exhibits~\cite{PORTER2004,Rucklidge2012}. 
When $Q_{zz}Q_{zw}>0$, the subspace $(z_1,z_2,w_1)\in\mathbb{R}^3$ is attracting~\cite{GUCKENHEIMER1992}, there is no persistent time dependence, and there are no Hopf bifurcations~\cite{PORTER2004}.
On the other hand, when $Q_{zz}Q_{zw}<0$, time-dependent solutions are possible as a consequence of Hopf bifurcations. 
For example, $z$-stripes and rhombs can both undergo Hopf bifurcations when $Q_{zz}Q_{zw}<0$~\cite{Proctor_Jones_1988,PORTER2004}. 

Triadic interactions can also form between three waves separated by $120^\circ$, each with the same wavenumber, as seen in \cref{figure:1b} and \cref{figure:1c}. 
The amplitude equations governing this type of 3WI are
\begin{subequations}
\begin{align}
\dot{z}_1 & = \mu z_1 +Q_{zhex} \conj{z}_3\conj{z}_5 +z_1 \left(A |z_1|^2 +B_{60} |z_3|^2 +B_{60} |z_5|^2\right), \label{z1dotsmallhex:1} \\
\dot{z}_3 & = \mu z_3 +Q_{zhex} \conj{z}_1\conj{z}_5 +z_3 \left(B_{60} |z_1|^2 +A |z_3|^2 +B_{60} |z_5|^2\right),\label{z3dotsmallhex:1}\\
\dot{z}_5 & = \mu z_5 +Q_{zhex} \conj{z}_1\conj{z}_3 +z_5 \left(B_{60} |z_1|^2 +B_{60} |z_3|^2 +A |z_5|^2\right), \label{z5dotsmallhex:1}
\end{align}
\end{subequations}
where $Q_{zhex}$ is the coefficient of the quadratic term arising from the hexagonal 3WI, $A$ is the self-interaction coefficient and $B_{60}$ is the coefficient for cubic coupling between modes separated by $60^\circ$. 
This system respects the symmetry group $\mathcal{D}_6 \ltimes T^2$~\cite{Hoyle2006}, which acts on the amplitudes as
\begin{subequations}
    \begin{align}
        \kappa_{hex} :& \ (z_1,z_3,z_5) \rightarrow (z_3,z_1,z_5), \label{ksym:hex} \\
    T_{\boldsymbol{\phi}}: & \ (z_1,z_3,z_5) \rightarrow (z_1 e^{i \phi_1}, z_3 e^{i \phi_3}, z_5 e^{-i(\phi_1+\phi_3)}), \hspace{0.3cm}  \boldsymbol{\phi} = (\phi_1,\phi_3), \ \phi_1, \phi_3 \in [0,2\pi), \label{Tsym2:hex}\\
R_{60}: & \ (z_1,z_3,z_5) \rightarrow (\bar{z}_3, \bar{z}_5,\bar{z}_1), \label{Rsym2:hex} 
    \end{align}
\end{subequations}
where we have defined the $60^\circ$ rotational symmetry $R_{60}$ in the anticlockwise direction.
Due to this rotation symmetry, only one quadratic coefficient appears in the system \cref{z1dotsmallhex:1}--\cref{z5dotsmallhex:1}.
All coefficients of the system are real due to the symmetry $R_{180} = (R_{60})^3$.

Like for the rhombic triad case \cref{z1dot:1}--\cref{wdot:1}, stripes solutions have only one non-zero amplitude in \cref{z1dotsmallhex:1}--\cref{z5dotsmallhex:1}, and hexagons have three amplitudes equal in magnitude (\cref{simple patterns}{\color{siaminlinkcolor}c}). 
Since the system only has one quadratic coefficient, $Q_{zhex}$, no persistent time-dependent dynamics are possible.

Expanding on these ideas, we consider waves influenced by both rhombic and hexagonal triadic interactions.
This results in having six wavevectors (and their negatives) on one circle and three on the other~\cite{Subramanian2026}, combining \cref{figure:1a,figure:1b,figure:1c}. 
We refer to patterns involving all of these waves as superlattice patterns; one example can be seen in \cref{simple patterns}{\color{siaminlinkcolor}d}. 
The nine wavevectors are defined by the relations
\begin{subequations}
\begin{align}
    \boldsymbol{q_1} & = \boldsymbol{k_1} + \boldsymbol{k_2}, & 
    \boldsymbol{q_2} & = \boldsymbol{k_3} + \boldsymbol{k_4}, &
    \boldsymbol{q_3} & = \boldsymbol{k_5} + \boldsymbol{k_6}, \label{rhombicrel} \\  
    \boldsymbol{q_1}& + \boldsymbol{q_2} + \boldsymbol{q_3} = 0, &
    \boldsymbol{k_1}& + \boldsymbol{k_3} + \boldsymbol{k_5} = 0, &
    \boldsymbol{k_2}& + \boldsymbol{k_4} + \boldsymbol{k_6} = 0, \label{hexrel}
\end{align}
\end{subequations}
where the first row shows the rhombic relations, and the second the hexagonal relations.

The 3WIs in \cref{rhombicrel,hexrel} do not require that the resulting pattern be periodic~\cite{Iooss2022,Subramanian2026}.
However, restricting the value of~$q$ can ensure that all of the wavevectors lie exactly on a hexagonal lattice~\cite{Dionne_1992,Dionne_1997,Iooss2022}. 
To do this, we write the wavevectors on the $k=1$ circle as linear combinations of two hexagonal basis vectors $\eone$ and~$\etwo$, where $\eone$ points in the positive $k_x$ direction and $\etwo$ is angled $120^\circ$ (anticlockwise) from $\eone$~\cite{Dionne_1992}:
    \begin{align}
    \boldsymbol{k_1} & = -(a-b)\eone -a \etwo , \indent & \boldsymbol{k_2}  & =  b \eone +a \etwo, \nonumber  \\ 
    \boldsymbol{k_3}  & = -b \eone +(a-b) \etwo , \indent & \boldsymbol{k_4}  & = (a - b) \eone -b \etwo , \label{hexagonal_lattice}\\
    \boldsymbol{k_5}  & = a \eone +b \etwo, \indent & \boldsymbol{k_6}  & =  -a \eone -(a-b) \etwo, \nonumber
\end{align}
for $(a, b) \in \mathbb{Z}^2$, with $a>b>a/2>0$, $a$ and $b$ co-prime and $a+b$ not a multiple of three.
To ensure the length of these vectors is 1, we set
\begin{equation}
    \eone = \frac{1}{\sqrt{a^2-ab+b^2}}(1,0)  \indent \textrm{and} \indent
    \etwo = \frac{1}{\sqrt{a^2-ab+b^2}}\left(-\frac{1}{2}, \frac{\sqrt{3}}{2}\right). \label{basisvectors}
\end{equation}
As explained by~\cite{Subramanian2026} there are two choices for our wavevectors on the $k=q$ circle:
\begin{align}
    \begin{split}   
    \boldsymbol{q_1} &= \boldsymbol{k_1} + \boldsymbol{k_2} \\
    &= (2b-a)\eone, 
    \end{split}     &\textrm{with} \quad 
    q  = \frac{2b-a}{\sqrt{a^2-ab+b^2}}.  \label{qlength:a}\\
    \addlinespace
    \begin{split}
    \boldsymbol{q_1} &= \boldsymbol{k_1} - \boldsymbol{k_4} \\
    &= (b-a)\left(2\eone + \etwo \right),
    \end{split}  &\textrm{with} \quad 
    q  = \frac{\sqrt{3}\left(a-b\right)}{\sqrt{a^2-ab+b^2}}. \label{qlength:b} 
\end{align}
The two cases are equivalent after relabeling, if required.
The wavevectors $\boldsymbol{q_2}$ and $\boldsymbol{q_3}$ are also linear combinations of $\boldsymbol{s_1}$ and $\boldsymbol{s_2}$ and can be expressed similarly. 
The smallest pair $(a,b)=(3,2)$ corresponds to $q=1/\sqrt{7}$, computed using \cref{qlength:a}. 
If the PDE is solved on the periodic domain associated with this hexagonal lattice, or if this periodicity appears in a PDE solution in a larger domain, its small-amplitude dynamics will be described by the amplitudes of the eighteen modes with wavevectors $\boldsymbol{k_1}$, \dots, $\boldsymbol{q_3}$ and their negatives.
We refer to this as the \emph{eighteen-mode subspace} of the problem.

Four wave (and higher) interactions occur when a larger combination of the eighteen wavevectors adds up to zero.
We refer to the number of waves involved as the \emph{order} of the interaction, and interactions of a given order lead to terms of total degree one less than the order in the amplitude equations (3WIs lead to quadratic terms, etc.).
Some higher-order interactions are implied by \cref{rhombicrel,hexrel} (for example $\boldsymbol{q_1}=\boldsymbol{k_1}-\boldsymbol{k_4}-\boldsymbol{k_6}$). 
These generic interactions do not require a hexagonal lattice.
Other non-generic interactions only appear when all eighteen wavevectors lie exactly on a hexagonal lattice, using~\cref{hexagonal_lattice}.
For example, in the case of $q=1/\sqrt{7}$ (determined from \cref{qlength:a}) we have $\boldsymbol{k_1} = \boldsymbol{q_2} - 2\boldsymbol{q_3}$ (\cref{figure:1d}), which is a four-wave interaction (\hbox{4WI}). 

In general, considering all possible combinations of three different wavevectors, the lowest order non-generic interactions are, for any $(a,b)$ pair,
\begin{align}
    (2b-a) \boldsymbol{k_1} & = (a-b)\boldsymbol{q_2} -b\boldsymbol{q_3} & \textrm{for $q$ given by \cref{qlength:a},} & \indent \textrm{order} = 2b, \label{hot:a} \\
    \addlinespace
    (a-b)\boldsymbol{k_1} &=  (a-b)\boldsymbol{k_6} + b \boldsymbol{q_3} &\textrm{for $q$ given by \cref{qlength:b},} & \indent \textrm{order}=2a-b \label{hot:b},
\end{align}
where the order is determined by computing the absolute sum of the coefficients of the wavevectors. 
The lowest order of these interactions occurs for $(a,b)=(3,2)$ ($q=1/\sqrt{7}$); this is the only value of~$q$ resulting in non-generic 4WIs between the eighteen wavevectors. 

Expanding \cref{z1dot:1}--\cref{wdot:1} to include both the rhombic and hexagonal interactions, we obtain amplitude equations for $(z_1,\dots,z_6,w_1,w_2,w_3)\in\mathbb{C}^9$. Two of these equations, truncated at cubic order, are
\begin{subequations}
\begin{align}
\begin{split}
    \dev{z_1}{t}  =& \mu z_1 + Q_{zw}\conj{z}_2w_1 +Q_{zhex}\conj{z}_3\conj{z}_5 \\
    \begin{split}
    &{} + z_1 \Big(A |z_1|^2 + B_{\alpha} |z_2|^2 + B_{60} |z_3|^2 + B_{60-\alpha} |z_4|^2 + B_{60}|z_5|^2 \\
    & \hspace{1cm} {}+ B_{60+\alpha} |z_6|^2 + C_{90-\alpha/2} |w_1|^2 + C_{30+\alpha/2} |w_2|^2 + C_{30-\alpha/2}|w_3|^2 \Big)
    \end{split} \\
    & {} + K_1 z_4 z_6 w_1 + K_2 \conj{z}_2 \conj{w}_2 \conj{w}_3 + K_3 z_4 \conj{z}_5 \conj{w}_2 + K_4 \conj{z}_3 z_6 \conj{w}_3 \\
    &\color{red} {} + K_{zww} z_4 \conj{w}_1 w_2 + K_{ww} w_2 \conj{w}_3^2,
\end{split} \label{z1hex:1}\\
\begin{split}
    \dev{w_1}{t}  =& \nu w_1 + Q_{zz}z_1 z_2 +Q_{whex}\conj{w}_2\conj{w}_3 \\
    \begin{split}
    &{}+ w_1 \Big(E_{90-\alpha/2} |z_1|^2 + E_{90-\alpha/2} |z_2|^2 + E_{30-\alpha/2} |z_3|^2 + E_{30+\alpha/2} |z_4|^2  \\
    & \hspace{1cm} {} + E_{30+\alpha/2}|z_5|^2 + E_{30-\alpha/2} |z_6|^2 + D |w_1|^2 + F_{60} |w_2|^2 + F_{60}|w_3|^2 \Big)
    \end{split} \\
    & {} + L_{hex} z_2 \conj{z}_3 \conj{z}_5 + L_{hex} z_1 \conj{z}_4 \conj{z}_6  + L_1 \conj{z}_3 \conj{z}_4  \conj{w}_3 + L_1 \conj{z}_5 \conj{z}_6 \conj{w}_2 \\
    & \color{red} {} + L_{wzz} \conj{z}_1 z_4 w_2 + L_{wzz} \conj{z}_2 z_5 w_3 + L_{wwz} \conj{z}_3 \conj{w}_1 w_3 + L_{wwz} \conj{z}_6 \conj{w}_1 w_2 \\
    &\color{red} {} + L_{wz} z_4 w_3^2 + L_{wz} z_5 w_2^2,
\end{split} \label{w1hex:1}
\end{align}
\end{subequations}
where the red terms (final line of \cref{z1hex:1} and final two lines of \cref{w1hex:1}) arise from the 4WIs. 
These terms are only present when $q=1/\sqrt{7}$: other values of~$q$ produce higher order non-generic interactions, leading to red terms that do not appear in the amplitude equations truncated at cubic order.
As before, the subscripts of $B$, $C$, $E$ and $F$ denote the angle between wavevectors: for $q=1/\sqrt{7}$, $\alpha \approx 22^\circ$. 

The remaining amplitude equations can be written similarly, by using the following transformations:
\begin{subequations}
\begin{align}
    \kappa :& (z_1, \dots, z_6, w_1,w_2,w_3) \mapsto 
    (z_2,z_1,z_6,z_5,z_4,z_3,w_1,w_3,w_2), \label{ksym:2} \\
    \addlinespace
        R_{60} :& (z_1, \dots, z_6,w_1,w_2,w_3) \mapsto (\conj{z}_3,\conj{z}_4,\conj{z}_5,\conj{z}_6,\conj{z}_1,\conj{z}_2,\conj{w}_2,\conj{w}_3,\conj{w}_1). \label{Rsym:2}
\end{align}
A translation symmetry similar to \cref{Tsym} can also be derived:
\begin{align}
    \begin{split}
    T_{\boldsymbol{\phi}}: & (z_1, \dots,w_3) \mapsto (z_1 e^{i( (-a+b)\phi_1 -a \phi_2)},z_2 e^{i (b \phi_1 + a \phi_2)},z_3 e^{i( -b \phi_1 + (a-b)\phi_2)}, \\ & \hspace{2.7cm} z_4 e^{i ((a-b)\phi_1 -b \phi_2)}, z_5 e^{i( a \phi_1 +b \phi_2)},z_6 e^{i (-a \phi_1 - (a-b)\phi_2)},  \\ & \hspace{2.7cm}w_1 e^{i(-a + 2b)\phi_1},w_2 e^{i((a-2b)\phi_1 + (a-2b)\phi_2)},w_3 e^{i(2b-a) \phi_2}), \\  & \hspace{3.1cm}\boldsymbol{\phi} = (\phi_1,\phi_2), \ \phi_1, \phi_2 \in [0,2\pi). \label{Tsym:2}
    \end{split}
\end{align}
\end{subequations}
The full system of equations is displayed in \hbox{\cref{app wnlt}}. 

\begin{table}
\centering
\setlength{\leftmargin}{0.35cm}
\begin{tabular}{|c|c|>{\centering\arraybackslash}p{6.5cm}|c|}
\hline
Name  & Number of Peaks & $(z_1,z_2,z_3,z_4,z_5,z_6,w_1,w_2,w_3)$ & \cref{fig:example patterns} \\
\hline \hline
$z$-stripes &\makecell{  $P_1 = 2$, $P_q=0$} & $(z,0,0,0,0,0,0,0,0), \ z \in \mathbb{R} $ & -
\\
\rule{0pt}{12pt}
$w$-stripes & \makecell{$P_1 = 0$, $P_q=2$} & $(0,0,0,0,0,0,w,0,0), \ w \in \mathbb{R} $ & (a)
\\
\rule{0pt}{12pt}
$z$-hexagons & \makecell{$P_1 = 6$, $P_q=0$} &
$(z,0,z,0,z,0,0,0,0), \ z \in \mathbb{R} $ & (b)
\\
\rule{0pt}{12pt}
\makecell{$w$-hexagons} & \makecell{$P_1 = 0$, $P_q=6$} & 
$(0,0,0,0,0,0,w,w,w), \ w \in \mathbb{R} $ & (c) \\
\rule{0pt}{12pt}
rhombs & $P_1 = 4$, $P_q = 2$
& $(z_1,z_2,0,0,0,0,w_1,0,0), \ z_1,z_2,w_1 \in \mathbb{R} $ & (d)\\
\rule{0pt}{12pt}
superhexagons & $P_1 = 12$, $P_q = 6$
& 
$(z,z,z,z,z,z,w,w,w), \ z,w \in \mathbb{R} $ & (e)\\
\rule{0pt}{12pt}
stars & $P_1=12$, $P_q=6$ & 
\makecell{$(z_1,z_2,z_1,z_2,z_1,z_2,w_1,w_1,w_1)$,} \ $z_1,z_2,w_1 \in \mathbb{R}$ & (f) \\
\hline
\end{tabular}
\caption{Definitions of simple equilibrium patterns.
The pattern $w$-hexagons include off-critical $w$-hexagons, and stars are equivalent to asymmetric superhexagons, and have $\mathrm{sign}(z_1) \neq \mathrm{sign}(z_2)$.
We define $P_1$ ($P_q)$ as the number of peaks in the $k=1$ ($k=q$) annulus in its Fourier spectrum: for details, see \cref{patternclassification}.}
\label{tab:equilibriacriteria}
\end{table} 

This system of ODEs can produce a large number of equilibria; however, only the simplest of these are typically stable~\cite{Subramanian2026}.
In our model PDE (\cref{model PDE}), we are interested in comparing the stability regions of patterns, computed both using the ODE amplitude equations and found as PDE solutions. 
Therefore, for our ODE analysis we will focus only on the simple equilibrium patterns that we have found as stable solutions in the~\hbox{PDE}.
These patterns are: stripes, hexagons, rhombs, superhexagons and asymmetric superhexagons (see \cref{simple patterns} and \cref{fig:example patterns} below). 
We perform a linear stability analysis on \cref{z1hex:1}--\cref{w1hex:1} (and the additional associated amplitude equations) for this selection of simple equilibria by computing the $18 \times 18$ Jacobian matrix and determining its eigenvalues numerically. 
For stripes and hexagons, we repeat this for both wavelengths $k=1$ and $k=q$, since different wavenumbers have different stability criteria. 
When discussing these patterns, we differentiate between these two wavenumber cases by using a prefix of either $z$- or $w$- before the type of pattern. 
The rhombic solutions have three non-zero amplitudes: two equal $z$-amplitudes and the corresponding $w$-amplitude generated from the sum of the first two wavevectors.
Superhexagons have all $z$-amplitudes equal and all $w$-amplitudes equal. 
The final pattern is asymmetric superhexagons, which bifurcate off the superhexagon branch, breaking the $60^\circ$~rotational symmetry and resulting in $z_1\neq z_2$ where $z_1$ and $z_2$ are the amplitudes of each hexagonal sub-lattice on the $k=1$ circle. 
A summary of the simple patterns we are considering and the number of non-zero amplitudes is given in \cref{tab:equilibriacriteria}. 
Examples of each of these patterns found in the model PDE introduced below, and their Fourier spectra, are shown in \cref{fig:example patterns}.
\begin{figure}
    \centering
    \includegraphics[width=0.7\linewidth]{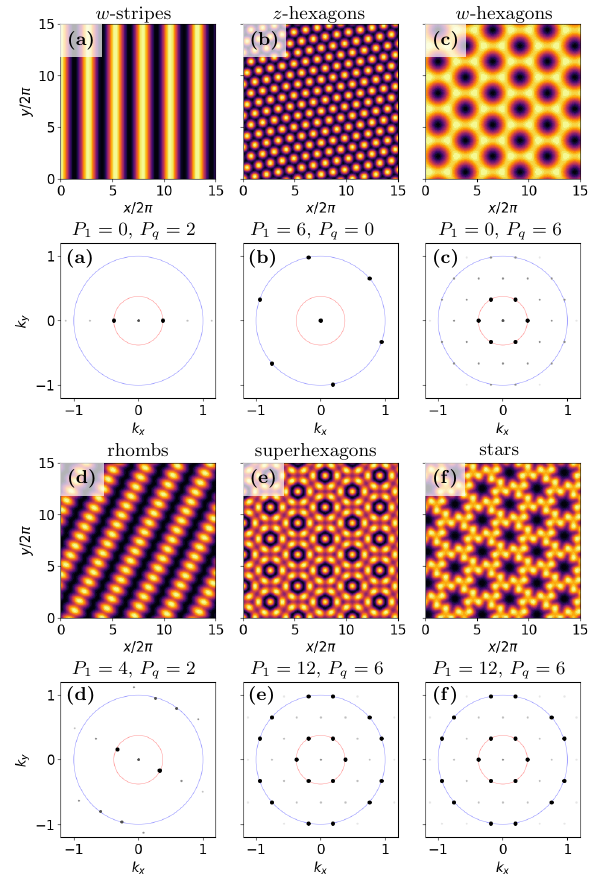}
     \caption{Solutions of the PDE \cref{PDE:2} showing examples of simple equilibrium patterns
     and their Fourier spectra (omitting $z$-stripes).
     We show a $15 \times 15$ section of each solution computed in an approximately $42 \times 42$ domain. 
     (a,b,c,f) are for $q=1/\sqrt{7}$, $\sigma_0 =-2$, $Q_1 = -0.9$, $Q_2=-2.75$, $Q_3 = -3.5$, $C_1 = -2.75$, $C_2 = -7.75$ and $C_3 = -16.5$.
     (d)~has $Q_1 = -0.7$ with all other values the same.
     The values used in (e) are $Q_1 =-1.24$, $Q_2=-2$, $Q_3=-0.8$, $C_1=-1$, $C_2=-5$ and $C_3=-15$ with the same values of $q$ and $\sigma_0$.
     The values of $\mu$ and $\nu$ are given in terms of $(r,\chi)$, where $(\mu,\nu) = (r\cos \chi,r\sin \chi)$:
     (a)~$(0.5,106^\circ)$, (b)~$(0.3,74^\circ)$, (c)~$(0.325,102^\circ)$, (d)~$(0.45,110^\circ )$, (e)~$(0.025,115^\circ)$ and (f)~$(0.025,175^\circ)$. }
    \label{fig:example patterns}
\end{figure}

We use the visual appearance of the patterns to guide nomenclature.
For example, the $w$-hexagons in \cref{fig:example patterns}{\color{siaminlinkcolor}c} have small but non-zero $z$ amplitudes, so they are technically superhexagons, but we will refer to them as $w$-hexagons nonetheless.
Asymmetric superhexagons have a different appearance depending on the relative signs of $z_1$ and $z_2$ (assuming real amplitudes)~\cite{Subramanian2026}.
When $z_1$ and $z_2$ have the same sign, asymmetric superhexagons interpolate between symmetric superhexagons and $z$-hexagons.
On the other hand, when $z_1$ and $z_2$ have opposite sign, as in  \cref{fig:example patterns}{\color{siaminlinkcolor}f}, asymmetric superhexagons look like ``stars''.
We only found asymmetric superhexagons to be stable in this second case, so we refer to these solutions as stars throughout the paper.

We also find PDE solutions that are generated by six equal-amplitude modes close to (but not on) the $k=q$ circle, without any modes on the $k=1$ circle.
We refer to these as ``off-critical $w$-hexagons'', and they are found when $\nu>0$, when there is a small band of unstable wavenumbers close to $k=q$.
We therefore distinguish between superhexagons, with six equal non-zero $z$~amplitudes and three equal non-zero $w$~amplitudes, and off-critical $w$-hexagons, with three equal non-zero $w$~amplitudes and the $z$~amplitudes equal to zero.
These are present as distinct equilibria in the amplitude equations truncated at cubic order, provided the red terms in \cref{z1hex:1}--\cref{w1hex:1} are absent (i.e., $q\neq1/\sqrt{7}$): it is the $K_{ww}$ term from the 4WIs in the $z$-amplitude equations that forces $w$-hexagons to have non-zero $z$-amplitudes.
We include off-critical $w$-hexagons in the $q=1/\sqrt{7}$ case (and drop the words ``off-critical'') by setting the red terms to zero, without changing the values of the other coefficients since these do not change much with a slightly different wavenumber.

A discussion of the full range of equilibrium solutions of the amplitude equations is in~\cite{Subramanian2026}, but most of these are unstable, so we compute only the amplitudes of the simple patterns, using the information in \cref{tab:equilibriacriteria} to write cubic polynomials for the $z$ and $w$~amplitudes.
Stripes and hexagons are straightforward, and rhombs involve solving a single cubic polynomial numerically.
Superhexagons and stars involve solving two and three (respectively) coupled cubic polynomials. We use Paramotopy~\cite{Paramotopy}, an extension of the software Bertini~\cite{Bertini}, to solve these coupled polynomials.
The solutions are then substituted into the Jacobian to determine their stability.

As described in \cref{introduction}, we hope to find spatiotemporal chaos in the PDE when the ODE amplitude equations have Hopf bifurcations leading to time dependence and no stable simple equilibria.
Hopf bifurcations from $z$-stripes are found in \cref{z1dot:1,z2dot:1,wdot:1} provided that $Q_{zz}Q_{zw}<0$~\cite{PORTER2004}.
In the nine complex amplitude equations \cref{z1hex:1,w1hex:1}, $w$-stripe equilibria can undergo a Hopf bifurcation independent of the sign of $Q_{zz}Q_{zw}$ provided that $q=1/\sqrt{7}$ and $K_{ww}L_{wz}<0$~\cite{Subramanian2026}, so this $w$-stripe Hopf bifurcation is a consequence of the 4WIs.
Other Hopf bifurcations are possible: for example, superhexagons and stars can undergo Hopf bifurcations but as we have not found a simple bifurcation criterion, we cannot rule these Hopf bifurcations out when $Q_{zz}Q_{zw}>0$ and $K_{ww}L_{wz}>0$.
Nonetheless, in our numerical results we only find evidence of time dependent dynamics when at least one of the quantities $Q_{zz}Q_{zw}$ or $K_{ww}L_{wz}$ is negative.

\section{Model PDE and Weakly Nonlinear Analysis} \label{model PDE}
We consider the following PDE: 
\begin{equation}
    \pdev{u}{t} = \mathcal{L}u + Q_1u^2 +Q_2 u \nabla^2 u +Q_3 |\boldsymbol{\nabla} u|^2 +C_1 u^3 +C_2 u^2 \nabla^2 u + C_3 u|\boldsymbol{\nabla} u|^2, \label{PDE:2}
\end{equation}
where the linear operator $\mathcal{L}$ is defined below in terms of the relationship between the wavenumber $k$ and the linear growth rate $\sigma$:
\begin{subequations}
\begin{equation}
    \sigma = \frac{k^2\left[A(k)\mu +B(k)\nu\right]}{q^4\left(1-q^2\right)^3} + \frac{\sigma_0}{q^4}\left(1-k^2\right)^2\left(q^2-k^2\right)^2, \label{growthrate:1}
\end{equation}
where
\begin{align}
    A(k)& = \left(k^2\left(q^2-3\right)-2q^2+4\right)\left(q^2-k^2\right)^2q^4, \label{A} \\
    B(k)& = \left(k^2\left(3q^2-1\right)+2q^2-4q^4\right)\left(1-k^2\right)^2. \label{B}
\end{align}
\end{subequations}
The linear operator $\mathcal{L}$ is obtained from \cref{growthrate:1} by replacing $k^2$ with $-\nabla^2$. This is the same linear operator considered by~\cite{Rucklidge2012,Subramanian2026}. 
Expression~\cref{growthrate:1} is similar to the expression for the linear growth rate found in the Swift--Hohenberg~\cite{Swift1977} and the Lifshitz--Petrich~\cite{Lifshitz_Petrich_1997} equations.
The Lifshitz--Petrich operator was extended by~\cite{Rucklidge2012} to allow for the growth rates of the critical wavenumbers to be controlled independently.
We have set $\sigma(1) = \mu$ and $\sigma(q)=\nu$ so that our modes have the same growth rates as in the ODE systems \cref{z1dot:1}--\cref{wdot:1} and \cref{z1hex:1}--\cref{w1hex:1}. 
We are free to control~$\sigma_0$, which is the growth rate of the $k=0$ mode.
Making $\sigma_0$ more negative narrows the band of unstable wavenumbers, reducing the influence of off-critical wavenumbers contributing to the solution, which is helpful to eliminate defects.
\Cref{fig:growthrate} shows a typical example of~$\sigma(k)$.

Nonlinearity in \cref{PDE:2} has been retained up to cubic order as higher order terms do not contribute to our truncated ODE system \cref{z1hex:1}--\cref{w1hex:1}.
We only include terms that preserve the $E(2)$ (Euclidean group) symmetries of the plane: translation, rotation and reflection, restricting to terms with spatial derivatives no larger than second order.  
The terms $u\nabla^2u$, $|\boldsymbol{\nabla}u|^2$, $u^2\nabla^2u$ and $u|\boldsymbol{\nabla}u|^2$ represent an extension to the Lifshitz--Petrich equation: they break the variational structure of the PDE when $Q_2 \neq 2Q_3$ or $C_2 \neq C_3$, which allows time-dependent solutions. 
The terms $u^2\nabla^2 u$ and $u|\boldsymbol{\nabla} u|^2$ extend the model investigated by~\cite{Rucklidge2012}. 

We proceed with a weakly nonlinear analysis  about the base state $u=0$, and calculate the coefficients in \cref{z1hex:1}--\cref{w1hex:1} as functions of the PDE parameters \cref{PDE:2}.  
The details of the derivation and the full list of expressions of the ODE coefficients can be found in \hbox{\cref{app wnlt}}. 
The expressions for the coefficients relevant to the Hopf bifurcations from $z$-stripes and (only in the case $q=1/\sqrt{7}$) $w$-stripes are
{\allowdisplaybreaks
\begin{subequations}
    \begin{align}
        Q_{zw} =& \ 2Q_1 - Q_2\left(1+q^2\right) + q^2Q_3, \label{Qzw:1} \\
        \addlinespace
    Q_{zz} =& \ 2Q_1 - 2Q_2 + Q_3 \left(2-q^2\right), \label{Qzz:1} \\
    \addlinespace
    \begin{split}
    \color{red} K_{ww} =  & \ \color{red} \frac{-22148 Q_1^2 + 9940 Q_1 Q_2 + 6328 Q_1 Q_3 - 968 Q_2^2 - 1258 Q_2 Q_3 - 371 Q_3^2}{5184 \sigma_0}
    \\
    & \color{red} + 3C_1 -\frac{3}{7}C_2 -\frac{2}{7}C_3, 
    \end{split}\label{K_ww:1} \\
    \addlinespace
     \begin{split}
     \color{red} L_{wz} = & \color{red} \frac{-22148 Q_1^2 + 19432 Q_1Q_2 - 12656 Q_1 Q_3 - 3296 Q_2^2 + 3500 Q_2 Q_3 - 575 Q_3^2}{5184\sigma_0} \\
     & \color{red} + 3C_1 - \frac{9}{7} C_2 + \frac{4}{7}C_3. 
 \end{split}\label{Lwz:1}
    \end{align}
\end{subequations}}%
The quadratic coefficients, $Q_{zz}$ and $Q_{zw}$, are independent of the cubic PDE parameters, whereas $K_{ww}$ and $L_{wz}$ depend on both the quadratic and cubic PDE parameters.
The two coefficients $K_{ww}$ and $L_{wz}$ only appear when $q=1/\sqrt{7}$ and so are colored red.

Weakly nonlinear analysis requires the solution $u$ to be small, which in turn usually requires the quadratic coefficients in the PDE to be small.
However, the spatiotemporal chaos we seek relies on three-wave interactions, and so requires significant quadratic coefficients.
We let the coefficients $Q_1$, $Q_2$ and~$Q_3$ be order one, and include their contributions to the cubic ODE coefficients, allowing the solution~$u$ also to be order one. 
Nonetheless, we show below that the weakly nonlinear theory is a good predictor of the behavior of the~\hbox{PDE}.

\FloatBarrier

 \section{Pattern Classification} \label{patternclassification}
We perform numerical simulations for a grid of $(\mu,\nu)$ parameters and identify the solution type using the classification method described below (with more detail in \cref{app:pattern classification}).
The classification of each pattern is based on examining both its Fourier power spectrum and the size and spatial distribution of its time derivative.

\begin{figure}
\centering
\includegraphics[width=143mm]{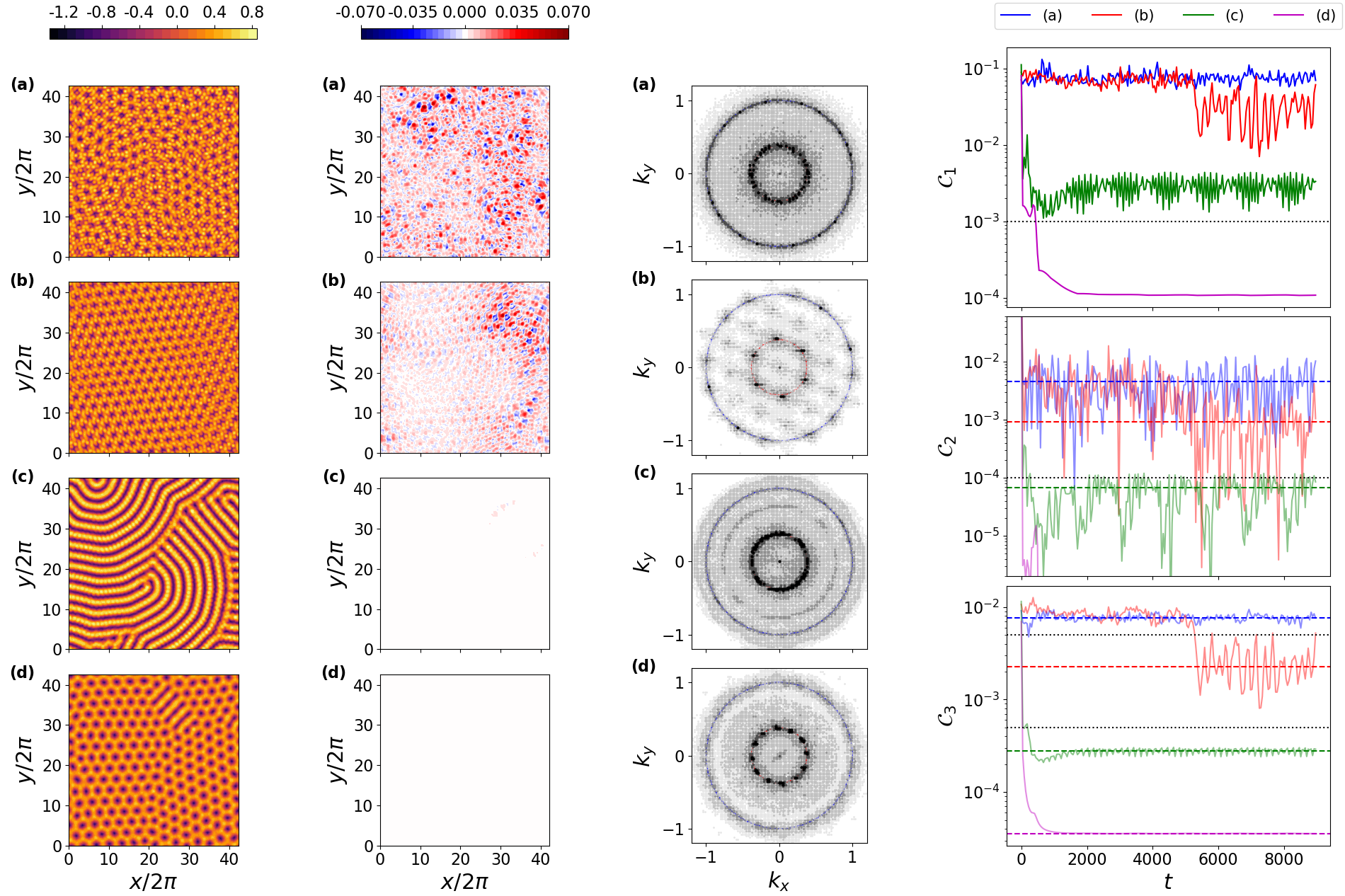}
\caption{A collection of patterned states (a)--(d) and their classification criteria for different values of $\mu$ and $\nu$, which are given in terms of $(r,\chi)$, where $(\mu,\nu)=(r\cos\chi,r\sin\chi)$. 
(a)~$(0.4,70^\circ)$, 
(b)~$(0.3,75^\circ)$, 
(c)~$(0.5,90^\circ)$ and 
(d)~$(0.5,140^\circ)$.
The other PDE parameters are $Q_1 = -1.06$, $Q_2 = -2$, $Q_3 = -0.8$, $C_1 = -1.4$, $C_2 = -5$, $C_3 = -15$, $\sigma_0=-2$ and $q=1/\sqrt{7}$. 
The first column shows the solution~$u(x,y,t)$ at the final computed time and 
the second column shows $\partial u/\partial t$ at the final time.
The third column shows the final Fourier power spectrum.
The annuli around $k=1$ and $k=q$ are filled in~(a) and contain fuzzy peaks in~(b). 
The final column shows the quantities $\mathcal{C}_1$ (time dependence), $\mathcal{C}_2$ (time derivative variation) and $\mathcal{C}_3$ (spatial change) for each of the patterned states (a)--(d) in blue, red, green and magenta. 
The colored dashed lines show the average values of $\mathcal{C}_2$ and $\mathcal{C}_3$. 
The black dotted lines show the classification thresholds, given in \hbox{\cref{app:pattern classification}}. 
The average value of $\mathcal{C}_2$ for (d) is approximately $10^{-8}$ so is not seen in the figure. 
The classifications are:
(a)~STC: fast-time dependence, large time variations and large spatial change with both annuli filled; 
(b)~TC: fast-time dependence, large time variations and small spatial change without filled annuli; 
(c)~fast $w$-stripes: fast-time dependence, small time variations and no spatial change; and
(d)~slow $w$-hexagons: slow-time dependence, small time variations and no spatial change.}
\label{Three_Criteria}
\end{figure}

Examples of the Fourier power spectra are shown in \cref{fig:example patterns} and \cref{Three_Criteria} (third column). 
In \cref{fig:example patterns}{\color{siaminlinkcolor}b} ($z$-hexagons), there are six sharp peaks in the Fourier spectrum on the circle $k=1$, and
in \cref{fig:example patterns}{\color{siaminlinkcolor}d} (rhombs), there are four sharp peaks on the circle $k=1$ and two on the circle $k=q$.
In the figures, we use larger and darker markers to represent larger Fourier amplitudes.
Typically, our simulations return patterns with markers clustered around the two circles $k=1$ and $k=q$.
The Fourier peaks are sharp for pure patterns, as in \cref{fig:example patterns}, but are more spread out if there are defects in the pattern or spatiotemporal chaos, as in \cref{Three_Criteria}.

Our automated method of identifying patterns relies on counting peaks in the Fourier spectrum close to the two circles.
We consider narrow annuli around each circle, which allows us to classify patterns with critical and slightly off-critical wavenumbers. 
We discretize the annuli into $5^\circ$~segments and compute the maximum amplitude of the modes with wavevectors in each segment.
The maximum amplitude of a segment is classified as a \emph{peak} if it is larger than the maximum amplitudes of the four neighboring segments (two on each side) and larger than a given threshold.
The first condition helps classify patterns with defects, and the threshold helps distinguish off-critical $w$-hexagons from superhexagons.
We set the threshold to be one third of the largest amplitude across both circles. 
We denote by $P_1$ (resp.~$P_q$) the number of peaks contained in the annulus around the circle $k=1$ (resp.~$k=q$).
\Cref{tab:equilibriacriteria} gives the values we use for $P_1$ and $P_q$ for each pattern.
\Cref{app:pattern classification} gives further details about the classification of patterns with defects and spatiotemporal chaos: 
patterns with defects, as well as solutions having patches of patterns with different orientations, have \emph{fuzzy} peaks in their Fourier spectra, as in \cref{Three_Criteria}{\color{siaminlinkcolor}d}, and spatiotemporal chaos is characterized by having persistent time dependence and both annuli in the Fourier spectrum being filled, as in \cref{Three_Criteria}{\color{siaminlinkcolor}a}.

We characterize each solution also by its time derivative $\partial u/\partial t$ and the spatial distribution of $\partial u/\partial t$.
In terms of just the time dependence, solutions can be equilibria, slowly varying or fast varying.
The time derivative can be concentrated only in one place or can be more spread out across the domain.
To distinguish between these possibilities, we introduce the following three metrics.

The first metric, $\mathcal{C}_1(t)$, is the maximum of $|\partial u/\partial t|$ over the domain scaled to the maximum of~$|u|$ over the domain:
\begin{equation}
    \mathcal{C}_1(t) \coloneqq 
    \frac{1}{\underset{(x,y)}{\max} \, \left|u(x,y,t)\right|} \, \underset{(x,y)}{\max} \left| \pdev{u(x,y,t)}{t}\right| ,  \label{crtieria:1}
\end{equation}
where $\partial u/\partial t$ is computed using finite differences.
The value $\tilde{\mathcal{C}}_1 = \mathcal{C}_1(t_{\textrm{final}})$, where $t_{\textrm{final}}$ is the final computed time, is used to determine if we have an equilibrium, a slowly varying~(see \cref{Three_Criteria}{\color{siaminlinkcolor}d}) or a fast varying~(see \cref{Three_Criteria}{\color{siaminlinkcolor}a,b,c}) solution.

We also use the rate of change of $\mathcal{C}_1(t)$ to define a second metric:
\begin{equation}
\mathcal{C}_2(t) \coloneqq \frac{1}{\Delta t} \bigl|\mathcal{C}_1\left(t+\Delta t\right)- \mathcal{C}_1(t)  \bigr|, \label{criteria:2}
\end{equation}
where $\Delta t$ is the time step used in our computations. We average~$\mathcal{C}_2(t)$ over the final $80\%$ of the simulation to reduce the effect of transients, and collect the resulting quantity, $\bar{\mathcal{C}}_2$.
The metric $\bar{\mathcal{C}}_2$ identifies patterns with persistent significant time dependence (see \cref{Three_Criteria}{\color{siaminlinkcolor}a,b}).

Finally, we use a metric~$\mathcal{C}_3$ to distinguish between patterns where the evolution is focused in one place (for example, at a slowly moving defect) and patterns where the evolution is spread across the whole domain. This metric is similar to~$\mathcal{C}_1$ but uses an average instead of a maximum:
\begin{equation}
    \mathcal{C}_3(t) \coloneqq 
    \frac{1}{\underset{(x,y)}{\max} \, \left|u(x,y,t)\right|} \,
    \underset{(x,y)}{\textrm{avg}}\left| \pdev{u(x,y,t)}{t}\right|, \label{criteria_3}
\end{equation}
taking the scaled spatial average of the time derivative.
As with~$\mathcal{C}_2$, we then compute the time average $\bar{\mathcal{C}}_3$, disregarding the first $20\%$ of values. 
We found that the averaging method used in \cref{criteria_3} is more robust than the root mean square and the mean average deviation, in particular in the presence of outliers, where the pattern remains steady except for a small number of defects (see \cref{Three_Criteria}{\color{siaminlinkcolor}c}).
And, as with~$\tilde{\mathcal{C}}_1$, we have two thresholds, which separate patterns with no spatial change (\cref{Three_Criteria}{\color{siaminlinkcolor}c,d}), patterns with small spatial change (\cref{Three_Criteria}{\color{siaminlinkcolor}b}), and patterns with large spatial change (\cref{Three_Criteria}{\color{siaminlinkcolor}a}).
We also found this approach preferable to using a correlation length argument~\cite{Sethna2006}, since our method allows for both spatiotemporal chaos (STC, \cref{Three_Criteria}{\color{siaminlinkcolor}a}) and temporal chaos (TC, \cref{Three_Criteria}{\color{siaminlinkcolor}b}) to be classified in the same way.

\FloatBarrier

\section{Numerical Results} \label{numerical}
To solve our PDE \cref{PDE:2} numerically, we use the fourth-order Runge--Kutta exponential time differencing method~(ETD4RK) introduced by~\cite{COX2002430}, which solves the linear part of the PDE to machine precision.
The stiffness induced by 8th order spatial derivatives \cite{BEYLKIN1998362} is handled satisfactorily by the method.
The nonlinear terms are approximated using a 4th-order Runge--Kutta method; we restrict these terms to second order spatial derivatives to avoid stiffness issues. 
To avoid cancellation errors when $|\sigma(k)\Delta t|<0.1$, we use the 5-term Taylor series approximant instead of the true ETD4RK coefficients for wavenumber~$k$~\cite{Kassam2005}.

For our simulations we use a periodic domain of $16\times28/\sqrt{3}\approx16\times16.17$ repetitions of the longer wavelength ($k=q$) pattern, which is approximately $42.33 \times 42.77$ of the shorter wavelength ($k=1$) pattern when $q=1/\sqrt{7}$.
The domain is chosen so that waves with all of the wavevectors in \cref{hexagonal_lattice,qlength:a} fit exactly and is approximately square.
The domain is large enough that the density of modes allows many interacting triads, potentially leading to more complex dynamics.
We use the SciPy~\cite{SciPy2020} fast Fourier transform~(FFT) over $384\times384$ Fourier modes, which is about 9 grid points per short wavelength. 
At each time-step we remove the contributions from modes with physical wavenumber larger than $3.5$ for de-aliasing, although we found these modes to be insignificant. 

We discretize our parameter space of linear growth rates $\mu$ and~$\nu$ into a circular grid to allow for a greater density of parameter values for small $\mu$ and~$\nu$. 
This is the region where we expect weakly nonlinear theory to provide the best approximation of the PDE dynamics. 
To form the circular grid, we take 13 circles of different radii, from $0.01$ to~$0.5$, and discretize every~$5^\circ$ in angle. 
We omit the region where both $\mu\leq0$ and $\nu\leq0$, as this is where the trivial solution is stable, resulting in a total of 689 grid points. 
The solution at each value of $\mu$ and $\nu$ is classified using the method from \cref{patternclassification} and verified manually.

\FloatBarrier

We set $q=1/\sqrt{7}$, so both three-wave and four-wave interactions are present between waves with wavenumber~$k=1$ and $k=q$.
We use the weakly nonlinear approximations found in \cref{model PDE} to predict the stability in the~\hbox{PDE} of the simple equilibria: stripes, hexagons, rhombs, superhexagons and stars. 
We include off-critical $w$-hexagons in our stability calculations, taking all red terms in \cref{z1hex:1}--\cref{w1hex:1} to be zero for these.

The stability calculations reveal that there can be regions in parameter space where there are no stable simple equilibria. 
We combine this with the known locations of Hopf bifurcations from $z$-stripes and $w$-stripes as the basis for our search for time dependence and spatiotemporal chaos in the~\hbox{PDE}.
When~$Q_{zz}Q_{zw}<0$, we expect 3WIs to be the driving force behind any time-dependent dynamics, whereas, when~$K_{ww}L_{wz}<0$ we expect 4WIs to fulfill this role.
We avoid parameter choices close to $Q_2=2Q_3$ and $C_2=C_3$ since these yield variational dynamics and prevent persistent time dependence.
 
\begin{figure}
\centering
\begin{tikzpicture}[scale=1]
\def\radius{1.1}
\draw[->] (-4.5,0) -- (4.5,0) node[right] {$Q_{zz}Q_{zw}$};
\draw[->] (0,-4) -- (0,4.2) node[above] {$K_{ww}L_{wz}$};
\node (3WI1) [circle , draw, thick, minimum size=\radius cm, fill=white] at (4.2,2){}; 
\node (3WI2) [circle , draw, thick, minimum size=\radius cm, fill=white] at (-3.5,1.5){};
\node (4WI1) [circle , draw, thick, minimum size=\radius cm, fill=white] at (3.2,3){}; 
\node (4WI2) [circle , draw, thick, minimum size= \radius cm,fill=white] at (3.2,-2){};
\node (34WI) [circle, draw, thick ,minimum size = \radius cm, fill=white] at (-2,-1.5){};
\draw[thick,
        postaction={decorate},
        decoration={markings, mark=at position 0.4 with {\arrow[scale=2]{>}},mark=at position 0.75 with {\arrow[scale=2]{>}}}]
        (3WI1) -- (3WI2);
\draw[thick, red,
        postaction={decorate},
        decoration={markings, mark=at position 0.45 with {\arrow[scale=2,color=red]{>}},mark=at position 0.8 with {\arrow[scale=2,color=red]{>}}}]
        (4WI1) -- (4WI2);
\draw[thick,
        postaction={decorate},
        decoration={markings, mark=at position 0.3 with {\arrow[scale=2]{>}},mark=at position 0.75 with {\arrow[scale=2]{>}}}]
        (4WI2) -- (34WI);
\node [above=5pt,rotate =4.3] at (-1.3,1.6) {Change $Q_1$}; 
\node [above=5pt,rotate =4.2] at (-1.3,0.8) {\cref{3WIs}}; 
\node [above=5pt,rotate =-5.8] at (1.3,-1.9) {Change $Q_1$}; 
\node [above=5pt,rotate =-6] at (1.3,-2.6) {\cref{3and4WI}}; 
\node [above=5pt] at (4.4,0.5) {\color{red} Change $C_1$}; 
\node [above=5pt] at (4.5,0.15) {\cref{4WIs}};
\node [scale=1.2] at (-0.2,-0.2) {0};
\def\radiusnew{1.08}
\begin{scope}[shift={(3WI2)},scale=0.5]
  \def\angleStart{72.5}
  \def\angleEnd{87.5}
  \pgfmathsetmacro\xStart{\radiusnew*cos(\angleStart)}
  \pgfmathsetmacro\yStart{\radiusnew*sin(\angleStart)}
  \pgfmathsetmacro\xEnd{\radiusnew*cos(\angleEnd)}
  \pgfmathsetmacro\yEnd{\radiusnew*sin(\angleEnd)}
  \fill[blue!30] 
    (0,0) -- (\xStart,\yStart) 
    arc[start angle=\angleStart, end angle=\angleEnd, radius=\radiusnew] 
    -- cycle;
\end{scope}
\begin{scope}[shift={(4WI2)},scale=0.5]
  \def\angleStart{70}
  \def\angleEnd{145}
  \pgfmathsetmacro\xStart{\radiusnew*cos(\angleStart)}
  \pgfmathsetmacro\yStart{\radiusnew*sin(\angleStart)}
  \pgfmathsetmacro\xEnd{\radiusnew*cos(\angleEnd)}
  \pgfmathsetmacro\yEnd{\radiusnew*sin(\angleEnd)}
  \fill[blue!30] 
    (0,0) -- (\xStart,\yStart) 
    arc[start angle=\angleStart, end angle=\angleEnd, radius=\radiusnew] 
    -- cycle;
\end{scope}
\begin{scope}[shift={(34WI)},scale=0.5]
  \def\angleStart{70}
  \def\angleEnd{155}
  \pgfmathsetmacro\xStart{\radiusnew*cos(\angleStart)}
  \pgfmathsetmacro\yStart{\radiusnew*sin(\angleStart)}
  \pgfmathsetmacro\xEnd{\radiusnew*cos(\angleEnd)}
  \pgfmathsetmacro\yEnd{\radiusnew*sin(\angleEnd)}
  \fill[blue!30] 
    (0,0) -- (\xStart,\yStart) 
    arc[start angle=\angleStart, end angle=\angleEnd, radius=\radiusnew] 
    -- cycle;
\end{scope}
\begin{scope}[shift={((5.2,3.6)},scale=0.25]
\node [] at (0,0){\includegraphics[width=25mm]{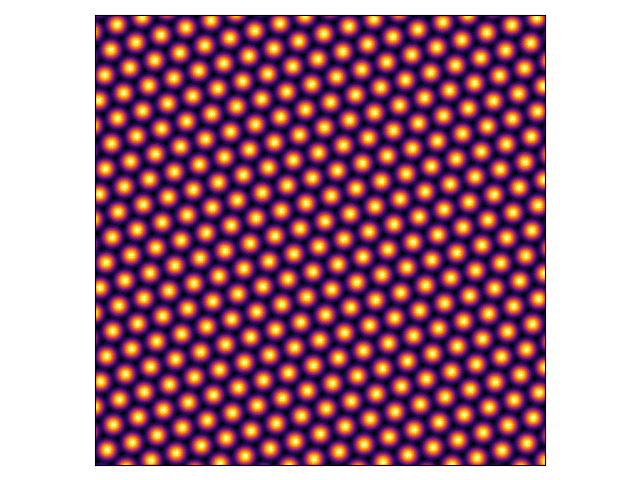}};
\end{scope}
\begin{scope}[shift={(5.2,-2.2)},scale=0.25]
\node[] at (0,0){\includegraphics[width=25mm]{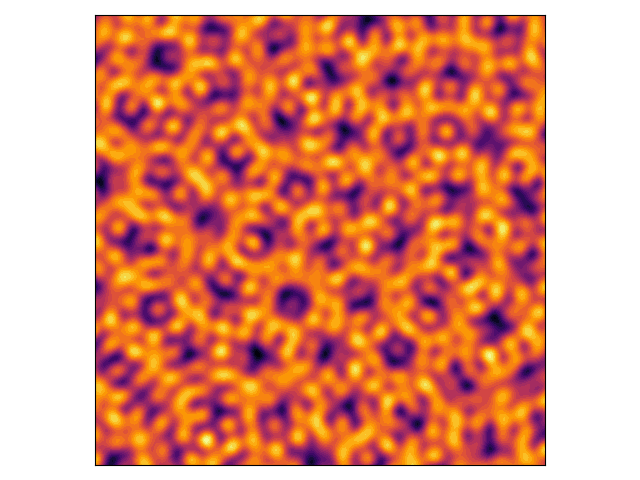}};
\end{scope}
\begin{scope}[shift={(-4,-2.2)},scale=0.25]
\node[] at (0,0){\includegraphics[width=25mm]{Pattern_Classification/Uchaoszoom.png}};
\end{scope}
\begin{scope}[shift={(-4,3.6)},scale=0.25]
\node[] at (0,0){\includegraphics[width=25mm]{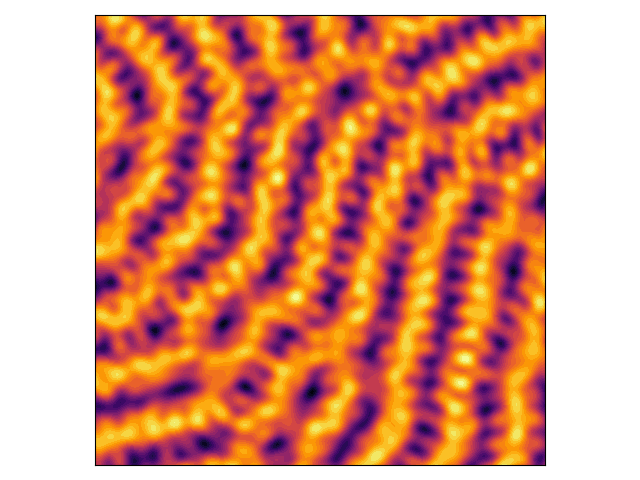}};
\end{scope}
\end{tikzpicture}
\caption{Schematic depiction of the changes we make to the PDE parameters in each set of numerical simulations.
\Cref{3WIs} isolates the 3WIs and \cref{4WIs} the 4WIs. 
Both 3WIs and 4WIs drive time dependence in \cref{3and4WI}. 
The purple region in each circle shows the approximate size of the PDE chaotic region in $(\mu,\nu)$-space for each quadrant; the largest chaotic region occurs for $K_{ww}L_{wz}<0$, with very little dependence on the sign of~$Q_{zz}Q_{zw}$. 
The chaotic solutions in the lower two quadrants also have more spatial disorder than those in the top left quadrant. 
No time dependence (after transients) is found in the top right quadrant.}
\label{numerical schematic}
\end{figure}

For each set of simulations, we change the value of one PDE parameter to drive either $Q_{zz}Q_{zw}$ or $K_{ww}L_{wz}$ through zero, as shown in \cref{numerical schematic}.
We begin with both $Q_{zz}Q_{zw}>0$ and $K_{ww}L_{wz}>0$, (top right quadrant of \cref{numerical schematic}), with PDE parameters chosen so that there is always at least one stable equilibrium in the ODE amplitude equations for every value of~$(\mu,\nu)$ that we consider.
Without this requirement, we found that the PDE easily produced unbounded solutions. 
We also choose PDE parameters so that when at least one of $Q_{zz}Q_{zw}$ or $K_{ww}L_{wz}$ is negative (other quadrants in \cref{numerical schematic}), there are ranges of $(\mu,\nu)$ where none of the simple equilibria are stable in the ODEs, pointing toward the possibility of time-dependent dynamics in the~\hbox{PDE}.
The requirement for having at least one stable equilibrium in the ODEs in the top right quadrant turns out to be sufficient for avoiding unbounded PDE solutions in all quadrants.

In \cref{3WIs}, we vary $Q_1$ to change $Q_{zz}Q_{zw}$ from positive to negative to display the effect of 3WIs on the dynamics of the~\hbox{PDE}. 
\Cref{4WIs} shows the transition of $K_{ww}L_{wz}$ from positive to negative by varying~$C_1$, to display the effect of 4WIs (note $Q_{zz}Q_{zw}$ does not change value in these simulations). 
Finally, \cref{3and4WI} continues on from the end state in \cref{4WIs} to explore the case when both 3WIs and 4WIs drive the time dependence: we vary~$Q_1$ to reduce the value of $Q_{zz}Q_{zw}$ from positive to negative. 

In each of \cref{3WIs,4WIs,3and4WI}, we present the weakly nonlinear stability predictions for each simple equilibrium pattern in \cref{tab:equilibriacriteria}. 
We compare this with the PDE solutions: almost all of the PDE equilibria that we find are one of the five simple patterns, possibly with defects.
We also find a small number of PDE equilibria that are not simple patterns but that can still be described in terms of $z$ and $w$ amplitudes: we class these as superlattice solutions, and they are explored in more depth in~\cite{Subramanian2026}. 

\subsection{Chaos Driven by Three-Wave Interactions} \label{3WIs}
Using \cref{Qzw:1}--\cref{Lwz:1}, we choose a range of PDE parameters that starts with $Q_{zz}Q_{zw}>0$ and ends with $Q_{zz}Q_{zw}<0$, while keeping $K_{ww}L_{wz}>0$, starting in the top right quadrant in \cref{numerical schematic} and ending in the top left quadrant.
This choice leads to a Hopf bifurcation from $z$-stripes within the $(\mu,\nu)$ grid~\cite{PORTER2004,Subramanian2026} and so we anticipate finding time-dependent dynamics driven by the 3WIs.
Having 4WIs and $K_{ww}L_{wz}<0$ is necessary for the Hopf bifurcation from $w$-stripes, so this bifurcation is not present here.

The starting point for these numerical simulations is $Q_1=-1.4$, $Q_2=-2.75$, $Q_3=-3.5$, $C_1=-2.75$, $C_2=-7.75$, $C_3=-16.5$, $\sigma_0=-2$ and $q=1/\sqrt{7}$.
We increase $Q_1$ from $-1.4$ to $-0.6$ in increments of $0.1$.
The quantity $Q_{zz}Q_{zw}$, initially positive, changes sign at $Q_1 \approx -1.32$.
The sign of $K_{ww}L_{wz}$ does not change. 
We start the first set of simulations with random initial conditions for all~$(\mu,\nu)$.
We start subsequent sets of simulations with a new value of~$Q_1$; the initial condition for each $(\mu,\nu)$ pair is the final state at that value of $(\mu,\nu)$ and the previous value of~$Q_1$.

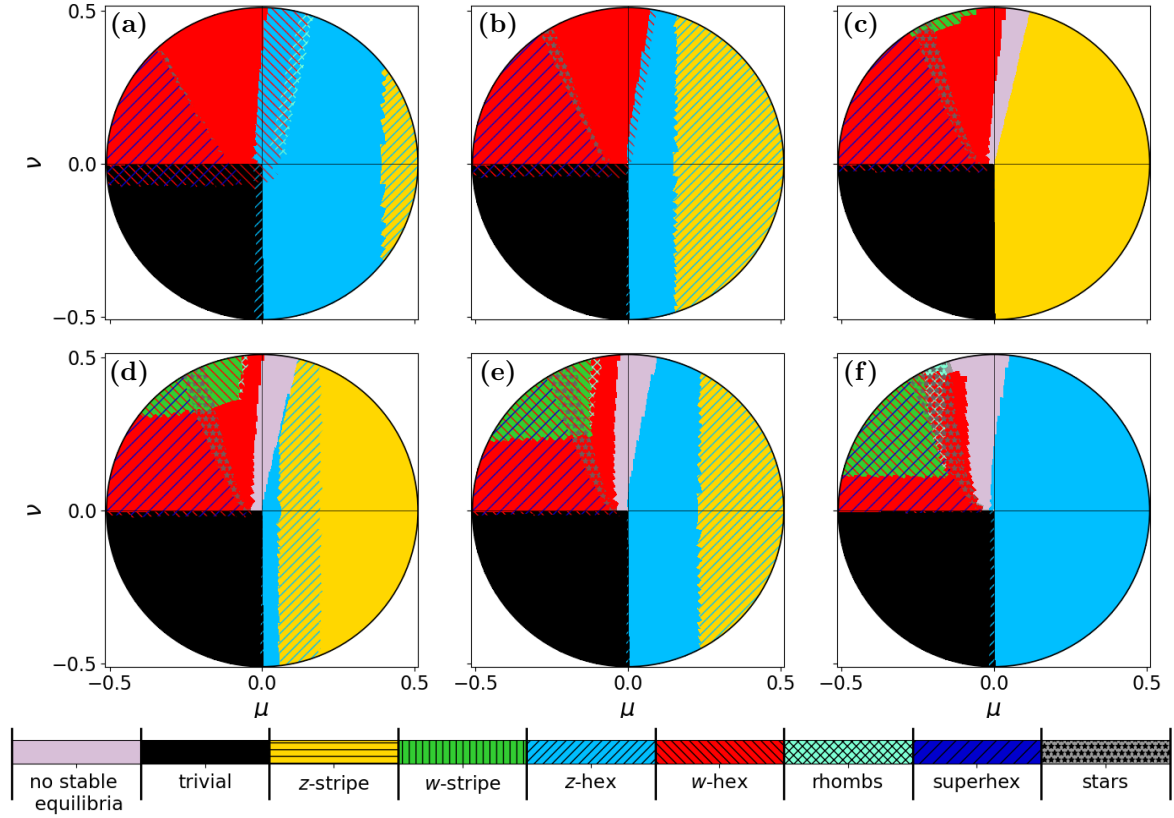
\begin{figure}
\centering
 \begin{tikzpicture}
\node[anchor=south west,inner sep=0] (image) at (0,0) {\includegraphics[width=\textwidth]{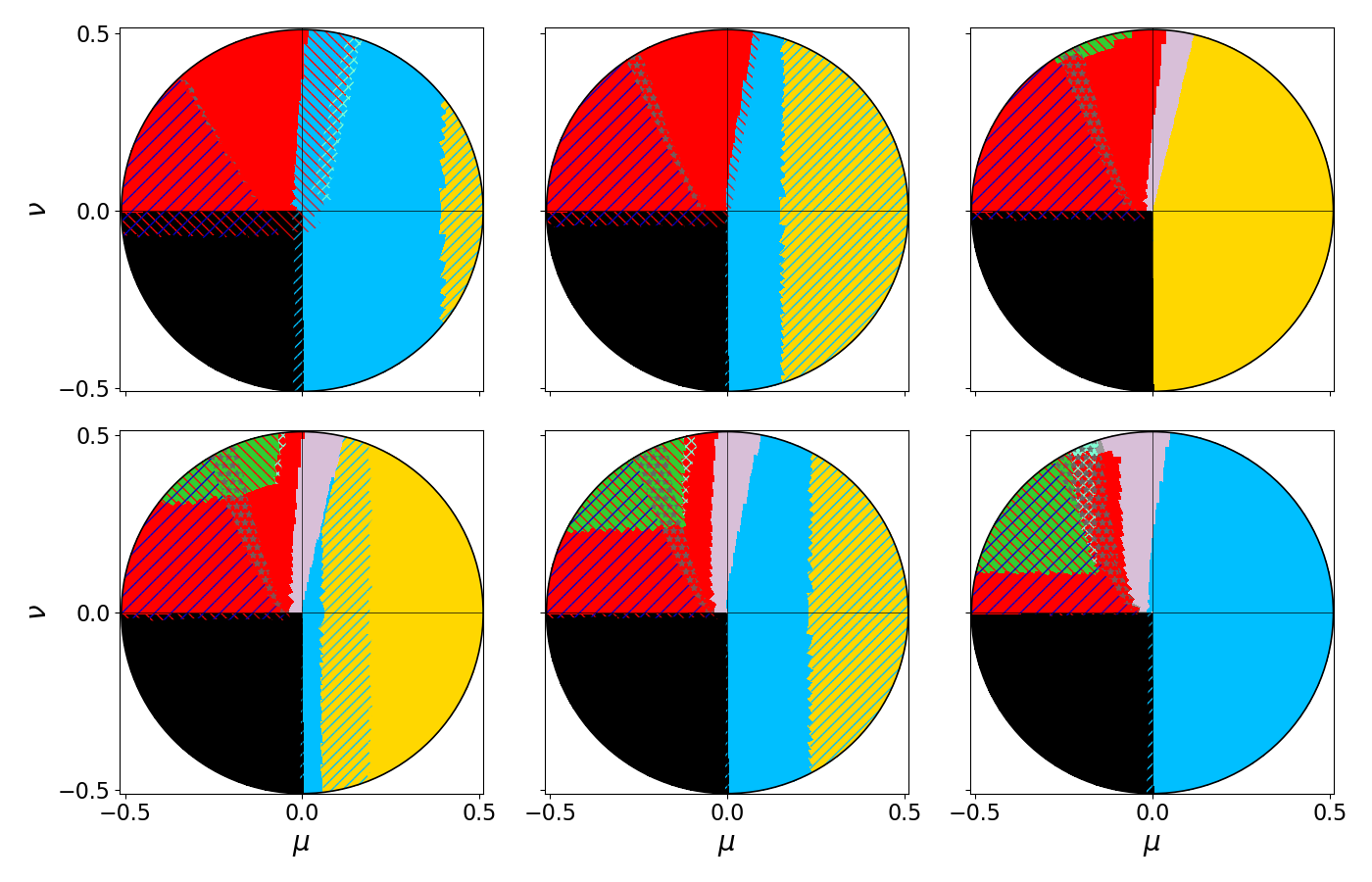}};
\begin{scope}[x={(image.south east)},y={(image.north west)}]
  \node[fill=none] at (0.108,0.943) {\small \textbf{(a)}};
  \node[fill=none] at (0.42,0.943) {\small \textbf{(b)}};
    \node[fill=none] at (0.727,0.943) {\small \textbf{(c)}};
  \node[fill=none] at (0.108,0.485) {\small \textbf{(d)}};
    \node[fill=none] at (0.42,0.485) {\small \textbf{(e)}};
  \node[fill=none] at (0.727,0.485) {\small \textbf{(f)}};
\end{scope}
\end{tikzpicture}
\vskip -0.35cm
\includegraphics[width=\textwidth]{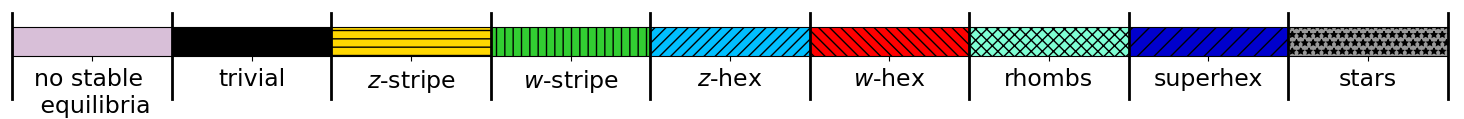}
\caption{Weakly nonlinear predictions for simple equilibrium patterns, with $Q_2=-2.75$, $Q_3=-3.5$, $C_1=-2.75$, $C_2=-7.75$, $C_3=-16.5$ and $q=1/\sqrt{7}$, with different values of $Q_1$: 
(a)~$-1.4$, (b)~$-1.2$, (c)~$-1.01$, (d)~$-0.9$, (e)~$-0.8$, (f)~$-0.6$. 
As $Q_1$ is increased, $Q_{zz}Q_{zw}$ changes from positive to negative at $Q_1 \approx -1.32$. 
Hatched markings denote regions of bistability, and each pattern has hatched markings with different orientations (see colorbar).
The color of the hatched markings corresponds to the pattern. 
We use the off-critical $w$-hexagon analysis to differentiate between $w$-hexagons and superhexagons.
The lavender region in (c)--(f) is where there are no stable simple equilibria.}
\label{3WI:WNLT}
\end{figure}

The weakly nonlinear theory predictions for stable stripes, hexagons, rhombs, super\-hexagons and stars 
are shown in \cref{3WI:WNLT}.
Each color indicates a different stable equilibrium pattern. 
Hatched markings show regions of bistability, e.g., yellow with light blue striped hatched markings corresponds to both $z$-stripes and $z$-hexagons being stable. 
With this choice of $Q_2$ and $Q_3$, $Q_{zhex}$~is zero when $Q_1=-1$, so we use $Q_1=-1.01$ to avoid this degeneracy in \cref{3WI:WNLT}{\color{siaminlinkcolor}c}.
The region of stable $z$-hexagons is bounded by a saddle-node bifurcation and a pitchfork bifurcation to rectangles~\cite{Buzano1983}.
These both occur when $\mu$ is $\mathcal{O}\left(Q_{zhex}^2\right)$~\cite{Hoyle2006}, so for $Q=-1.01$, the stability region for $z$-hexagons is very small.

The lavender regions in \cref{3WI:WNLT}{\color{siaminlinkcolor}(c)--(f)} correspond to values of $\mu$ and $\nu$ where none of the simple equilibria are stable.
For this choice of $Q_2$, $Q_3$, $C_1$, $C_2$ and $C_3$, the rightmost edges of the lavender regions correspond roughly to the Hopf bifurcation from $z$-stripes.
We have not exhaustively classified the dynamics in this region, but we have found examples of other equilibria, traveling solutions, and oscillatory and chaotic solutions.
This region is relatively narrow; we have not been able to find other parameter values that give a wide lavender region while maintaining $K_{ww}L_{wz}>0$.

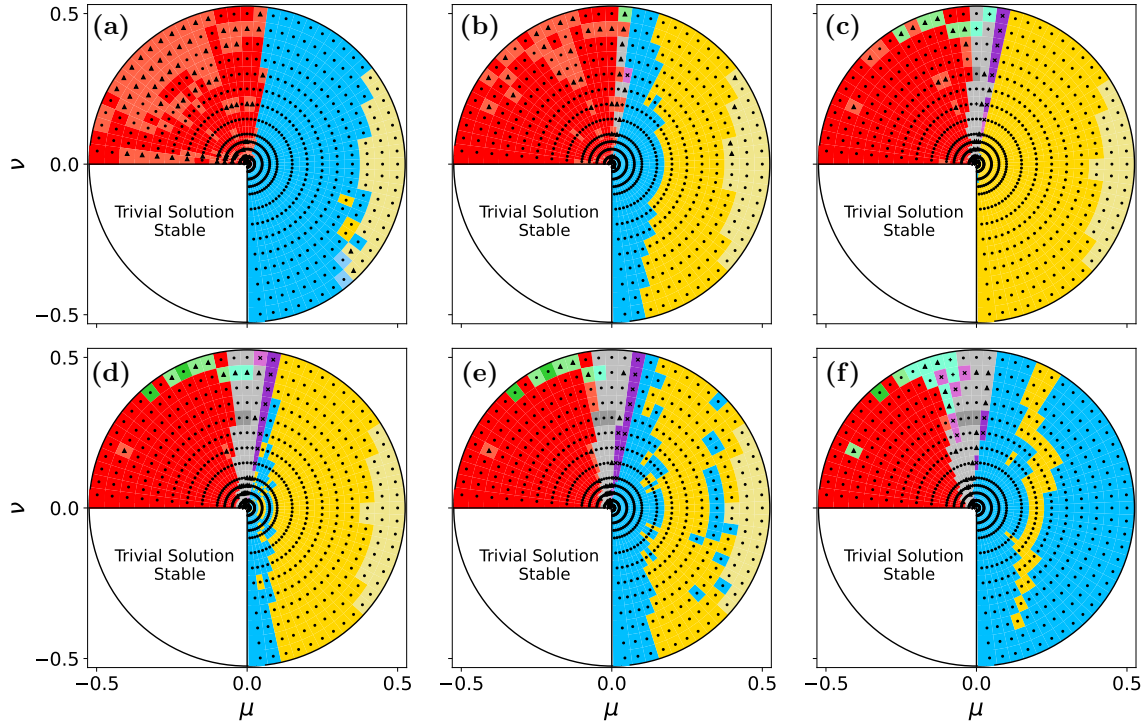
\begin{figure}
\centering
\begin{tikzpicture}
\node[anchor=south west,inner sep=0] (image) at (0,0)
{\includegraphics[width=\textwidth]{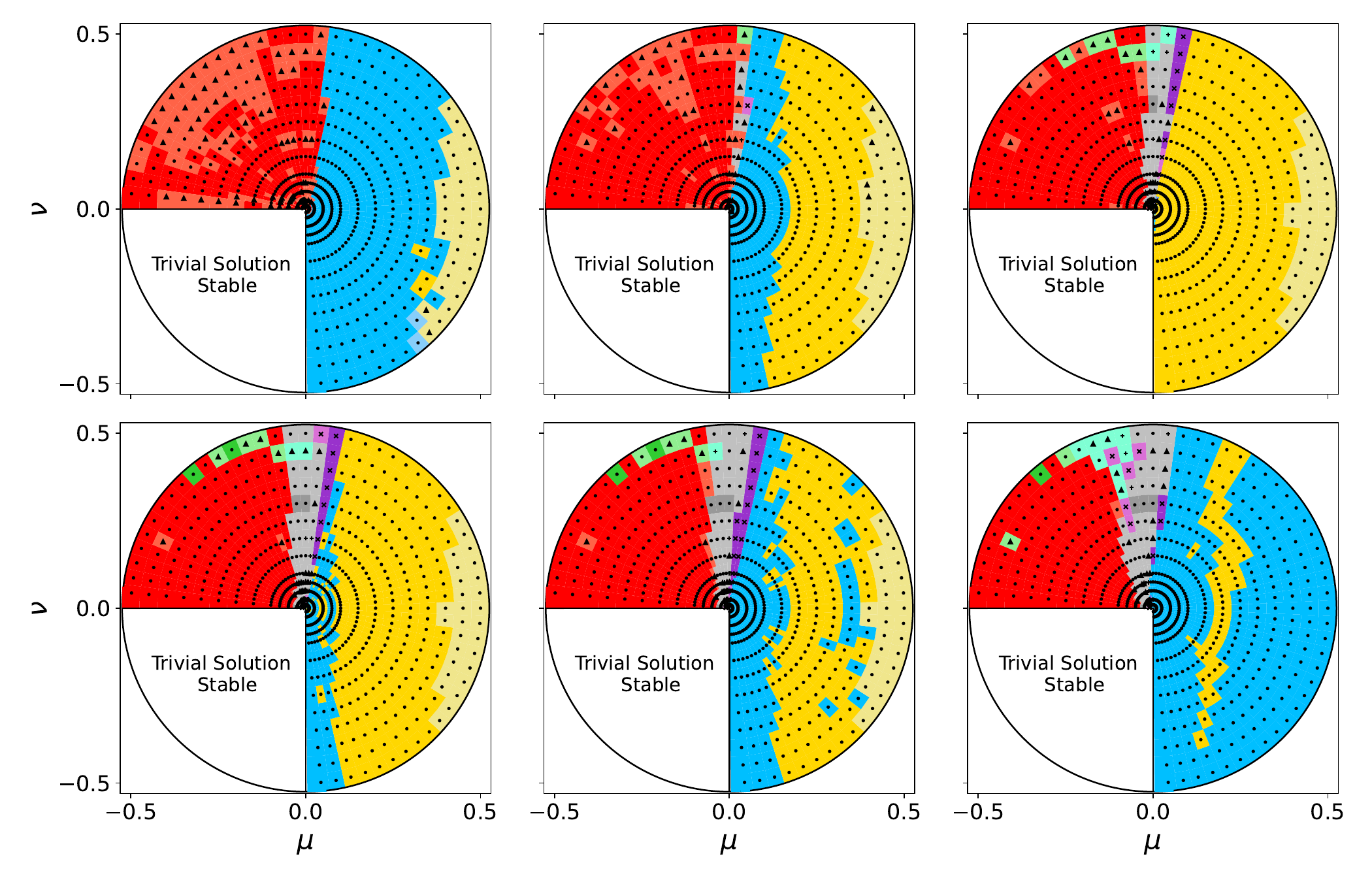}};
\begin{scope}[x={(image.south east)},y={(image.north west)}]
  \node[fill=none] at (0.108,0.943) {\small \textbf{(a)}};
  \node[fill=none] at (0.42,0.943) {\small \textbf{(b)}};
    \node[fill=none] at (0.727,0.943) {\small \textbf{(c)}};
  \node[fill=none] at (0.108,0.486) {\small \textbf{(d)}};
    \node[fill=none] at (0.42,0.486) {\small \textbf{(e)}};
  \node[fill=none] at (0.727,0.486) {\small \textbf{(f)}};
\end{scope}
\end{tikzpicture}
\vskip -0.2cm
\caption{Bifurcation set showing the patterns observed with varying $(\mu,\nu)$ within the PDE \cref{PDE:2}.
The lower rows are zoomed versions of the upper rows.
The PDE parameters are the same as in \cref{3WI:WNLT}, with $\sigma_0 = -2$.
Each black marker corresponds to an individual simulation classified according to \cref{patternclassification}:
{\large $\bullet$}~represents equilibria, {\large$\blacktriangleup$} for slow-time-dependent patterns, $+$~for fast-time-dependence with limited  spatial change, and $\times$~for fast-time-dependence with significant spatial change.
The lighter of the two shades of color correspond to defects or modulation.
The data for this figure is available from~\cite{Pinkney_data}. }
\label{3WI:PDE}
\end{figure}

\begin{figure}
\centering
\begin{tikzpicture}
\node[anchor=south west,inner sep=0] (image) at (0,0)
{\includegraphics[width=\textwidth]{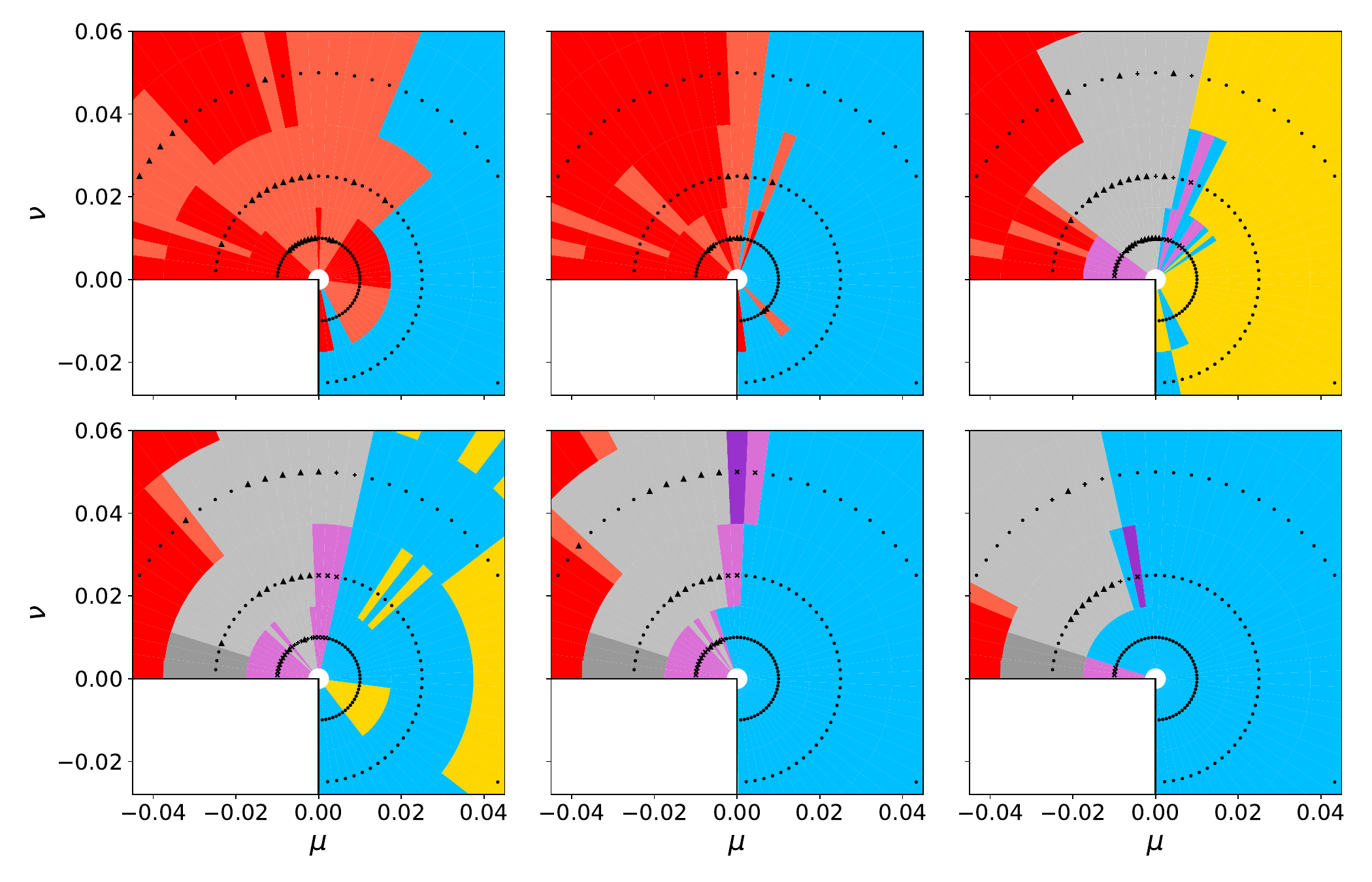}};
\begin{scope}[x={(image.south east)},y={(image.north west)}]
  \node[fill=none] at (0.12,0.579) {\small \textbf{(a)}};
  \node[fill=none] at (0.425,0.579) {\small \textbf{(b)}};
    \node[fill=none] at (0.727,0.579) {\small \textbf{(c)}};
  \node[fill=none] at (0.12,0.125) {\small \textbf{(d)}};
    \node[fill=none] at (0.425,0.125) {\small \textbf{(e)}};
  \node[fill=none] at (0.727,0.125) {\small \textbf{(f)}};
\end{scope}
\end{tikzpicture}
\vskip -0.35cm
\includegraphics[width=\textwidth]{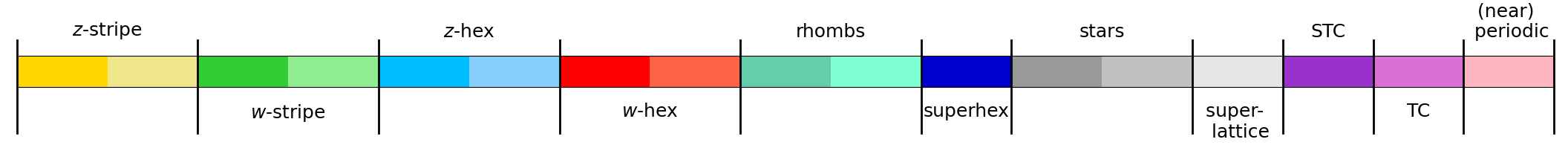}
\caption{Zoomed in bifurcation set of \cref{3WI:PDE} for small $\mu$ and $\nu$.
The data for this figure is available from~\cite{Pinkney_data}. }
\label{3WI:PDEzoom}
\end{figure}

The results of the corresponding PDE simulations are shown in \cref{3WI:PDE,3WI:PDEzoom}.
As predicted, when $Q_{zz}Q_{zw}>0$ (\cref{3WI:PDE}{\color{siaminlinkcolor}a}) we do not find any cases of chaotic dynamics.
The only solutions with time dependence have slow evolution of defects within the patterned state.
These defects persist in the solution after thousands of time units. 
We found only small regions of chaotic dynamics (STC and TC in purple and light purple respectively in \cref{3WI:PDE}) when $Q_{zz}Q_{zw}<0$.
These regions of chaotic dynamics in the PDE lie roughly within (but do not fill) the corresponding regions of no stable equilibria in the ODE.
The regions of stability of $z$-stripes, $z$-hexagons and $w$-hexagons match well between the ODEs and PDE.
The regions of stars do not agree, but this is because of their bistability with $w$-hexagons, which dominate at the first value of~$Q_1$.
Superhexagons are found only in the ODEs for a similar reason.
The classification of stars in the PDEs includes asymmetric stars, with unequal $z$ and $w$ amplitudes.

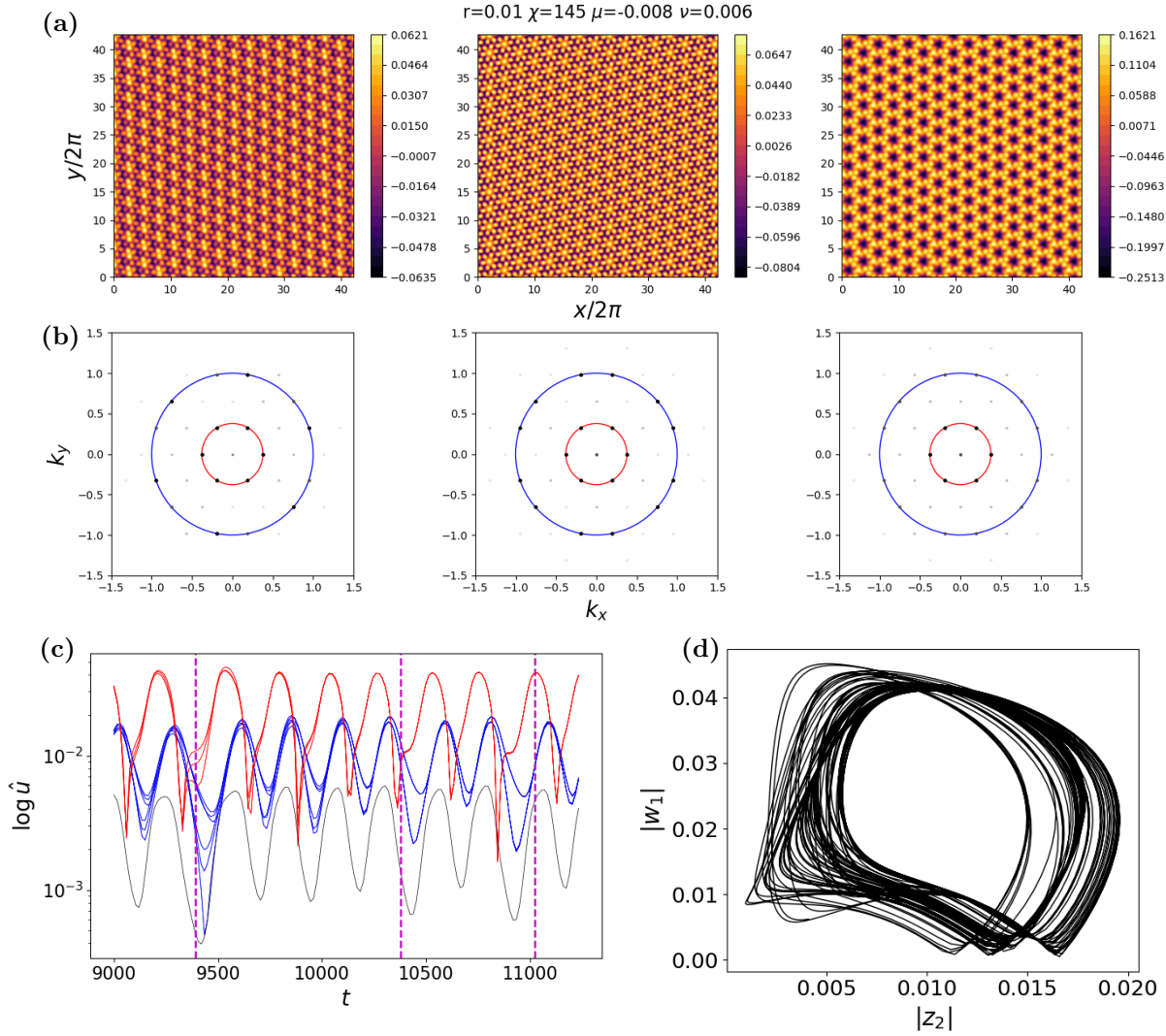
\begin{figure}
\captionsetup[subfloat]{labelformat=empty} 
\centering
\begin{tikzpicture}
\node[anchor=south west,inner sep=0] (image) at (0,0)
{\includegraphics[width=153mm]{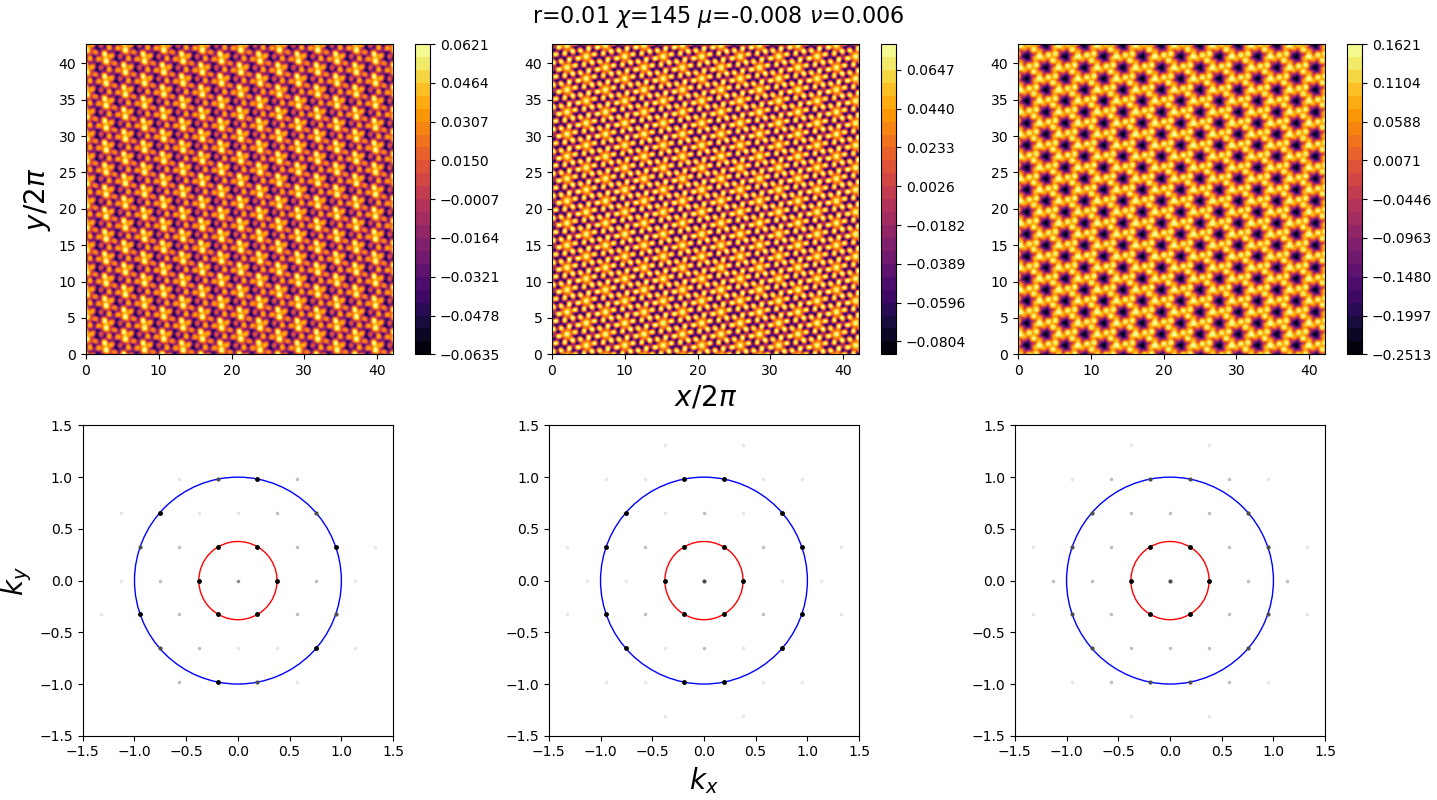}};
\begin{scope}[x={(image.south east)},y={(image.north west)}]
  \node[fill=none] at (0.013,0.96) {\small \textbf{(a)}};
  \node[fill=none] at (0.013,0.46) {\small \textbf{(b)}};
\end{scope}
\end{tikzpicture}
\begin{tikzpicture}
\node[anchor=south west,inner sep=0] (image) at (0,0)
{\hspace{-1.3cm}\includegraphics[width=86mm]{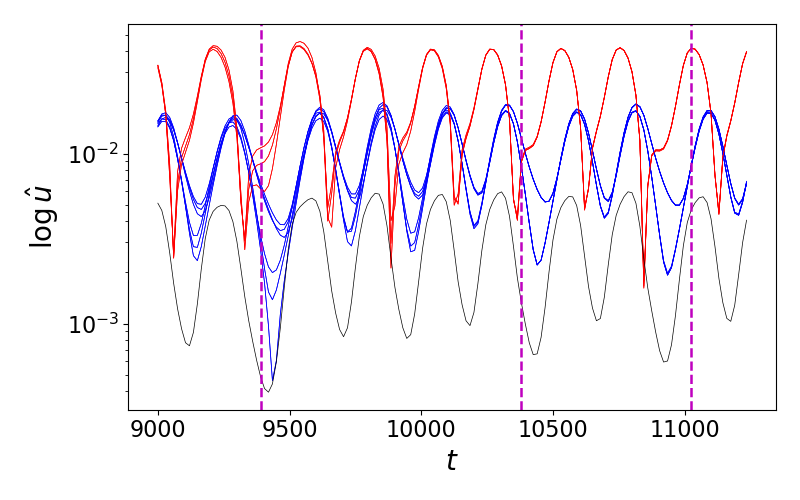}};
\node[anchor=south west,inner sep=0] (image2) at (7.4,-0.2)
{\includegraphics[width=74mm]{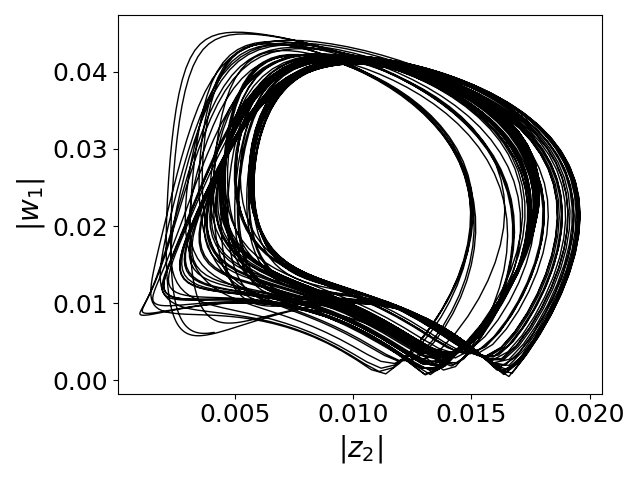}};
\begin{scope}[x={(image.south east)},y={(image.north west)}]
  \node[fill=none] at (-0.05,0.96) {\small \textbf{(c)}};
  \node[fill=none] at (1.15,0.96) {\small \textbf{(d)}};
\end{scope}
\end{tikzpicture}
\caption{Time evolution of a temporally chaotic solution of the~\hbox{PDE}.
The parameters are as in \cref{3WI:PDE}{\color{siaminlinkcolor}d}, with $r=0.01$ and $\chi = 145^\circ$ where $(\mu,\nu) = (r\cos \chi, r \sin \chi)$.
The three patterned states in~(a) are the solution at different times, indicated by the dashed magenta lines in~(c).  
The larger contributions to the patterned state are represented by a darker and larger marker in the Fourier spectra in~(b). 
For these three patterned states, the $k=q$ modes are approximately equal, whereas the $k=1$ modes are unequal. 
Note the colorbar is different for each pattern; this is to emphasize how the patterned state is evolving.
The red, blue, black curves in~(c) correspond to the evolution of the $k=q$, $k=1$, $k=0$ modes respectively. 
Small amplitude modes are omitted from the plot. 
A phase portrait of the $z_2$ and $w_1$ amplitudes is shown in~(d).}
\label{smallchaos}
\end{figure}

Interestingly, the widest region of temporally chaotic (TC) dynamics occurs for very small $\mu$ and~$\nu$, with individual parameter values having~\hbox{STC}.
Some of the TC examples in the PDE for small $(\mu,\nu)$ have exactly the six modes on the $k=q$ circle and the twelve modes on the $k=1$ circle used in the weakly nonlinear theory.
One such example is shown in \cref{smallchaos}. 

\FloatBarrier

\subsection{Chaos Driven by Four-Wave Interactions} \label{4WIs}
For our next set of results, we consider the effect of the non-generic four-wave interactions (unique to the value $q=1/\sqrt{7}$) on the dynamics of the~\hbox{PDE}. 
In this case, we fix all parameters except~$C_1$, with $Q_{zz}Q_{zw}$ independent of~$C_1$.
We fix $Q_{zz}Q_{zw}$ to be positive in order to focus on the impact of the time dependence arising from the 4WIs. 
Proceeding as we did in \cref{3WIs}, we begin with $C_1=-2.4$ in the $Q_{zz}Q_{zw}>0$ and $K_{ww}L_{wz}>0$ quadrant (\cref{numerical schematic}) with a small amplitude, random initial condition. 
Increasing $C_1$ to~$-1$ allows us to transition into the $K_{ww}L_{wz}<0$ regime, with $K_{ww}L_{wz}=0$ at $C_1\approx-2.29$.
There is a Hopf bifurcation from $w$-stripes when \hbox{$K_{ww}L_{wz}<0$~\cite{Subramanian2026}} and so we anticipate finding time-dependent dynamics driven by the 4WIs.

\begin{figure}
\centering
 \begin{tikzpicture}
\node[anchor=south west,inner sep=0] (image) at (0,0) {\includegraphics[width=\textwidth]{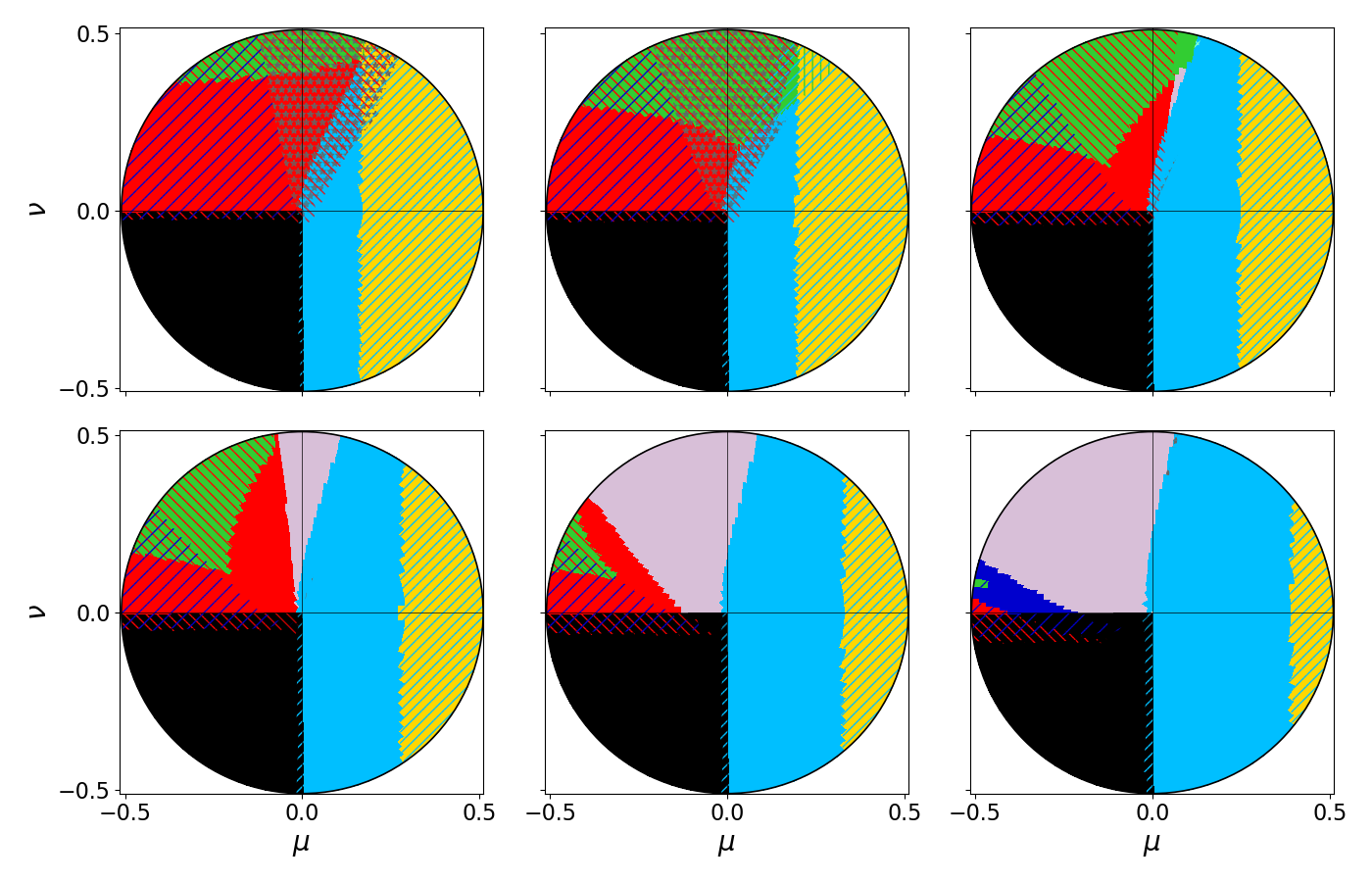}};
\begin{scope}[x={(image.south east)},y={(image.north west)}]
  \node[fill=none] at (0.108,0.943) {\small \textbf{(a)}};
  \node[fill=none] at (0.42,0.943) {\small \textbf{(b)}};
    \node[fill=none] at (0.727,0.943) {\small \textbf{(c)}};
  \node[fill=none] at (0.108,0.485) {\small \textbf{(d)}};
    \node[fill=none] at (0.42,0.485) {\small \textbf{(e)}};
  \node[fill=none] at (0.727,0.485) {\small \textbf{(f)}};
\end{scope}
\end{tikzpicture}
\vskip -0.2cm
\includegraphics[width=\textwidth]{QzzQzwnegKwwLwzpos/ODEcolourbar_all_hatchings.png}
\caption{Weakly nonlinear predictions for $Q_1=-1.06$, $Q_2=-2$, $Q_3=-0.8$, $C_2=-5$ and $C_3=-15$. 
The value of $C_1$ is different for each subfigure: (a)~$-2.4$, 
(b)~$-2.0$, 
(c)~$-1.6$, 
(d)~$-1.4$, 
(e)~$-1.2$, 
(f)~$-1.0$. 
$K_{ww}L_{wz}$~changes sign at $C_1 \approx -2.29$. 
A large region with no stable simple equilibria (lavender) emerges for $K_{ww}L_{wz}<0$.}
\label{Kruns:ODE}
\end{figure}

The stability region of the simple equilibria using weakly nonlinear theory is shown in \cref{Kruns:ODE}. 
The most striking feature is how the regions of stability of $w$-stripes and $w$-hexagons shrink as $C_1$ is increased, which results in large lavender regions where none of the simple equilibria are stable.
In addition, the star equilibria, which are clearly present in \cref{Kruns:ODE}{\color{siaminlinkcolor}a,b}, are almost entirely absent in the other four panels.
There is an increasingly large lavender region of no stable simple equilibria in \cref{Kruns:ODE}{\color{siaminlinkcolor}c--f}.
Unlike in \cref{3WIs}, the lavender region is not bounded by a Hopf bifurcation from a stripe solution.

\begin{figure}
\centering
\begin{tikzpicture}
\node[anchor=south west,inner sep=0] (image) at (0,0)
{\includegraphics[width=\textwidth]{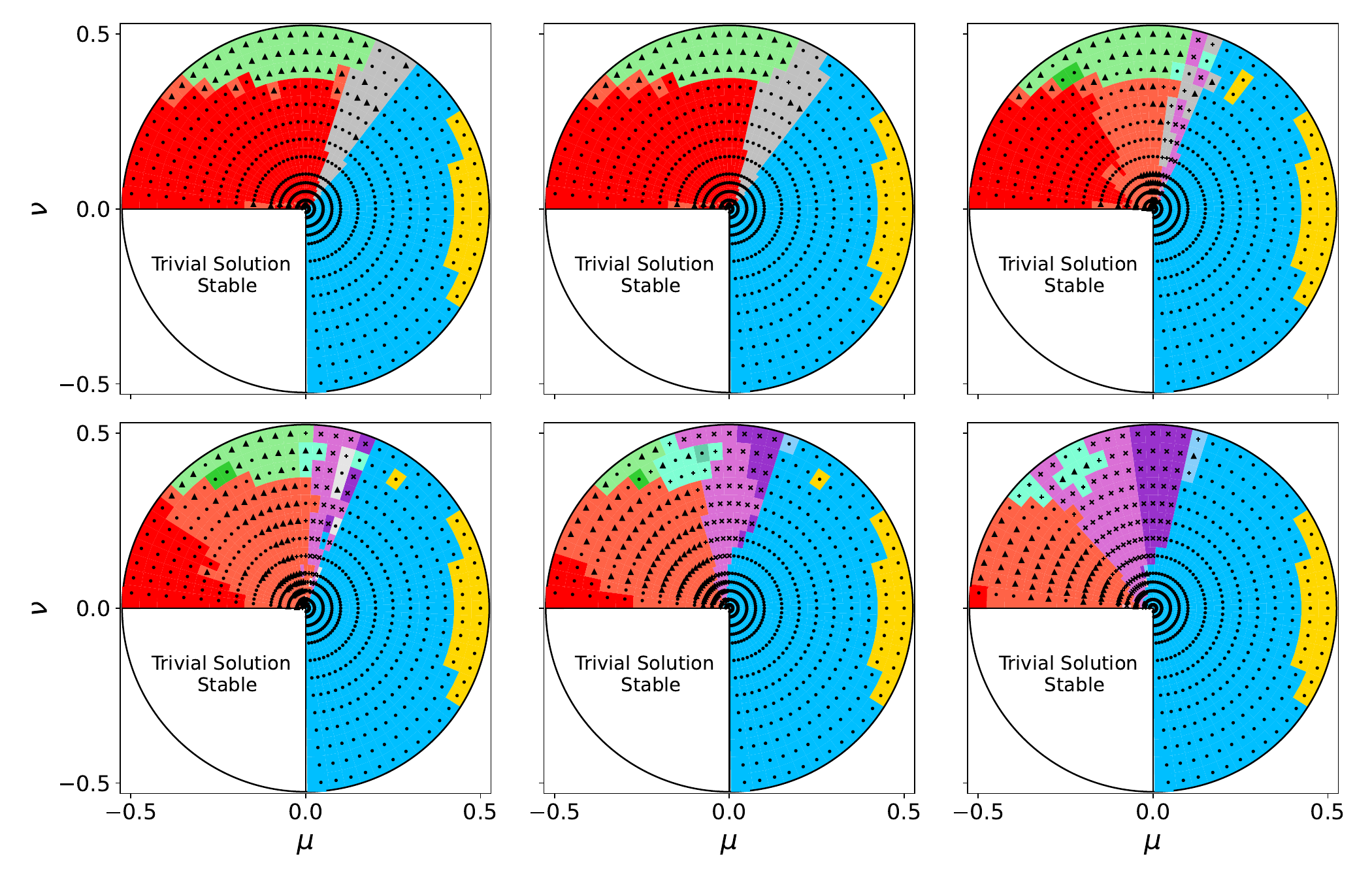}};
\begin{scope}[x={(image.south east)},y={(image.north west)}]
  \node[fill=none] at (0.108,0.943) {\small \textbf{(a)}};
  \node[fill=none] at (0.42,0.943) {\small \textbf{(b)}};
    \node[fill=none] at (0.727,0.943) {\small \textbf{(c)}};
  \node[fill=none] at (0.108,0.486) {\small \textbf{(d)}};
    \node[fill=none] at (0.42,0.486) {\small \textbf{(e)}};
  \node[fill=none] at (0.727,0.486) {\small \textbf{(f)}};
\end{scope}
\end{tikzpicture}
\vskip -0.35cm
\includegraphics[width=\textwidth]{QzzQzwnegKwwLwzpos/PDEcolourbar_new_chaos.png}
\caption{Bifurcation set showing the patterns observed with varying $(\mu, \nu)$ within the PDE \cref{PDE:2}. The PDE parameters are the
same as in \cref{Kruns:ODE}, with $\sigma_0=-2$.}
\label{4WI:PDE}
\end{figure}

\begin{figure}
\centering
\begin{tikzpicture}
\node[anchor=south west,inner sep=0] (image) at (0,0)
{\includegraphics[width=\textwidth]{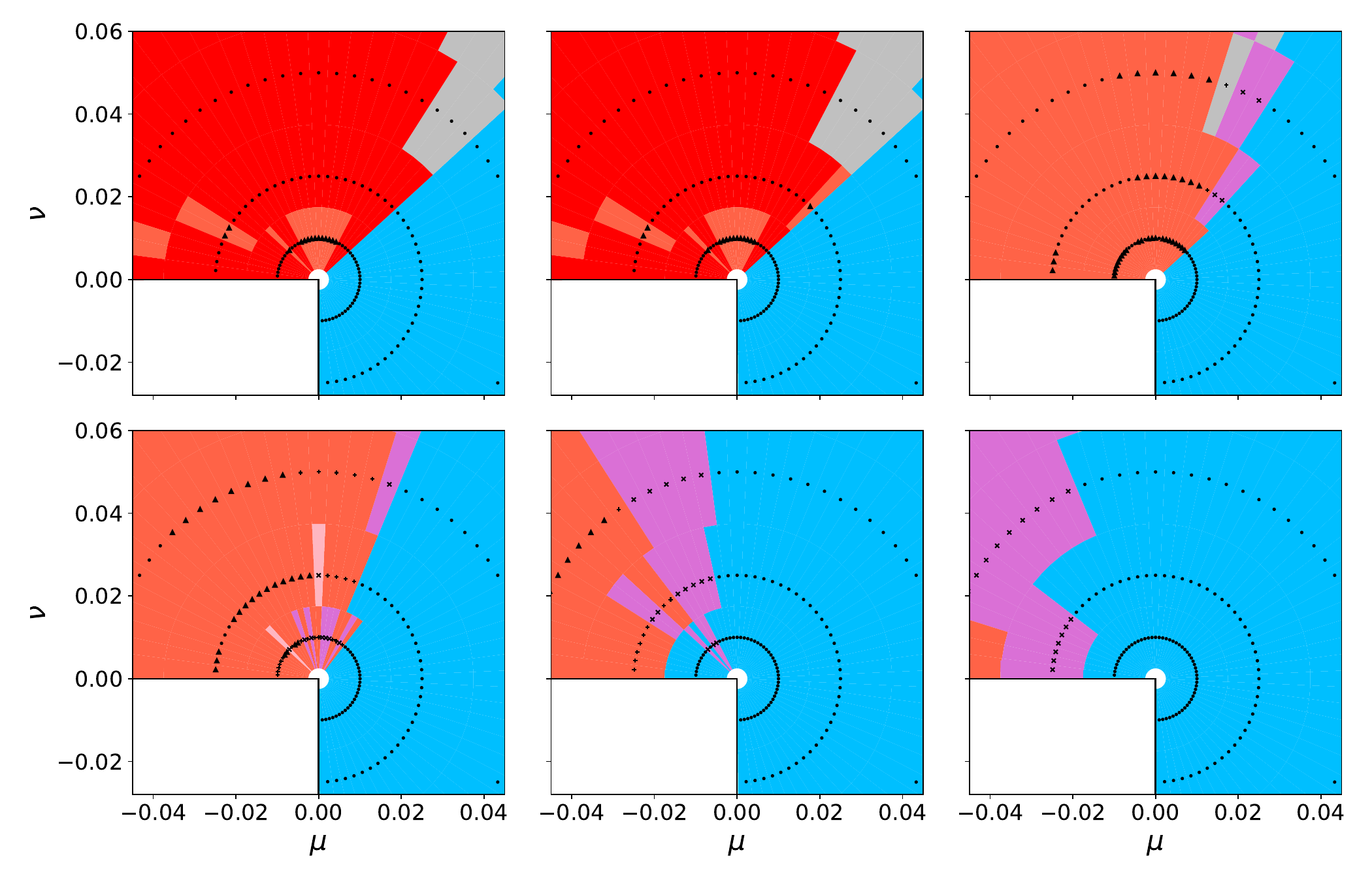}};
\begin{scope}[x={(image.south east)},y={(image.north west)}]
  \node[fill=none] at (0.12,0.579) {\small \textbf{(a)}};
  \node[fill=none] at (0.425,0.579) {\small \textbf{(b)}};
    \node[fill=none] at (0.727,0.579) {\small \textbf{(c)}};
  \node[fill=none] at (0.12,0.125) {\small \textbf{(d)}};
    \node[fill=none] at (0.425,0.125) {\small \textbf{(e)}};
  \node[fill=none] at (0.727,0.125) {\small \textbf{(f)}};
\end{scope}
\end{tikzpicture}
\vskip -0.05cm
\includegraphics[width=\textwidth]{QzzQzwnegKwwLwzpos/PDEcolourbar_new_chaos.png}
\caption{Zoomed in bifurcation set of \cref{4WI:PDE} for small $\mu$ and $\nu$.}
\label{4WI:PDE:zoom}
\end{figure}

Results from the PDE simulations are shown in \cref{4WI:PDE,4WI:PDE:zoom}.
We observe $w$-hexagons with defects persisting for a much larger region of parameter space than expected from the weakly nonlinear prediction of (off-critical) $w$-hexagons. 
However, we see no examples of defect-free $w$-hexagons outside of the predicted stable regions. 
In \cref{4WI:PDE}{\color{siaminlinkcolor}a}, the calculations are started from random initial conditions, and $z$- and $w$-stripes appear only in the parameter regions where they are stable in the ODEs, with an almost-exact match in the case of $w$-stripes.
The initial conditions for each subsequent panel come from the previous one, so the regions of $z$- and $w$-stripes are ``frozen-in''.
In the case of $w$-stripes, the initial region of their stability no longer matches the ODE region of stability for the later values of~$C_1$, so $w$-stripes progressively disappear from \cref{4WI:PDE}{\color{siaminlinkcolor}d--f}.
Similarly, in \cref{4WI:PDE}{\color{siaminlinkcolor}a}, stars appear only in the parameter regions where they are stable in the ODEs, and they progressively disappear as $C_1$ is increased.
In fact, upon closer inspection of these PDE star solutions, we find the amplitudes of the $k=1$ modes on each hexagonal lattice are not all equal but are similar, so these stars are slightly asymmetric.
We have not computed the amplitudes for asymmetric stars in the ODEs, so whilst this pattern looks very much like stars, its stability will differ from that of symmetric stars.

Regions of TC appear from \cref{4WI:PDE}{\color{siaminlinkcolor}c}, and increase in size as $C_1$ is increased, with STC appearing in \cref{4WI:PDE}{\color{siaminlinkcolor}d} and similarly increasing in size.
By $C_1=-1$ in \cref{4WI:PDE}{\color{siaminlinkcolor}f}, with the most negative value of $K_{ww}L_{wz}$, STC occupies a substantial portion of the bifurcation set, mainly for larger~$\nu$ and $\mu$ small and positive. 
The regions of TC and STC generally overlap the regions where the ODEs predict no stable simple equilibria. 
In \cref{4WI:PDE:zoom}{\color{siaminlinkcolor}e,f}, TC extends to the smallest values of $\mu$ and~$\nu$.

\begin{figure}
\centering
\begin{tikzpicture}
\node[anchor=south west,inner sep=0] (image) at (0,0)
{\includegraphics[width=153mm]{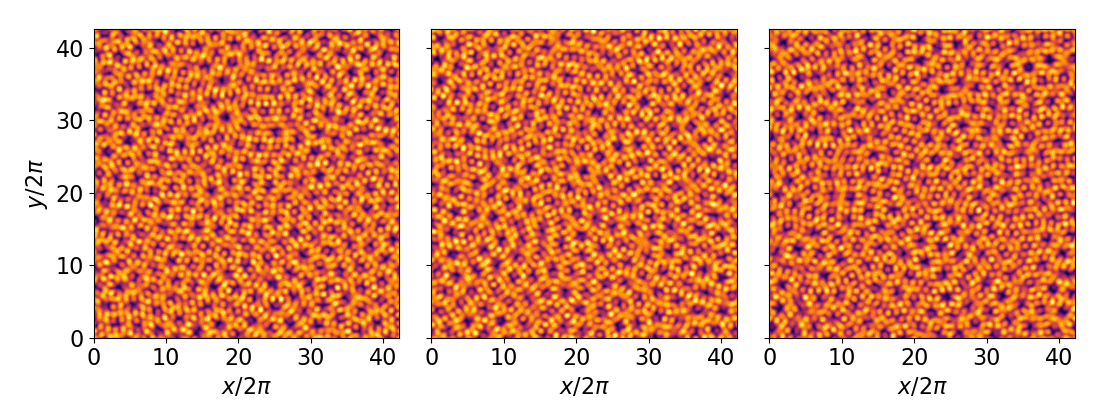}};
\begin{scope}[x={(image.south east)},y={(image.north west)}]
  \node[fill=none] at (0.013,0.93) {\small \textbf{(a)}};
\end{scope}
\end{tikzpicture}
\begin{tikzpicture}
\node[anchor=south west,inner sep=0] (image) at (0,0)
{\includegraphics[width=153mm]{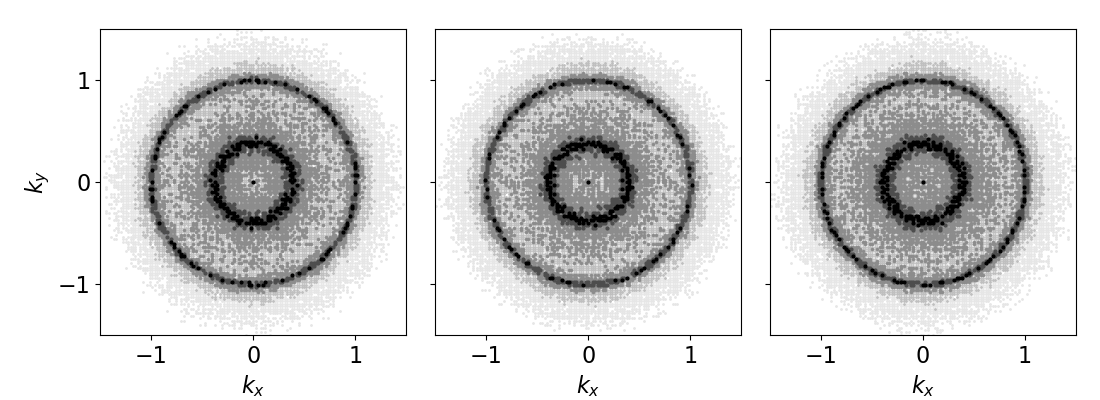}};
\begin{scope}[x={(image.south east)},y={(image.north west)}]
  \node[fill=none] at (0.013,0.93) {\small \textbf{(b)}};
\end{scope}
\end{tikzpicture}
\caption{Spatiotemporal chaos with $r=0.5$ and $\chi = 80^\circ$ where $(\mu,\nu) = (r\cos \chi, r \sin \chi)$, and with $C_1=-1$, as in \cref{4WI:PDE}{\color{siaminlinkcolor}f}.
These parameter values give $Q_{zz}Q_{zw}>0$ and $K_{ww}L_{wz}<0$. (a)~shows the patterned solution at different times and (b)~the corresponding Fourier spectrum. 
Each frame is separated by 90 time units. 
We note the density of modes clustered around both entire circles.}
\label{STC}
\end{figure}

We show an example of fully developed STC in \cref{STC}, with three panels at different times, with parameters chosen at the outer edge of the purple region in \cref{4WI:PDE}{\color{siaminlinkcolor}f}.
In the Fourier power spectra, annuli around both $k=1$ and $k=q$ are filled, and the pattern is continually changing in both space and time.
For smaller values of $\mu$ and~$\nu$, the number of modes contributing to spatiotemporally chaotic solutions tends to decrease, leading to the progressive simplification of the spatial structure, with larger patches of hexagons with the same orientation.

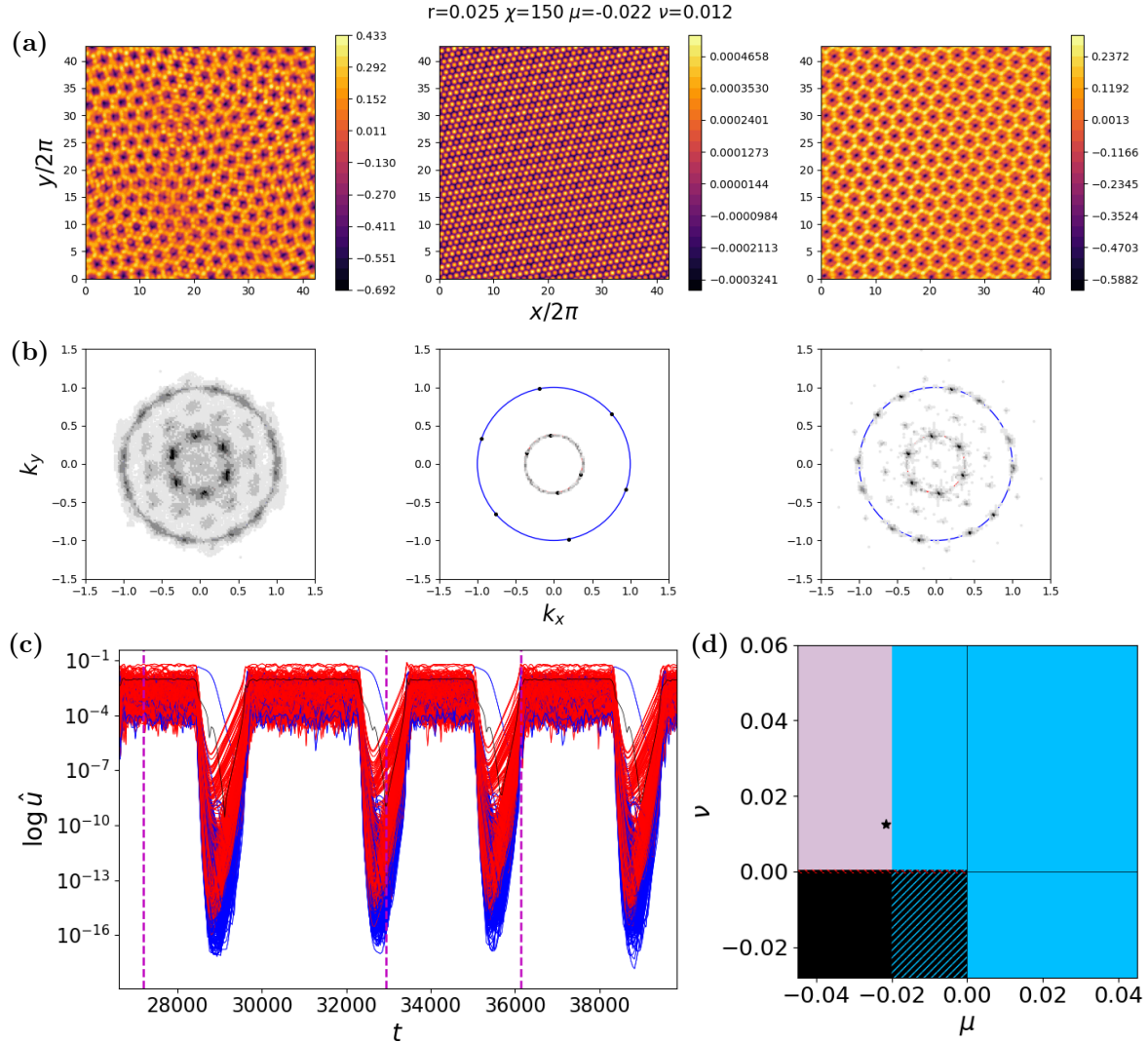
\begin{figure}
\captionsetup[subfloat]{labelformat=empty} 
\centering
\begin{tikzpicture}
\node[anchor=south west,inner sep=0] (image) at (0,0)
{\includegraphics[width=153mm]{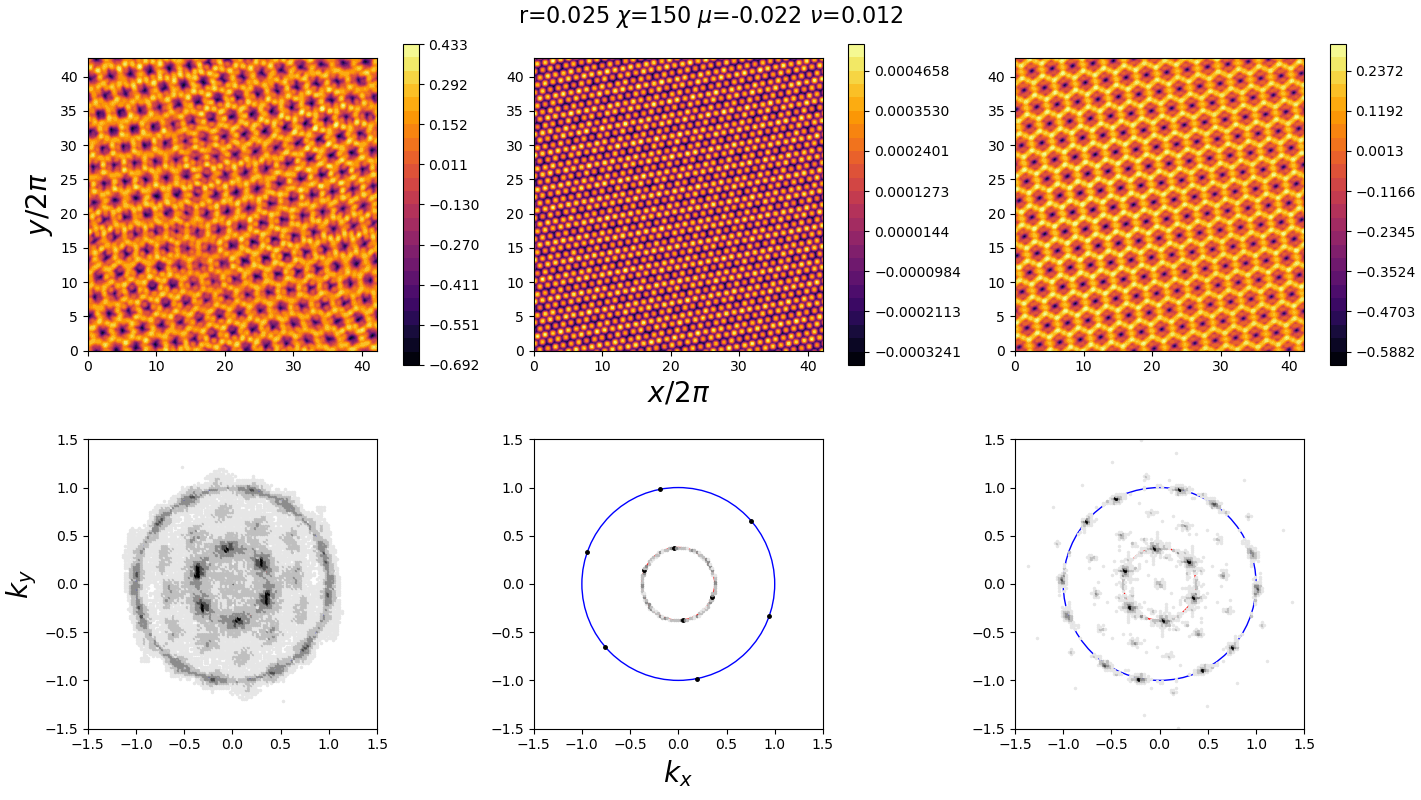}};
\begin{scope}[x={(image.south east)},y={(image.north west)}]
  \node[fill=none] at (0.013,0.93) {\small \textbf{(a)}};
  \node[fill=none] at (0.013,0.445) {\small \textbf{(b)}};
\end{scope}
\end{tikzpicture}
\vskip -0.2cm
\begin{tikzpicture}
\node[anchor=south west,inner sep=0] (image) at (0,0)
{\hspace{-1.5cm}
\includegraphics[width=95mm]{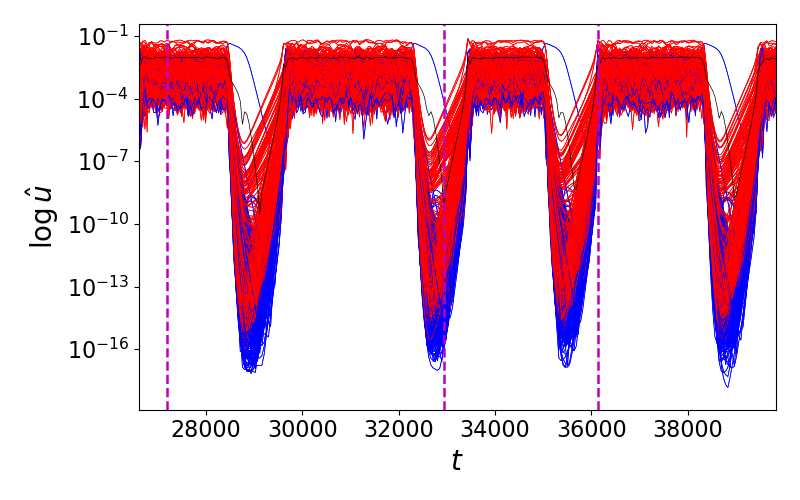}};
\node[anchor=south west,inner sep=0] (image2) at (8,0.36)
{\includegraphics[width=62mm]{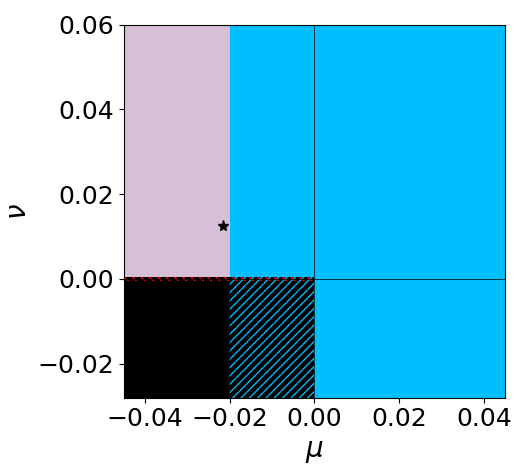}};
\begin{scope}[x={(image.south east)},y={(image.north west)}]
  \node[fill=none] at (-0.12,0.96) {\small \textbf{(c)}};
  \node[fill=none] at (1.02,0.96) {\small \textbf{(d)}};
\end{scope}
\end{tikzpicture}
\caption{Time evolution of an intermittent temporally chaotic solution of the PDE with $r=0.025$ and $\chi = 150^\circ$ where $(\mu,\nu) = (r\cos \chi, r \sin \chi)$, and with $C_1=-1$, as in \cref{4WI:PDE:zoom}{\color{siaminlinkcolor}f}. 
The three patterned states in~(a) show a solution at different values of time, indicated by the dashed magenta lines in~(c).  
Note the colorbar is different at each time. 
The larger contributions to the patterned state are represented by darker and larger markers in the Fourier spectrum in~(b).
The red, blue, black curves in~(c) correspond to the evolution of the $k=q$, $k=1$, $k=0$ modes respectively.
Panel~(d) shows the zoomed ODE stability predictions for these parameter values, the black star marking the values of $\mu$ and $\nu$ used in the PDE simulation.}
\label{smallchaos:4WI}
\end{figure}

We find examples of intermittent TC at the smallest values of $\mu$ and~$\nu$, with apparently heteroclinic connections between a temporally chaotic saddle and a $z$-hexagon pattern. 
One example is shown in \cref{smallchaos:4WI}, for parameter values close to the saddle-node bifurcation that limits the existence of $z$-hexagons (\cref{smallchaos:4WI}{\color{siaminlinkcolor}d}).
This solution displays a strong resemblance to type-I intermittency~\cite{Pomeau1980}.
In \cref{smallchaos:4WI}, we start with a temporally chaotic solution (left panels in \cref{smallchaos:4WI}{\color{siaminlinkcolor}a,b}). 
There is then a stage where the amplitudes of most modes decay exponentially, leaving only $z$-hexagons (center panels).
Trajectories get temporarily trapped near $z$-hexagons, but these are a transient, as the parameter values are outside the region of existence of $z$-hexagons.
As the $z$-hexagon modes decay, the dominant growing modes are on the $k=q$ circle and are aligned with the $z$-hexagon modes.
The reasoning for the growth of these $w$~modes is explored in \cref{onset of chaos}.
Trajectories return to the chaotic saddle, with a brief phase of resembling superhexagons (right panels).
These superhexagons do not contain the same $z$~modes as those in the $z$-hexagon phase, but instead the superhexagon pattern from the aligned $w$~modes and corresponding modes from 3WIs.
The time spent near the chaotic saddle varies from cycle to cycle (\cref{smallchaos:4WI}{\color{siaminlinkcolor}c}), according to how long it takes to find the $z$-hexagon trapping region again.
Similar to \cref{smallchaos}, the dominant modes in the PDE solution are the eighteen modes used in the ODEs (but rotated), but in this case, the peaks are fuzzy in the temporally chaotic phase and sharp in the $z$-hexagon phase (\cref{smallchaos:4WI}{\color{siaminlinkcolor}a,b}), as opposed to being sharp throughout the evolution in \cref{smallchaos}.
The presence of this kind of intermittent chaos is associated, in other problems, with the presence, at nearby parameter values, of spatiotemporal chaos~\cite{YANAGITA1995288,Rempel2007a,Rempel2007b}.

\FloatBarrier

\subsection{Chaos Driven by both Three-Wave and Four-Wave Interactions} \label{3and4WI}
So far, we have presented examples where the 3WIs and 4WIs separately promote time dependence.
In this section, we explore the case where both types of interaction promote time dependence by focusing on $Q_{zz}Q_{zw}<0$ and $K_{ww}L_{wz}<0$.
We begin in the $Q_{zz}Q_{zw}>0$ and $K_{ww}L_{wz}<0$ quadrant of \cref{numerical schematic} and change the value of $Q_1$ between simulations to decrease the value of $Q_{zz}Q_{zw}$ below zero.
We use the final states of \cref{4WI:PDE}{\color{siaminlinkcolor}f} as initial conditions for our first set of simulations, varying~$Q_1$ but keeping the other PDE parameters unchanged.

\begin{figure}
\centering
\vskip -0.1cm
\begin{tikzpicture}
\node[anchor=south west,inner sep=0] (image) at (0,0) 
{\includegraphics[width=0.82\textwidth]{QzzQzwnegKwwLwzpos/ODEcolourbar_all_hatchings.png}};
\begin{scope}[x={(image.south)},y={(image.north)}]
  \node[fill=none] at (-0.001,1.1) {\small Weakly nonlinear colorbar};
\end{scope}
\end{tikzpicture}
\vskip -0.3cm
\begin{tikzpicture}
\node[anchor=south west,inner sep=0] (image) at (0,0) 
{\includegraphics[width=102mm]{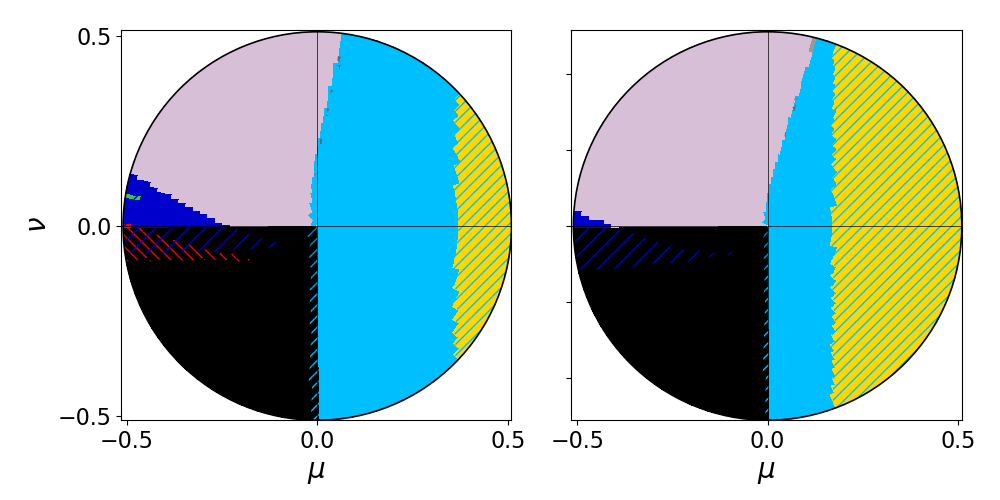}};
\begin{scope}[x={(image.south east)},y={(image.north west)}]
  \node[fill=none] at (0.315,1) {\small \textbf{(a)}};
  \node[fill=none] at (0.765,1) {\small \textbf{(b)}};
\end{scope}
\end{tikzpicture}
\vskip -0.1cm
\begin{tikzpicture}
\node[anchor=south west,inner sep=0] (image) at (0,0) 
{\includegraphics[width=102mm]{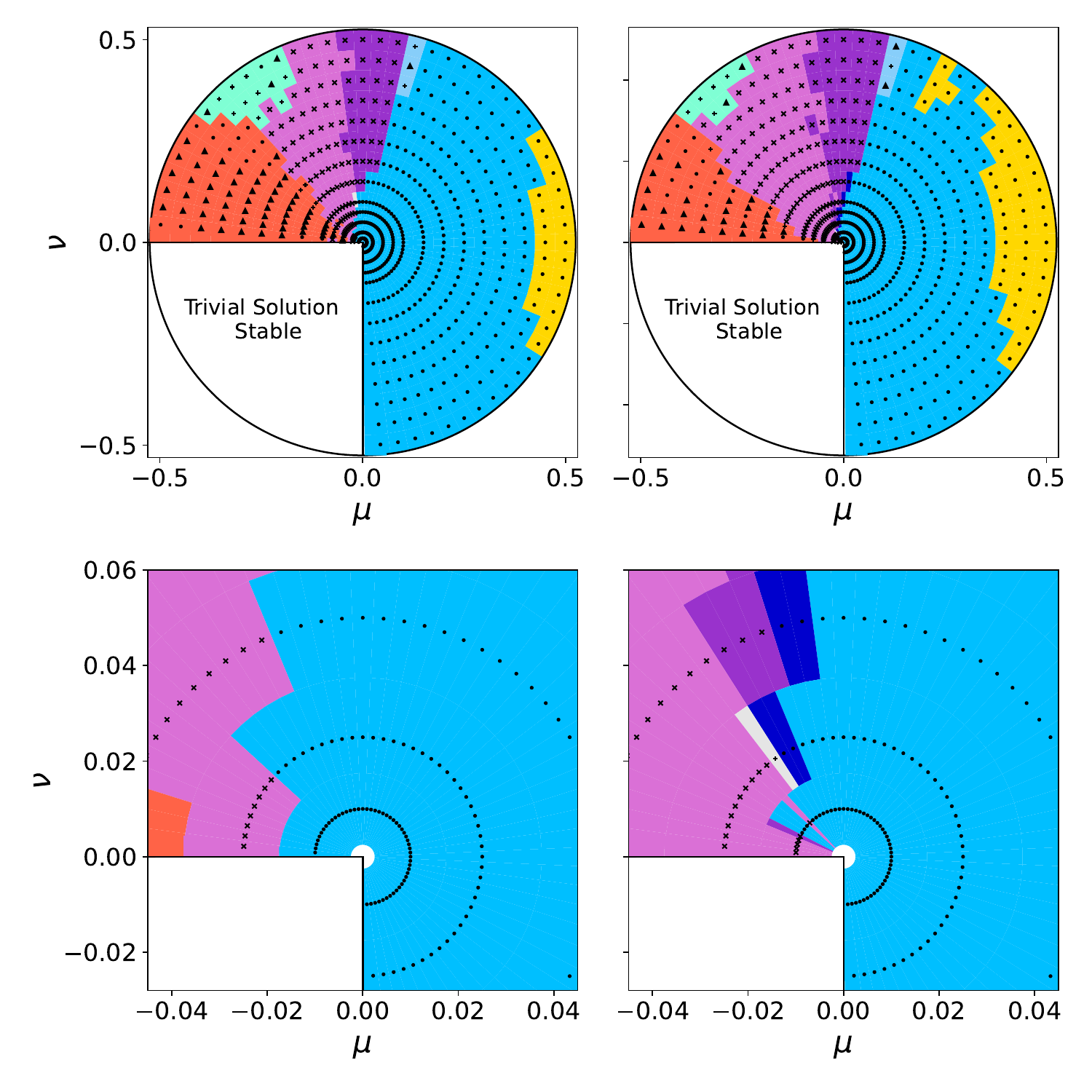}};
\end{tikzpicture}
\begin{tikzpicture}
\node[anchor=south west,inner sep=0] (image) at (0,0) 
{\includegraphics[width=0.85\textwidth]{QzzQzwnegKwwLwzpos/PDEcolourbar_new_chaos.png}};
\begin{scope}[x={(image.south)},y={(image.north)}]
  \node[fill=none] at (0.04,1) {PDE colorbar};
\end{scope}
\end{tikzpicture}
\caption{Weakly nonlinear predictions (top row) and PDE results (middle and bottom rows) for $Q_2=-2$, $Q_3=-0.8$, $C_1=-1$, $C_2=-5$, $C_3=-15$ and $\sigma_0=-2$. 
The values of $Q_1$ are: 
(a)~$-1.08$, (b)~$-1.24$. 
$Q_{zz}Q_{zw}$ changes sign for $Q_1 \approx -1.0857$. The region for $Q_{zz}Q_{zw}<0$ is extremely small; we observe very little change in the chaotic region as $Q_1$ is decreased.}
\label{QnegKnegruns}
\end{figure}

For the PDE parameter values in \cref{4WI:PDE}{\color{siaminlinkcolor}f}, there is only a limited range of $Q_1$ that has $Q_{zz}Q_{zw}<0$, and the magnitude of $Q_{zz}Q_{zw}$ does not change greatly over this range.
In our bifurcation sets, in \cref{QnegKnegruns}, we therefore show only two value of~$Q_1$, the first and last of the range we investigated.
\cref{QnegKnegruns} shows the stability predictions from the weakly nonlinear theory (top row) and PDE solutions (middle and lower rows) for these two values of~$Q_1$. 

The most significant changes are an increase in the size of the lavender region of absence of stable simple equilibria in the ODEs, and (more significantly) an increase in the size of the chaotic region for small $\mu$ and~$\nu$ in the~\hbox{PDE}, including examples of~\hbox{STC}.
The STC is qualitatively similar to that in \cref{STC}, though with larger patches of hexagons with the same orientation.
Among the examples of TC, there are also a few examples of intermittent chaos, which we describe in more detail below. 
We also observe a large region of $w$-hexagons with defects within the PDE results, left over, as argued above, from the initial results in \cref{4WI:PDE}{\color{siaminlinkcolor}a}.
There are a small number of examples of stable superhexagons (in dark blue in the lower panel of \cref{QnegKnegruns}{\color{siaminlinkcolor}b}), which do not overlap with the stable region of superhexagons in the ODEs, and which we have not found in the~PDE for any other parameter choices.

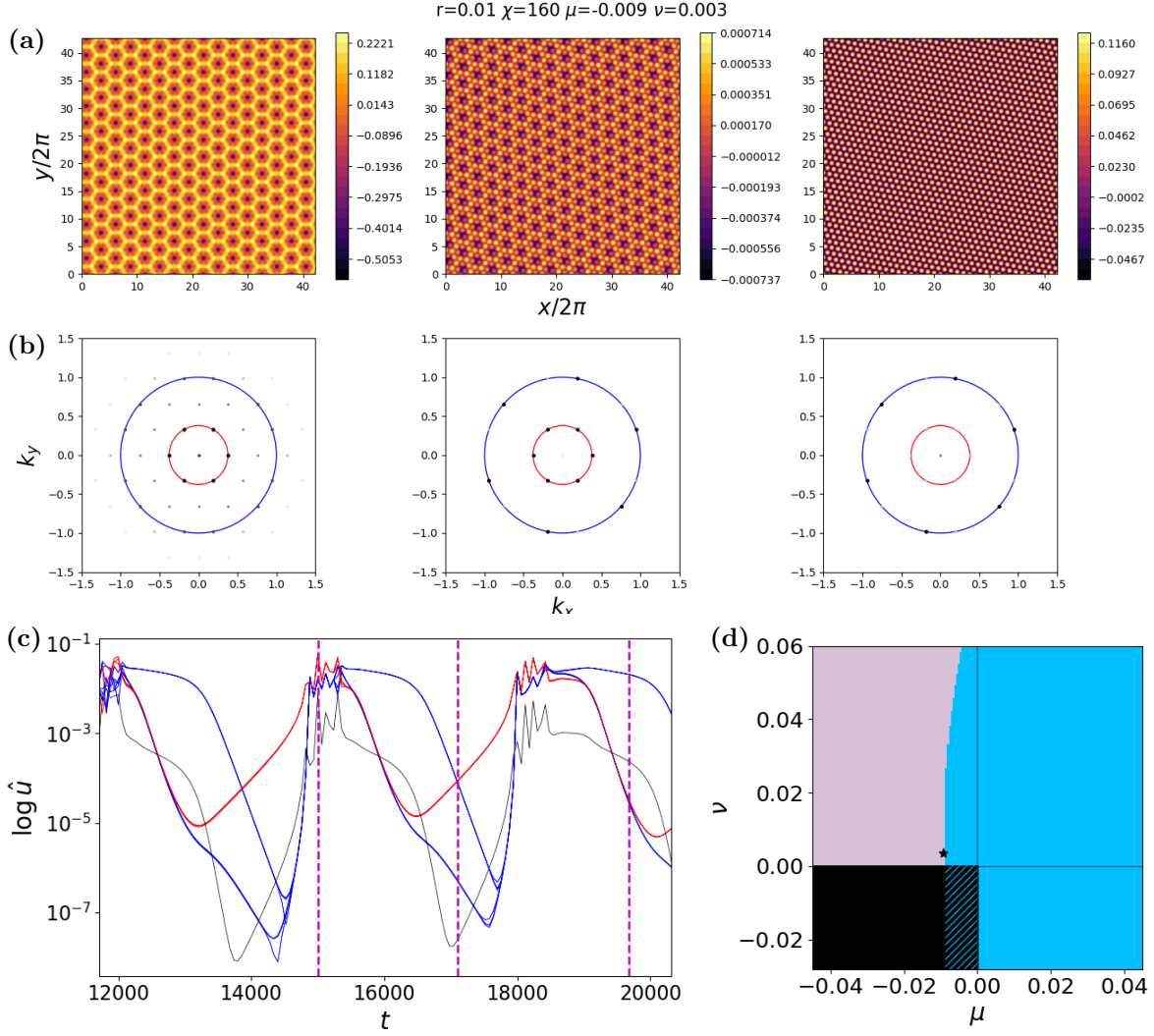
\begin{figure}
\captionsetup[subfloat]{labelformat=empty} 
\centering
\begin{tikzpicture}
\node[anchor=south west,inner sep=0] (image) at (0,0)
{\includegraphics[width=153mm]{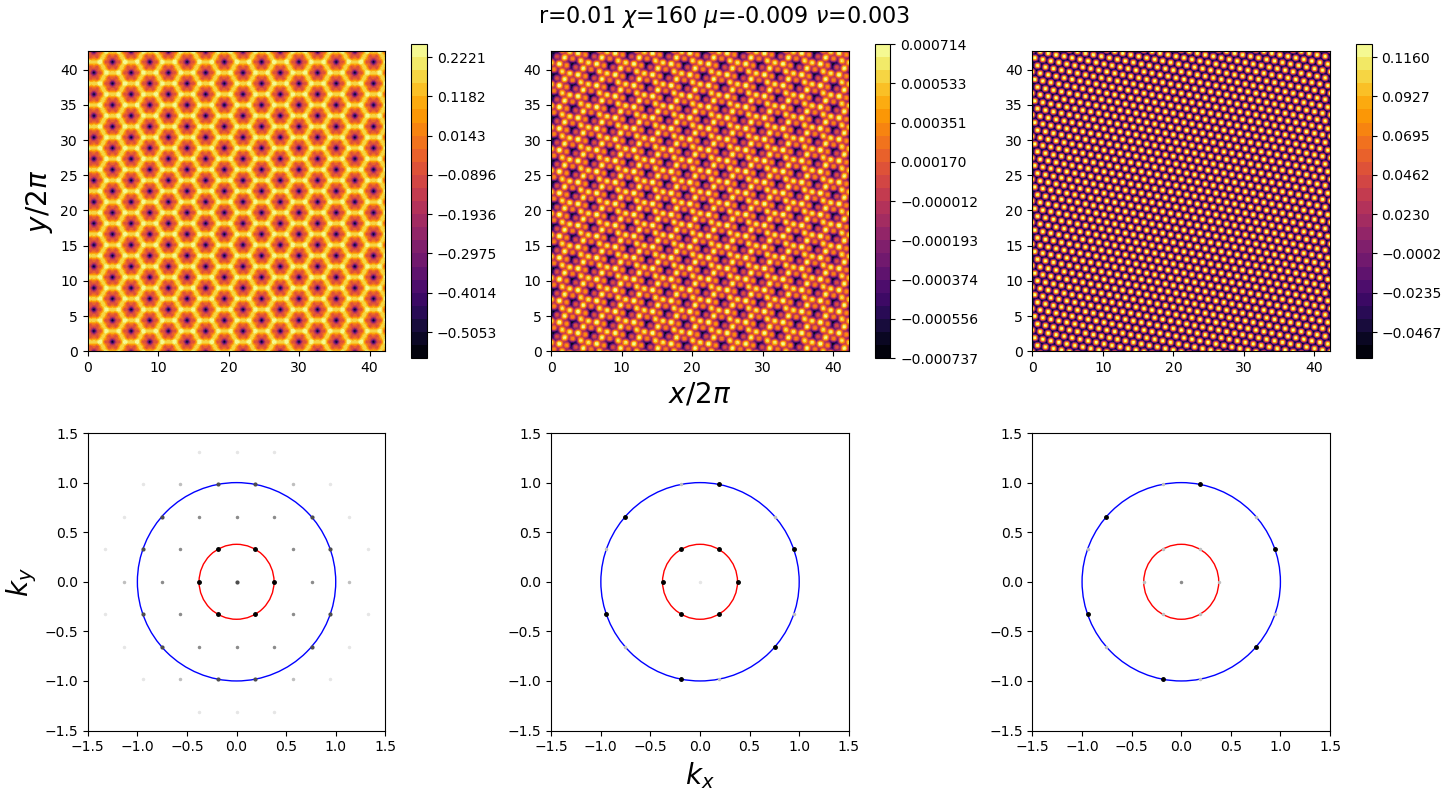}};
\begin{scope}[x={(image.south east)},y={(image.north west)}]
  \node[fill=none] at (0.013,0.93) {\small \textbf{(a)}};
  \node[fill=none] at (0.013,0.445) {\small \textbf{(b)}};
\end{scope}
\end{tikzpicture}
\vskip -0.2cm
\begin{tikzpicture}
\node[anchor=south west,inner sep=0] (image) at (-1.5,0){\hspace{-1.3cm}
\includegraphics[width=95mm]{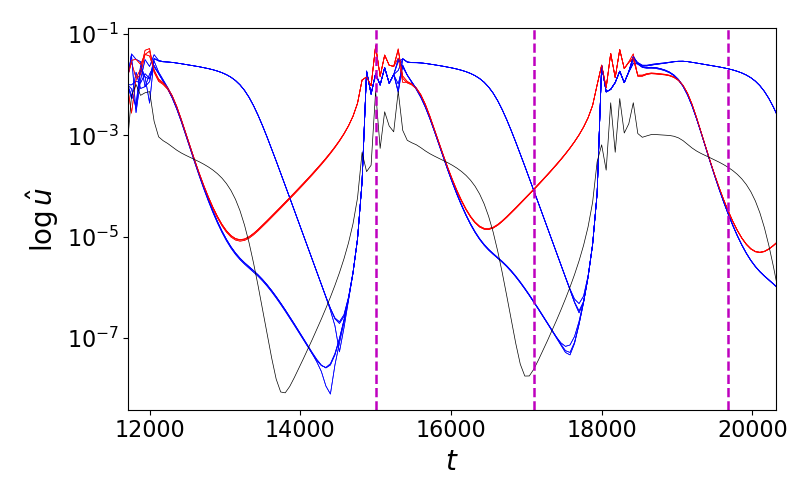}};
\node[anchor=south west,inner sep=0] (image2) at (7,0.36){\includegraphics[width=60mm]{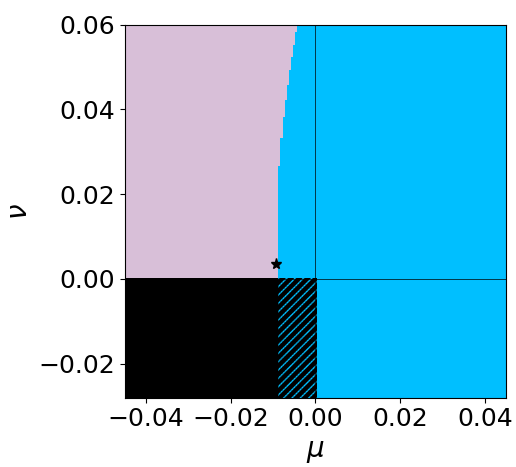 }};
\begin{scope}[x={(image.south east)},y={(image.north west)}]
  \node[fill=none] at (-0.11,0.94) {\small \textbf{(c)}};
  \node[fill=none] at (1.28,0.94) {\small \textbf{(d)}};
\end{scope}
\end{tikzpicture}
\caption{Time evolution of an intermittent temporally chaotic solution of the PDE with $r=0.01$ and $\chi = 160^\circ$ where $(\mu,\nu) = (r\cos \chi, r \sin \chi)$, and with $Q_1=-1.24$, as in \cref{QnegKnegruns}{\color{siaminlinkcolor}b}. 
The layout of the panels and the meaning of the colors are as in \cref{smallchaos:4WI}.}
\label{smallchaos:34WI}
\end{figure}

For small $\mu$ and~$\nu$, we see a type of chaotic solution (\cref{smallchaos:34WI}) that has features of both \cref{smallchaos}, where the dynamics are concentrated on the eighteen modes of the ODEs, and \cref{smallchaos:4WI}, with intermittency between chaos and $z$-hexagons.
The intermittency is similar to that in \cref{smallchaos:4WI}: there is a short period of chaotic dynamics that persists until the trajectory is trapped close to $z$-hexagons. 
These disappear and the chaos resumes.
This behavior exists just outside the existence and stability boundary of $z$-hexagons and looks similar to type~I intermittency~\cite{Pomeau1980}, as we also found in \cref{4WIs}.
Unlike in \cref{smallchaos:4WI}, there is no fuzziness in the Fourier spectrum, and (like \cref{smallchaos}), there are twelve (six) modes on the $k=1$ ($k=q$) circle.

\FloatBarrier

\section{Onset of Chaos} \label{onset of chaos}
So far, we have found examples of chaotic dynamics within the eighteen-mode subspace of the problem (\cref{smallchaos,smallchaos:34WI}), examples of TC where the eighteen modes are still visible but are fuzzy (\cref{smallchaos:4WI}), and examples of fully developed~\hbox{STC} (\cref{STC}).
We started with the hypothesis 3WIs could explain the transitions between these possibilities.
In this section, we test this hypothesis in more detail, and find that the original idea plays a role, but that there are additional interactions to take into account.

\begin{figure}
\centering
\begin{tikzpicture}
\node[anchor=south west,inner sep=0] (image) at (0,0)
{\includegraphics[width=0.99\linewidth]{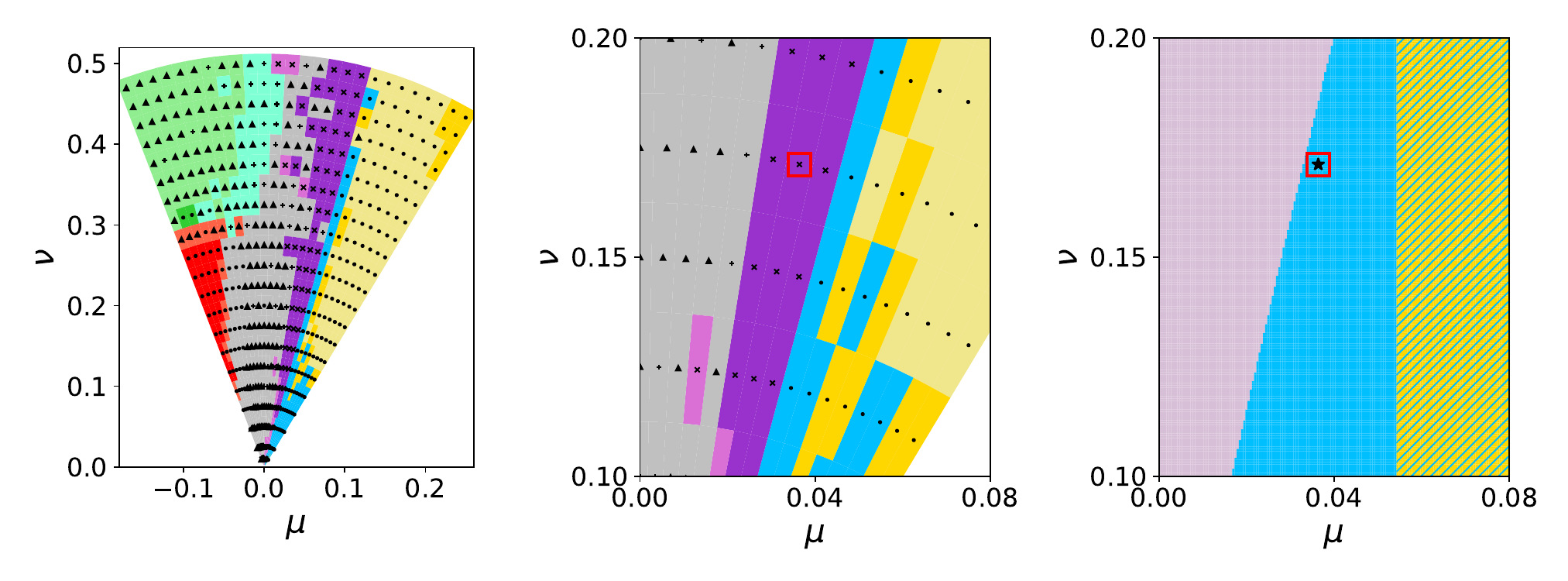}};
\begin{scope}[x={(image.south east)},y={(image.north west)}]
\node[fill=none] at (0.02,0.91) {\small \textbf{(a)}};
\node[fill=none] at (0.34,0.91) {\small \textbf{(b)}}; 
\node[fill=none] at (0.67,0.91) {\small \textbf{(c)}}; 
\end{scope}
\end{tikzpicture}
\caption{Bifurcation set showing the patterns observed with varying $(\mu,\nu)$ within the PDE \cref{PDE:2}. 
The PDE parameters and colors are as in \cref{3WI:PDE}{\color{siaminlinkcolor}d}.
We use a finer discretization of $(\mu,\nu)$ values that contains the observed region of chaos with the full range of $(\mu,\nu)$ values shown in~(a).
(b)~shows a zoomed section of (a).
(c)~shows a zoomed section of the weakly nonlinear stability predictions.
The marker contained within the red square in~(b) and (c) corresponds to the $(\mu,\nu)$ value we use to analyze the onset of chaos in \cref{onsettochaos,schematic:fourier,3WIswitharrows}.
}
\label{QnegKpossimruns7wedge}
\end{figure}
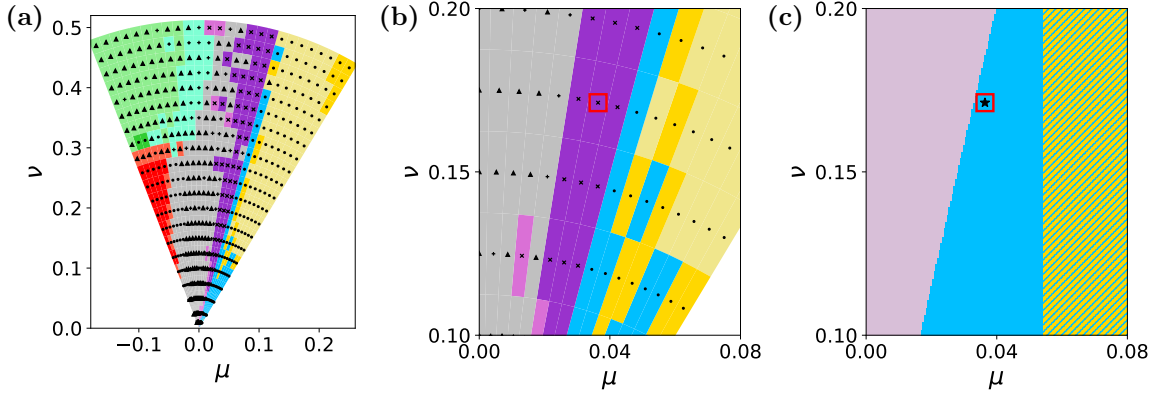

We choose PDE parameters respecting $Q_{zz}Q_{zw}<0$ and $K_{ww}L_{wz}>0$ (top left quadrant of \cref{numerical schematic}), so the time dependence is driven by 3WIs, and focus on the relationship between the nonlinear interactions and the onset of chaotic dynamics in the~\hbox{PDE}.
We solve the PDE for $q=1/\sqrt{7}$, $\sigma_0 = -2$, $Q_1 = -0.9$, $Q_2 = -2.75$, $Q_3 = -3.5$, $C_1=-2.75$, $C_2 = -7.75$ and $C_3 = -16.5$, the same conditions as in \cref{3WI:PDE}{\color{siaminlinkcolor}d}, with a finer discretization of~$(\mu,\nu)$, centered around the region where we previously found chaotic dynamics for these PDE parameters.
The new discretization takes radii separated by~$0.025$, with angles between $60^\circ$ and $110^\circ$ divided into $2^\circ$ increments, giving a total of 546 grid points.
Unlike in \cref{3WI:PDE}{\color{siaminlinkcolor}d}, the results shown in \cref{QnegKpossimruns7wedge} start from a small amplitude, random initial condition rather than from the end of a calculation at a different value of~$Q_1$. 

The PDE solutions reveal the same band of temporal chaos (light purple region) and spatiotemporal chaos (purple region) as in \cref{3WI:PDE}{\color{siaminlinkcolor}d}.
The onset of TC and STC occurs close to the stability boundaries of $z$-stripes and $z$-hexagons (\cref{QnegKpossimruns7wedge}{\color{siaminlinkcolor}a,b}).
We have investigated the transition from equilibrium solutions to STC starting from both $z$-stripes and $z$-hexagons.
The first of these shows the transitions more clearly, so we take a pure $z$-stripe solution (with its defects removed) from $r=0.175$ and $\chi = 74^\circ$ (where $(\mu,\nu) = (r\cos \chi , r \sin \chi))$ as an initial condition for $r=0.175$ and $\chi = 78^\circ$, which lies in the region of STC in \cref{QnegKpossimruns7wedge}{\color{siaminlinkcolor}a,b}.
These new values of $\mu$ and $\nu$ lie close to the lavender region and outside the stable $z$-stripes region of \cref{QnegKpossimruns7wedge}{\color{siaminlinkcolor}c} therefore we do not expect to remain at $z$-stripes.

\begin{figure}[!ht] 
\centering
\begin{tikzpicture}
\node[anchor=south west,inner sep=0] (image) at (0,0)
{\includegraphics[width=0.99\linewidth]{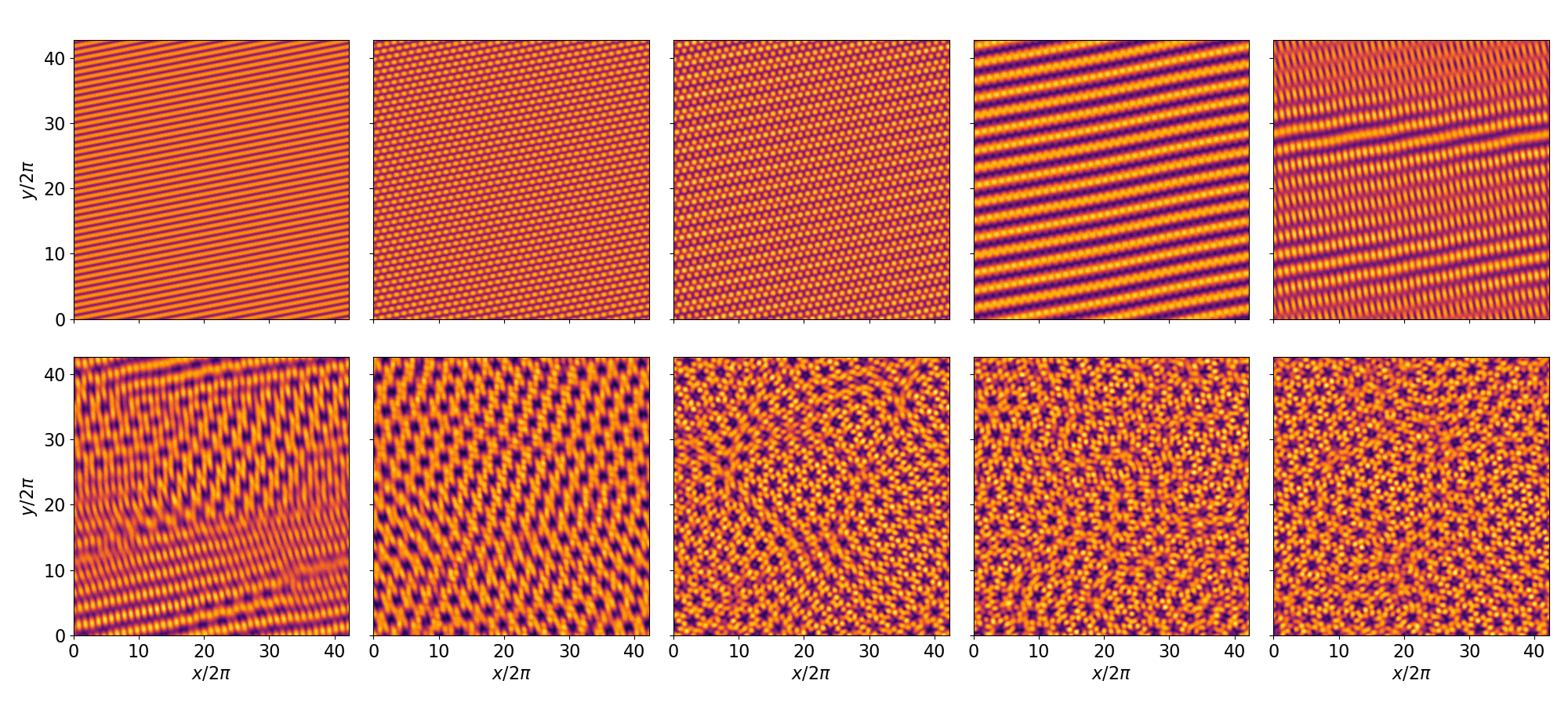}};
\begin{scope}[x={(image.south east)},y={(image.north west)}]
\node[fill=white,opacity=0.75] at (0.072,0.91) {\small \textbf{(a)}};
\node[fill=white,opacity=0.75] at (0.26,0.91) {\small \textbf{(b)}}; 
\node[fill=white,opacity=0.75] at (0.45,0.91) {\small \textbf{(c)}};
\node[fill=white,opacity=0.75] at (0.64,0.91) {\small \textbf{(d)}};
\node[fill=white,opacity=0.75] at (0.83,0.91) {\small \textbf{(e)}};
\node[fill=white,opacity=0.75] at (0.072,0.46) {\small \textbf{(f)}}; 
\node[fill=white,opacity=0.75] at (0.26,0.46) {\small \textbf{(g)}};  
\node[fill=white,opacity=0.75] at (0.45,0.46) {\small \textbf{(h)}};
\node[fill=white,opacity=0.75] at (0.64,0.46) {\small \textbf{(i)}}; 
\node[fill=white,opacity=0.75] at (0.83,0.46) {\small \textbf{(j)}}; \end{scope}
\end{tikzpicture}
\begin{tikzpicture}
\node[anchor=south west,inner sep=0] (image) at (0,0) 
{\includegraphics[width=0.99\linewidth]{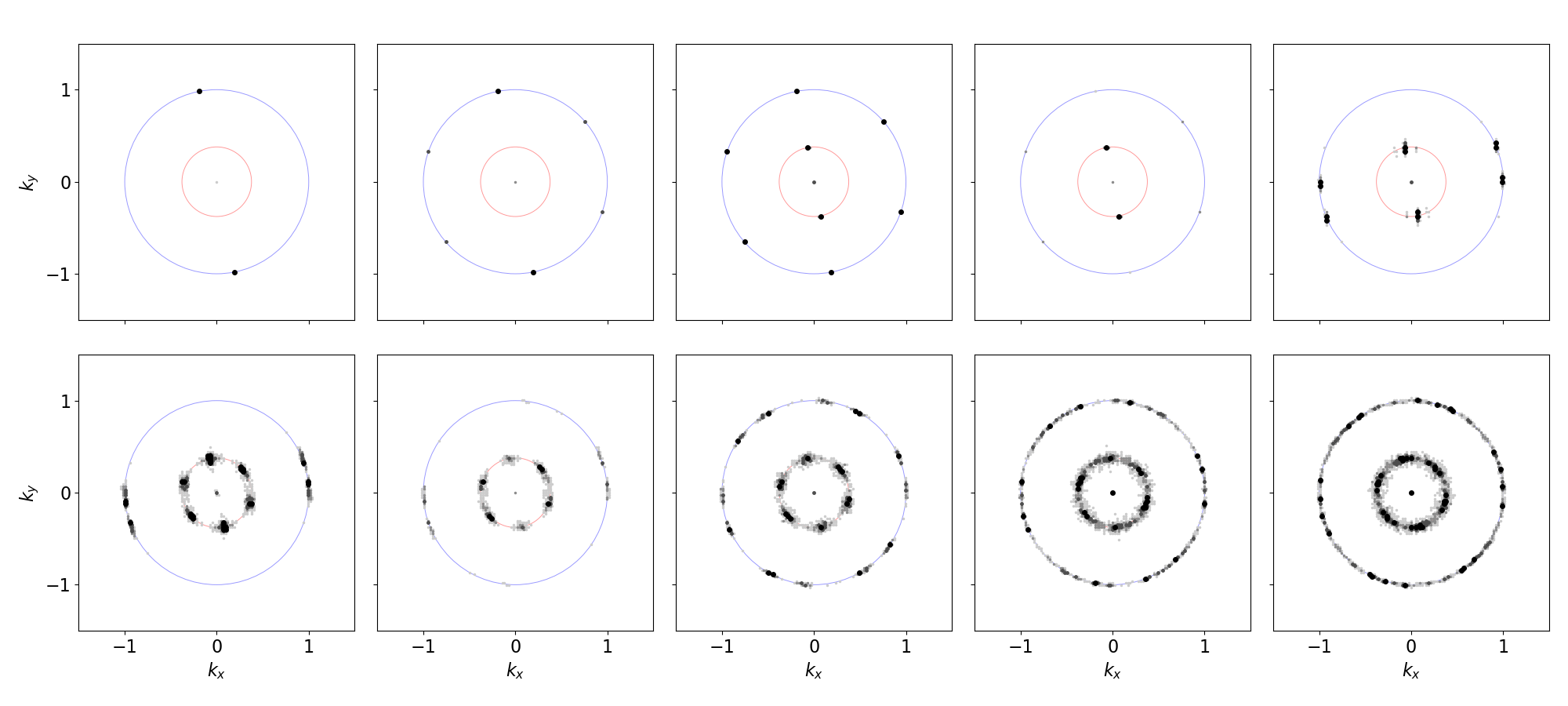}};
\begin{scope}[x={(image.south east)},y={(image.north west)}]
\node[fill=none] at (0.072,0.9) {\small \textbf{(a)}};
\node[fill=none] at (0.26,0.9) {\small \textbf{(b)}}; 
\node[fill=none] at (0.45,0.9) {\small \textbf{(c)}};
\node[fill=none] at (0.64,0.9) {\small \textbf{(d)}};
\node[fill=none] at (0.83,0.9) {\small \textbf{(e)}};
\node[fill=none] at (0.072,0.46) {\small \textbf{(f)}}; 
\node[fill=none] at (0.26,0.46) {\small \textbf{(g)}};  
\node[fill=none] at (0.45,0.46) {\small \textbf{(h)}};
\node[fill=none] at (0.64,0.46) {\small \textbf{(i)}}; 
\node[fill=none] at (0.83,0.46) {\small \textbf{(j)}};
\end{scope}
\end{tikzpicture}
\caption{A PDE solution (upper two rows) and its corresponding Fourier spectrum (lower two rows) at different times. 
The solution evolves between the following states:
(a)~$z$-stripes at $t=0$;
(b)~$z$-hexagons at $t=2112$; 
(c)~$z$-hexagons and aligned $w$-stripes at $t=2627$;
(d)~$w$-stripes and decaying $z$-hexagons at $t=2805$; 
(e)~modulated rhombs at $t=2871$, with details explained in \cref{3WIswitharrows}; 
(f)~patches of rhombs and $w$-hexagons at $t=2957$;
(g)~$w$-hexagons and multiple orientations of rhombs emerging at $t=3016$; 
(h)~temporal chaos at $t=3102$;
(i)~developing spatiotemporal chaos at $t=4791$;
(j)~spatiotemporal chaos at $t=6290$.
Here we have $r=0.175$ and $\chi = 78^\circ$ where $(\mu,\nu) = (r\cos \chi , r \sin \chi)$ with all other PDE parameters the same as \cref{QnegKpossimruns7wedge}. For clarity, only the largest Fourier components are plotted.}
\label{onsettochaos}
\end{figure}
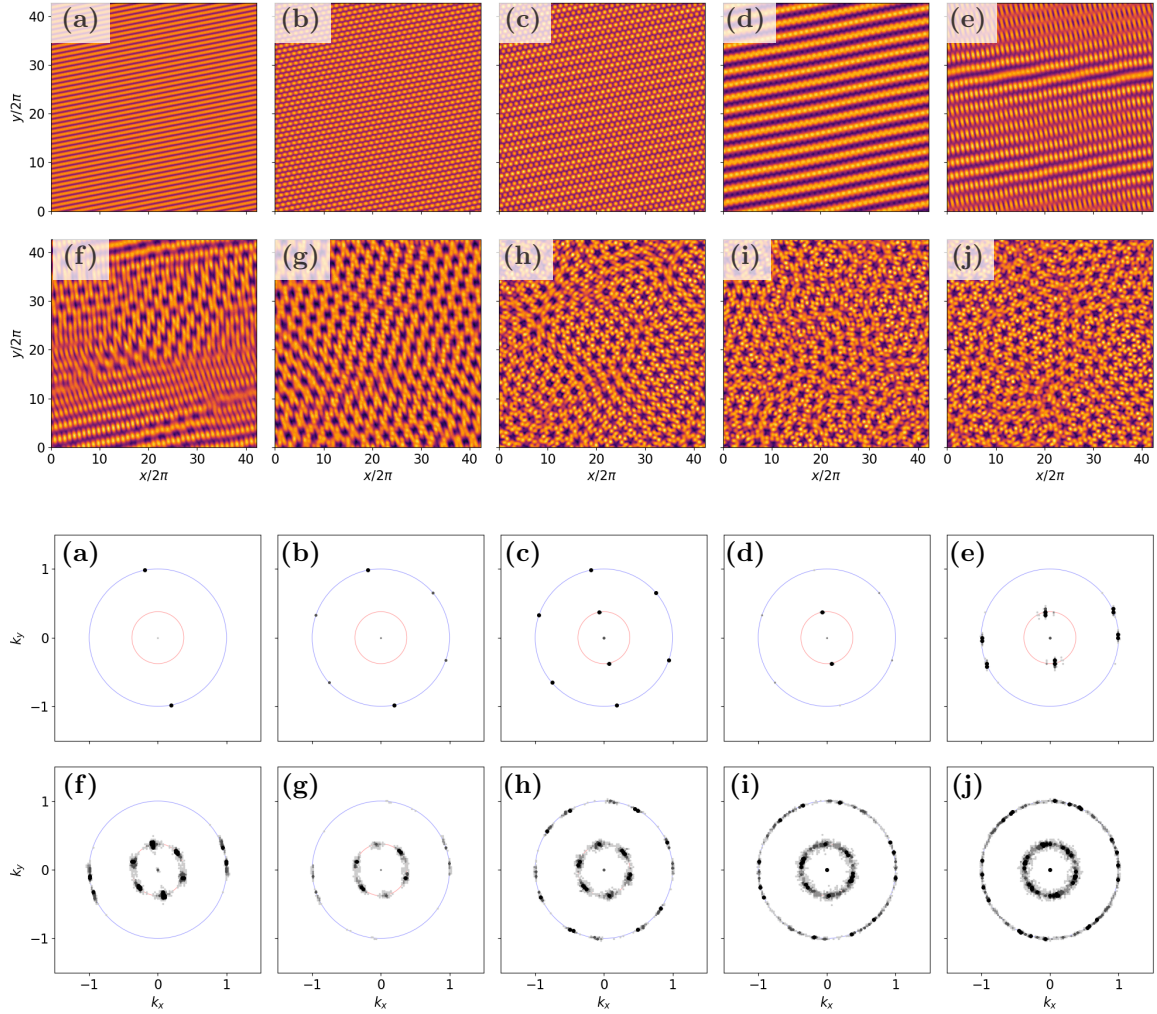

\begin{figure}
\centering
\begin{tikzpicture}[scale=0.78]
\draw[->] (0,0) -- (14,0) node[right] {$P_1$};
\draw[->] (0,0) -- (0,8.5) node[above] {$P_q$};  
\node (a) [mycircle, label=below:{\shortstack{\llap{\small{$z$-stripes }}\textbf{(a)}\\(2,0)}}] at (2,0){};  
\node (b) [mycircle, label=below:{\shortstack{\llap{\small{$z$-hex }}\textbf{(b)}\\(6,0)}}] at (6,0){};
\node (c) [mycircle, label=right:{\shortstack{\textbf{(c)}\rlap{ \small{$z$-hex + $w$-stripes}}\\(6,2)}}] at (6,2){};
\node (d) [mycircle, label=left:{\shortstack{\llap{\small{$w$-stripes }}\textbf{(d)}\\(0,2)}}] at (0,2){};
\node (e) [mycircle, label={[yshift=10pt] right:{\shortstack{\textbf{(e)}\rlap{ \small{rhombs}}\\(4f,2f)}}}] at (4,2.2){}; 
\node (fg) [mycircle, label=left:{\shortstack{\textbf{(f)},\textbf{(g)} \rlap{ \small{rhombs + $w$-hex}} \\(4f,6f)}}] at (4,6){};
\node (h) [mycircle, label=below:{\shortstack{\llap{\small{TC} }\textbf{(h)}\\(12f,6f)}}] at (12,6){};
\node (inf) [mycircle, label=above:{\shortstack{\textbf{(i)},\textbf{(j)}\rlap{ \small{STC}}}}] at (13,7.5){};
\draw[ultra thick,postaction={decorate},
        decoration={markings, mark=at position 0.5 with {\arrow[scale=1.3]{>}}}](a)--(b);
\draw[ultra thick,postaction={decorate},
        decoration={markings, mark=at position 0.5 with {\arrow[scale=1.3]{>}}}](b)--(c);
\draw[ultra thick,postaction={decorate},
        decoration={markings, mark=at position 0.5 with {\arrow[scale=1.3]{>}}}](c)--(d);
\draw[ultra thick,red,postaction={decorate},
        decoration={markings, mark=at position 0.5 with {\arrow[scale=1.3]{>}}}](d)--(e);
\draw[ultra thick,postaction={decorate},
        decoration={markings, mark=at position 0.5 with {\arrow[scale=1.3]{>}}}](e)--(fg);
\draw[ultra thick,postaction={decorate},
        decoration={markings, mark=at position 0.5 with {\arrow[scale=1.3]{>}}}](fg)--(h);
\draw[ultra thick,postaction={decorate},
        decoration={markings, mark=at position 0.5 with {\arrow[scale=1.3]{>}}}](h)--(inf);

\end{tikzpicture}
\caption{Schematic indicating the number of peaks on the $k=1$ ($P_1$) and $k=q$ ($P_q$) circles during the transition from equilibrium $z$-stripes to STC in \cref{onsettochaos}{\color{siaminlinkcolor}a--j}. 
The red line indicates that the transition to (4,2) involves four $k=1$ peaks that do not include the original two from $z$-stripes.
The letter ``f'' after a number indicates that those peaks are fuzzy.}
\label{schematic:fourier}
\end{figure}
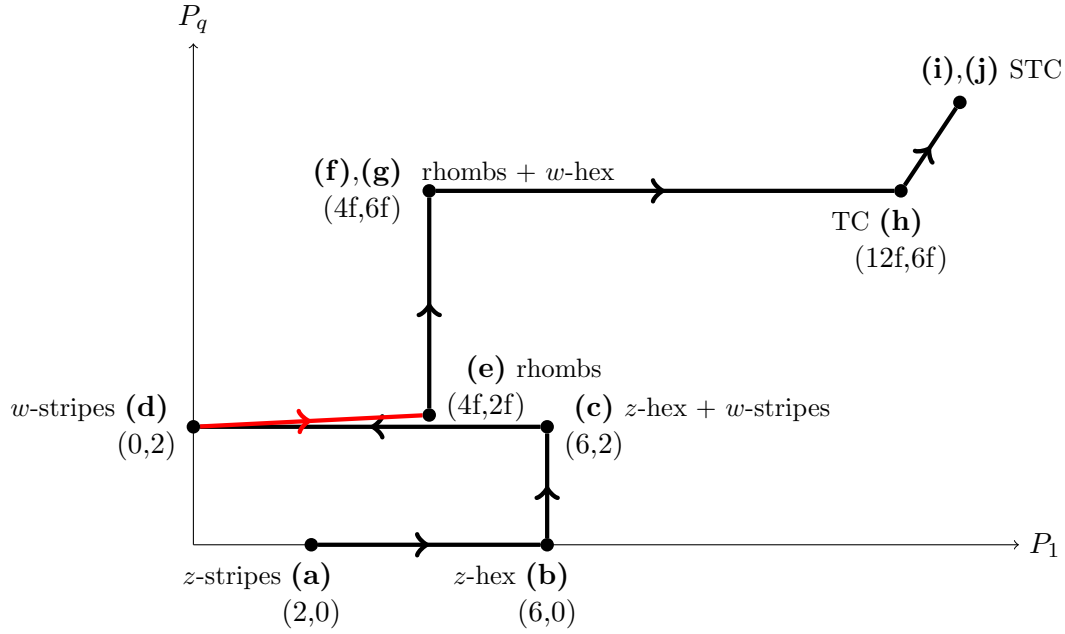

\begin{figure}
\centering
\begin{tikzpicture}
\node[anchor=south west,inner sep=0] (image) at (0,0) 
{\includegraphics[width=110mm]{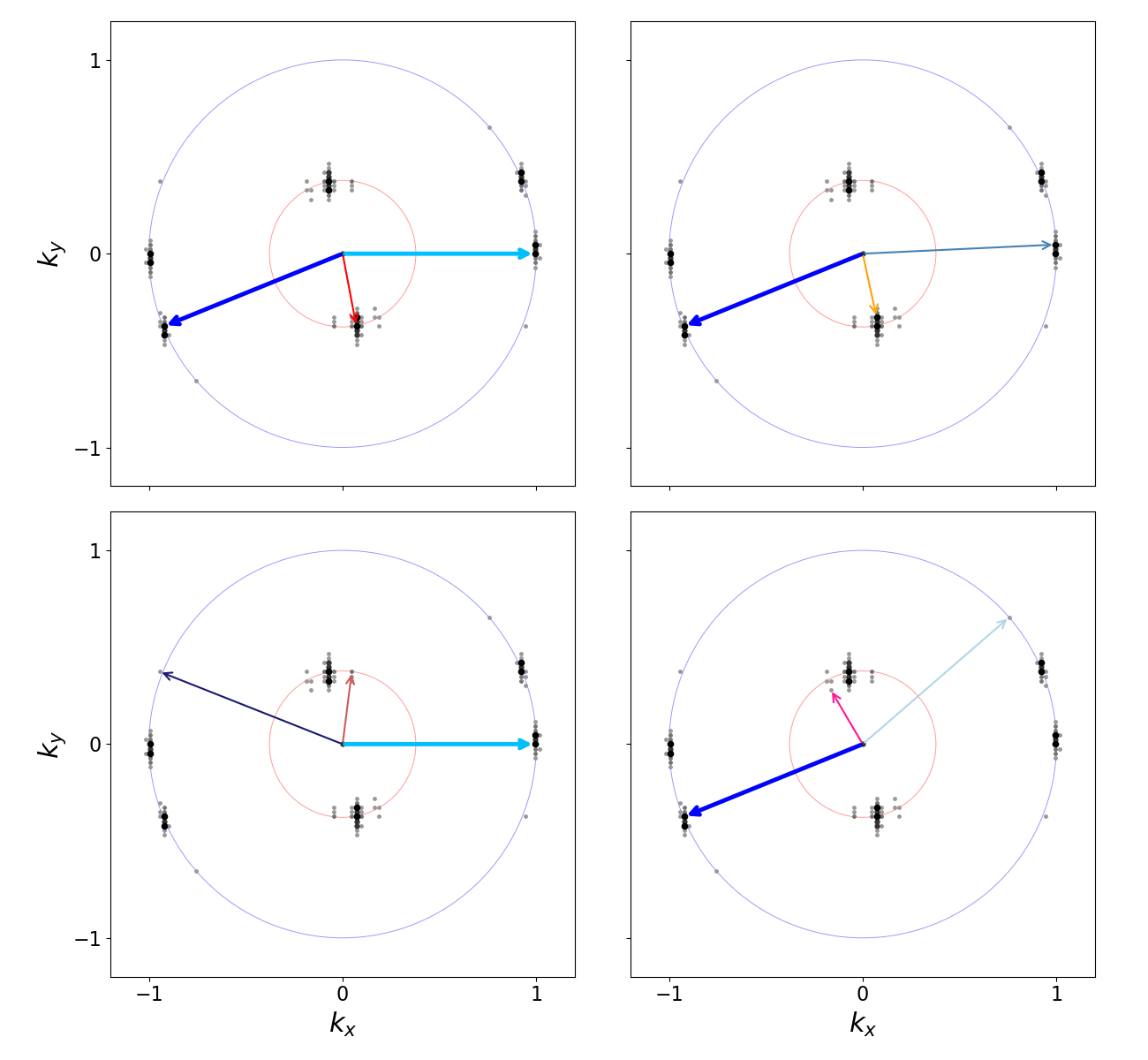}};
\begin{scope}[x={(image.south east)},y={(image.north west)}]
  \node[fill=none] at (0.13,0.95) {\small \textbf{(a)}};
  \node[fill=none] at (0.585,0.95) {\small \textbf{(b)}};
  \node[fill=none] at (0.13,0.485) {\small \textbf{(c)}};
  \node[fill=none] at (0.585,0.485) {\small \textbf{(d)}};  
\end{scope}
\end{tikzpicture}
\caption{Four sets of competing 3WIs in the Fourier power spectrum in \cref{onsettochaos}{\color{siaminlinkcolor}e}.
Triads are indicated by the three vectors in each panel. 
The bold wavevectors appear in more than one panel and are marked by the same color each time they appear.
See the text for more details.}
\label{3WIswitharrows}
\end{figure}
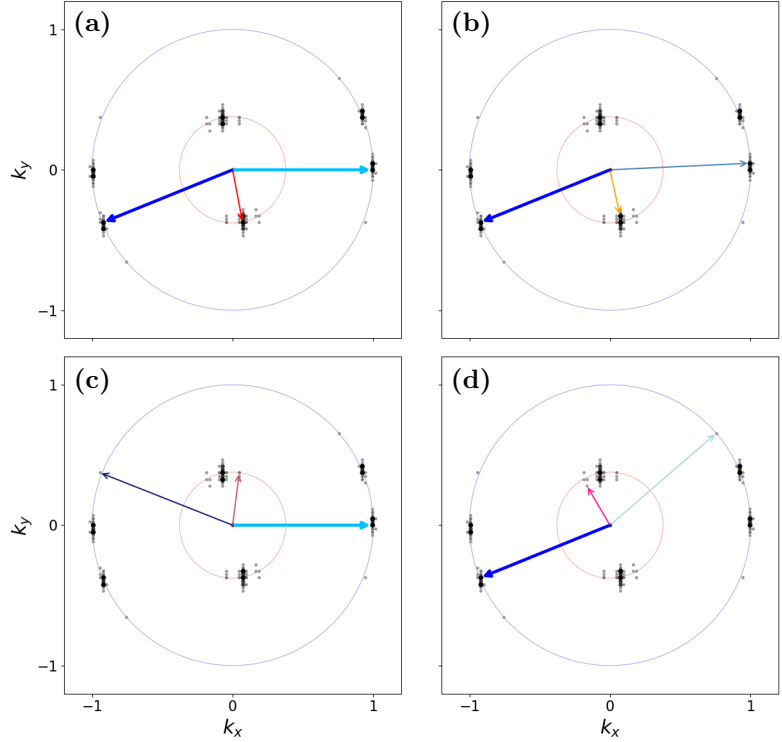

The pattern and its Fourier spectrum at different stages in the transition to STC is in \cref{onsettochaos}, with a schematic diagram giving the number of Fourier peaks on each circle at each of the different stages in \cref{schematic:fourier}.
We first describe the different stages before turning an analysis of most important 3WIs in \cref{3WIswitharrows}.

We start with $z$-stripes (\cref{onsettochaos}{\color{siaminlinkcolor}a}). 
The first modes to grow starting from $z$-stripes are those $60^\circ$ apart from the original stripe modes, leading to almost-equal amplitude $z$-hexagons, as shown in panel~(b). 
We then see a $w$-stripe pattern appear in panel~(c); these emerge at the original $z$-stripe orientation. 
The resulting $w$-stripe pattern dominates in panel~(d).
Next, a pair of modes (and their negatives) on the $k=1$ circle, $22^\circ$ degrees apart, emerge in panel~(e), leading to patches of rhombs with (somewhat) fuzzy Fourier peaks.
Four of the six remnant peaks of the $z$-hexagons are just visible in panel~(e), and the new rhombic modes on the $k=1$ circle split the $60^\circ$ arc between the remnant $z$-hexagon wavevectors into thirds of approximately $20^\circ$ each.
In panel~(f), four new fuzzy peaks appear on the $k=q$ circle, so there are now six peaks of similar amplitude with $k=q$ ($w$-hexagons) and four fuzzy peaks on the $k=1$ circle (rhombs).
In physical space, we see patches of $w$-hexagons and rhombs.
The four rhombs peaks decrease in amplitude in panel~(g), while two other pairs of rhombs peaks (and their negatives) start to emerge, linked to the other wavevectors in $w$-hexagons.
In panel~(h), the twelves peaks on the $k=1$ circle have grown to approximately equal amplitude, and six peaks on the $k=q$ circle are now quite fuzzy: the original six $w$~mode peaks are all supplemented on either side by fuzziness that is (at least roughly) aligned with the twelve $z$~modes.
This solution is TC, with continually evolving patches of $w$-hexagons.
STC develops over the next 3000 time units in panels~(i) and~(j), with both circles fully, but somewhat unevenly, occupied in panel~(j).

Several of the transitions just described can be understood in terms of the eighteen-mode amplitude equations, for example, the relative stability of $z$-stripes and $z$-hexagons.
Furthermore, a separate PDE simulation done in a small domain, large enough for only the eighteen superlattice modes to be present, reveals persistent chaotic oscillations between $z$-hexagons and stars (asymmetric superhexagons), as well as symmetry-broken versions of both of these.
The amplitude equations are chaotic at the same parameter values, though the details are different.
This persistent time dependence should therefore be expected in the PDE simulations after the $z$-hexagon stage in \cref{onsettochaos}{\color{siaminlinkcolor}b}, but in fact, in the large domain, the PDE moves out of the eighteen-mode subspace first by developing $w$-stripes aligned with the original $z$-stripes.

The fact that the $w$-stripes that emerge in \cref{onsettochaos}{\color{siaminlinkcolor}c} are aligned with the original $z$-stripes can be understood by calculating, using weakly nonlinear theory, the growth rate of a mode on the $k=q$ circle in the presence of large-amplitude $z$-hexagons.
We find that modes on the $k=q$ circle that are aligned with the six equal-amplitude $z$~modes in the $z$-hexagons have the largest growth rate, slightly larger than $w$~modes that would be involved in rhombic 3WIs.
Moreover, if the modes in the $z$-hexagons have unequal amplitudes (as in \cref{onsettochaos}{\color{siaminlinkcolor}b}), the fastest-growing $w$~mode aligns with the strongest $z$~mode.
This mechanism introduces modes on the $k=q$ circle that are different from the six $w$~modes within the eighteen mode restriction.
We also observe these aligned $w$~modes in the onset of intermittent chaos as discussed in \cref{4WIs}.

The development of fuzzy peaks in the Fourier spectrum, corresponding to having modulated patterns or patches of pattern with different orientations, can be understood in terms of the 3WIs.
We start by breaking the Fourier power spectrum in \cref{onsettochaos}{\color{siaminlinkcolor}e} into its component triads in \cref{3WIswitharrows}.
These triads involve not only the prominent rhombs peaks but also the much smaller peaks left over from the $z$-hexagons.
\Cref{3WIswitharrows}{\color{siaminlinkcolor}a} shows the triad involving the modes with largest amplitude: these are responsible for the overall rhombs pattern.
Panels (c) and~(d) show two additional triads that both compete with the triad in~(a).
(Recall that competing triads are those that each have two $k=1$ modes and one $k=q$ mode, with one $k=1$ mode in common, and that competing triads do not both fit within the original eighteen modes.)
The other $k=1$ wavevector in the triad in panel~(c) is one of the original hexagonal wavevectors seen in \cref{onsettochaos}{\color{siaminlinkcolor}b}.
Similarly, the other $k=1$ wavevector in the triad in panel~(d) is a different one of the original hexagonal wavevectors.
This can happen because the $22^\circ$ spacing between the twelve modes on the $k=1$ circle is approximately a third of~$60^\circ$.
However, the $22^\circ$ spacing is not exactly a third of~$60^\circ$, so the angles between the $k=1$ modes within each triad are not exactly the same, and the lengths of the vectors will not be exactly~1 and~$q$.
These discrepancies introduce multiple closely spaced peaks on the $k=q$ circle.
Once the other peaks on the $k=q$ circle appear with the development of $w$-hexagons (\cref{onsettochaos}{\color{siaminlinkcolor}f}), these closely spaced peaks develop into fuzzy peaks on the $k=q$ circle.
Finally, the fuzziness on $k=q$ is transferred to fuzziness on $k=1$ through 3WIs such at the one in \cref{3WIswitharrows}{\color{siaminlinkcolor}b}.
These interactions happen as the mode amplitudes are growing from low values, and in \cref{onsettochaos}{\color{siaminlinkcolor}e}, the modes that appear first are the ones leading to two wavelengths of modulation in the $y$~direction.

From \cref{onsettochaos}{\color{siaminlinkcolor}f--h}, all eighteen of the main modes develop strong fuzzy peaks, with evolving amplitudes characteristic of temporal chaos, following the 3WIs involving each of the fuzzy peaks on the $k=q$ circle. 
Over the next 3000 time units (\cref{onsettochaos}{\color{siaminlinkcolor}h--j}), further 3WIs cause first the $k=q$ circle to be more fully occupied, followed by the $k=1$ circle, resulting in spatiotemporal chaos in the end.

\section{Conclusions}
\label{sec:conclusions}
We have investigated how nonlinear interactions between waves with two wavenumbers can lead to spatiotemporal chaos.
We started with the hypothesis from~\cite{Rucklidge2012} that time dependence within the eighteen-mode amplitude equations coupled with the availability of modes with all orientations---consistent with large domain PDE simulations---could lead to competing 3WIs that would push the dynamics out of the eighteen-mode subspace and hence give rise to spatiotemporal chaos.
We only found evidence of time-dependent dynamics in the PDE~\cref{PDE:2} when at least one of the ODE Hopf criteria ($Q_{zz}Q_{zw}<0$ or $K_{ww}L_{wz}<0$) was satisfied, and when there were no simple stable equilibria in the amplitude equations.
The first of the criteria is the one well known from the theory of 3WIs~\cite{GUCKENHEIMER1992,PORTER2004,Rucklidge2012,CASTELINO2020}, and the second is a new criterion that is only relevant for the wavenumber ratio~$q=1/\sqrt{7}$~\cite{Subramanian2026}.

However, our detailed examination of the onset of STC revealed that this is more complicated than first hypothesized: we found that the initial transition out of the eighteen-mode subspace came from the growth of new $w$~modes \emph{aligned} with existing $z$~modes.
The original hypothesis was that competing triads would take the dynamics out of the eighteen-mode subspace, and the new $w$~modes associated with competing triads would \emph{not be aligned} with the $z$~modes within the eighteen-mode subspace.
The reason for the growth of aligned $w$~modes, rather than $w$~modes associated with competing triads, can seen from a stability analysis of $w$~modes with arbitrary angle in the presence of fully developed $z$-hexagons.
The linear analysis gave positive growth rates for both possible types of $w$~modes: the aligned $w$~modes had the largest growth rate, but the $w$~modes associated with competing triads were only a few degrees away from being aligned, and the difference in growth rates was relatively small, so both of these mechanisms ran side-by-side.

Exiting the eighteen-mode subspace is a key step in the development of complex spatial structure.
Once the dynamics made this exit, having both aligned and competing triad $w$~modes present led to fuzziness (see \cref{app:pattern classification}) in the Fourier spectrum of the solution.
This fuzziness allowed time-dependent solutions to become more spatially complex, with fuzzy peaks on the $k=q$ circle, leading to fuzzy peaks on the $k=1$ circle. 
This process sometimes stopped at temporal chaos with fuzzy peaks, typically associated with patterns with continually evolving defects, but we found many examples where it continued until the Fourier spectrum was distributed around both the $k=1$ and $k=q$ circles, a characteristic of spatiotemporal chaos.

We used the level of fuzziness in the Fourier spectra to differentiate between TC and STC as part of our pattern classification method.
The method also allowed for different patterns to be categorized on the basis of their time dependence as well as their spatial structure. 
The broad pattern class was determined through counting the number of peaks on the $k=1$ and $k=q$ circles in its Fourier spectrum. 
By computing a local Fourier transform, as outlined by~\cite{JANY2020}, patterns with defects and modulation were also able to be classified (details in \cref{app:pattern classification}).
We developed three metrics to analyze the time derivative of the pattern and so classify the pattern into equilibria, slow evolving or fast evolving time-dependent solutions.
One of these metrics distinguished between fast evolving time-dependent solutions with little spatial variation and \hbox{TC/STC}, which have a greater level of spatial variation.
Our pattern classification method enabled us to classify thousands of PDE simulations efficiently.

The eighteen-mode amplitude equations are relevant to any pattern forming system that exhibits 3WIs on two length scales, with the smaller wavenumber being less than half of the larger. 
Despite the weakly nonlinear approximation only being valid for small amplitude solutions, we found that the ODE predictions were a good qualitative guide to the PDE behavior even outside this limit, with the size and placement of the regions of stability for each equilibrium pattern in the PDE similar to the prediction.
Our results highlight the usefulness of the ODE system: in other applications, the weakly nonlinear calculation could be used to predict not only stable equilibrium patterns but also the potential for spatiotemporal chaos where no stable simple equilibria are predicted.

The largest regions of TC and STC were found for $K_{ww}L_{wz}<0$ and either sign of $Q_{zz}Q_{zw}$; the regions for $Q_{zz}Q_{zw}<0$ were only slightly larger those when $Q_{zz}Q_{zw}>0$.
To the best of our knowledge, no examples of STC have previously been reported for $q<\frac{1}{2}$ with $Q_{zz}Q_{zw}>0$ and $K_{ww}L_{wz}<0$.
The STC reported in a two-layer Turing model in~\cite{CASTELINO2020} had $q=1/\sqrt{7}$ and $Q_{zz}Q_{zw}<0$ (the value of $K_{ww}L_{wz}$ was not given).
The STC reported in~\cite{Rucklidge2012} in the same PDE~\cref{PDE:2} with $C_1=-1$, $C_2=0$ and $C_3=0$ had $q=0.66$, which allows for a wider range of 3WIs, and had both relevant pairs of quadratic coefficients of opposite sign.
The lavender regions (absence of stable simple equilibria) in the stability predictions are absent when $Q_{zz}Q_{zw}>0$ and $K_{ww}L_{wz}>0$, but present when $Q_{zz}Q_{zw}<0$; these regions are significantly larger when $K_{ww}L_{wz}<0$ than $K_{ww}L_{wz}>0$, therefore, it is not surprising that we found more examples of TC and STC for $K_{ww}L_{wz}<0$. 
We were unable to find PDE parameters resulting in $Q_{zz}Q_{zw}<0$ and $K_{ww}L_{wz}>0$ that gave a large region of absence of stable simple equilibria in the amplitude equations whilst maintaining bounded PDE solutions.
Although the $K_{ww}$ and $L_{wz}<0$ terms only exist in our ODEs for the precise case of $q=1/\sqrt{7}$, we have also found evidence of TC and STC for values of $q$ near this critical value, even with $Q_{zz}Q_{zw}>0$.
Details of these investigations will be presented elsewhere.

Most of the examples of STC were found for the larger values of~$\nu$ that we used, with $\mu$ small.
For small $\mu$ and~$\nu$, most of the chaotic solutions were~\hbox{TC}.
Some of these TC solutions had intermittent chaotic behavior, which resembled the type-I intermittency described in~\cite{Pomeau1980}.
We observed two different types of intermittent chaotic solutions, one involving just the eighteen modes from the amplitude equations and the other involving additional modes clustered around these eighteen modes.
In both of these cases the solution alternated intermittently between TC and $z$-hexagons, with the $z$-hexagons just outside their existence region. 
For the second case (involving fuzzy peaks), the first modes to grow in the transition back to TC involved $w$~modes aligned with the decaying $z$-hexagons.
These aligned $w$~modes were also the first modes to emerge outside of the eighteen mode subspace in the onset of~\hbox{STC} for other parameter values.
Therefore, the aligned $w$~modes play a crucial role in the chaotic dynamics of this system for both this intermittent~TC and for the onset of~\hbox{STC}.

In this paper, we have focused on the case $q<\frac{1}{2}$, where twelve modes on the outer circle and six on the inner form an eighteen-mode subspace.
With $q>\frac{1}{2}$, this is still possible, but there is also the possibility of twelve modes on the inner circle and six on the outer forming a different eighteen-mode subspace.
Each of these has its own pair of quadratic coefficients, so the mechanism we have described for exiting the eighteen-mode subspace is available to either of these, provided its pair of quadratic coefficients has opposite sign.
This supports the correlation between the presence of both pairs of quadratic coefficients with opposite sign and that of spatiotemporal chaos in the Faraday wave experiment~\cite{Rucklidge2012} and a coupled reaction--diffusion model~\cite{CASTELINO2020}.
The special value $q=2\sin15^\circ=\sqrt{2-\sqrt{3}}\approx0.5176$ leads to twelve modes on each circle and twelve-fold quasipatterns~\cite{Lifshitz_Petrich_1997,Rucklidge2012}, with a potential that exiting this twenty-four-mode subspace will similarly lead to~\hbox{STC}.
There is a related sixty-mode icosahedral subspace available in three dimensions with $q=\frac{1}{2}(\sqrt{5}-1)\approx0.6180$~\cite{Subramanian2016}.
We anticipate investigating the $q>\frac{1}{2}$ possibilities in more detail in future work.

The presence of Hopf bifurcations has been a significant focus of our discussion and the rationale for where we expected to find time dependence and regions of~\hbox{STC}.
There are more Hopf bifurcations in this system than just the ones from $z$-stripes and $w$-stripes~\cite{Subramanian2026}, though we have not focused on the periodic orbits created in these.
Analysis of unstable periodic orbits is an active research area; it has been shown that unstable periodic orbits serve as the underlying structure of chaotic dynamics~\cite{ChaosBook}.
Identifying unstable periodic orbits can be challenging in high dimensional systems. 
Our PDE has the advantage of being easy to work with and may be a useful model to use for establishing the effectiveness of new methods of identifying periodic orbits and their role in complex dynamics, such as the adjoint-based variational method~\cite{Azimi2022} and machine learning~\cite{Beck2025,beck2026}.

\FloatBarrier

\goodbreak

\appendix
\section{Weakly Nonlinear Analysis} \label{app wnlt}
Here we present the weakly nonlinear analysis of our PDE~\cref{PDE:2}, repeated here for convenience:
\begin{equation}
    \pdev{u}{t} = \mathcal{L}u + Q_1u^2 +Q_2 u \nabla^2 u +Q_3 |\boldsymbol{\nabla} u|^2 +C_1 u^3 +C_2 u^2 \nabla^2 u + C_3 u|\boldsymbol{\nabla} u|^2. \label{PDE:app}
\end{equation}
For conciseness, the analysis is presented for general~$q$ for only one rhombic triad \cref{z1dot:1,z2dot:1,wdot:1}.
We give the full set of eighteen amplitude equations~\cref{z1hex:1,w1hex:1} and their coefficients, written as functions of the PDE parameters, for $q=1/\sqrt{7}$.

We consider a small amplitude expansion of~$u$ close to onset:
\begin{equation}
    u = \varepsilon u_1 +\varepsilon^2 u_2 + \varepsilon^3 u_3 + \mathcal{O}(\varepsilon^4), \label{asym:1}
\end{equation}
and introduce the following scalings: 
\begin{equation}
    \mu \to \varepsilon^2 \mu, \indent \nu \to \varepsilon^2 \nu, \indent \pdev{}{t} \to \varepsilon \pdev{}{T_1} +\varepsilon^2 \pdev{}{T_2}, \label{scalings}
\end{equation} 
where $0<\varepsilon \ll 1$.
Using these scalings we split the linear operator into $\mathcal{O}(1)$ and $\mathcal{O}(\varepsilon^2)$ terms:
\begin{equation}
    \mathcal{L} = \LaTeXunderbrace{\frac{\sigma_0}{q^4} \left(1+\nabla^2\right)^2\left(q^2+\nabla^2\right)^2}_{\mathcal{L}_0}+\LaTeXunderbrace{\frac{-\nabla^2 \left[A\left(-\nabla^2\right)\mu +B\left(-\nabla^2\right)\nu\right]}{q^4\left(1-q^2\right)^3}}_{\mathcal{L}_2} \, \varepsilon^2,  \label{linearoperator}
\end{equation}
with functions $A$ and $B$ as defined in \cref{A}--\cref{B}.

Applying these expansions to the PDE \cref{PDE:2}, the $\mathcal{O}(\varepsilon)$ terms yield
\begin{equation}
    \mathcal{L}_0 u_1 = 0. \label{O(eps):1}
\end{equation}
For this to be satisfied, $u_1$ can be written a linear combination of modes with wavenumber $k=1$ or $k=q$ only, so
\begin{equation}
    u_1 = \sum_{j, |\boldsymbol{k_j}|=1} z_j(T_1,T_2)e^{i\boldsymbol{k_j}\cdot \boldsymbol{x}} +\sum_{j,|\boldsymbol{q_j}|=q} w_j(T_1,T_2) e^{i\boldsymbol{q_j} \cdot \boldsymbol{x}}. \label{u1:1}
\end{equation}
Since we are considering only one triad, we choose
\begin{equation}
u_1 = z_1 \expkone + z_2 \expktwo + w_1 \expq + c.c. \label{u1:2}
\end{equation}

At $\mathcal{O}(\varepsilon^2)$ we obtain
\begin{equation}
    \mathcal{L}_0 u_2 =\pdev{u_1}{T_1} - Q_1 u_1^2 -Q_2 u_1 \nabla^2 u_1 -Q_3 |\boldsymbol{\nabla} u_1|^2. \label{O(eps^2):1}
\end{equation}
Applying the Fredholm alternative and noting $\mathcal{L}_0$ is self-adjoint, we derive solvability constraints for finding a non-trivial solution for $u_2$: we require the right-hand side of \cref{O(eps^2):1} to be zero when projected onto the modes present in the solution for~$u_1$. 
The solvability constraint for $e^{i\boldsymbol{k_1}\cdot \boldsymbol{{x}}}$ yields 
\begin{equation}
    \pdev{z_1}{T_1} = \left[2Q_1-Q_2\left(q^2+1\right)+q^2Q_3\right] \conj{z}_2 w_1. \label{z1:quad}
\end{equation}
Similar solvability constraints can also be derived for $e^{i\boldsymbol{k_2}\cdot \boldsymbol{{x}}}$ and $e^{i\boldsymbol{q_1}\cdot \boldsymbol{{x}}}$.

Since equation \cref{O(eps^2):1} is linear in $u_2$, the solution for $u_2$ will consist of a complementary function: a solution to the homogeneous equation $\mathcal{L}_0 u_2=0$, and a particular solution: a solution for the inhomogenous part. 
The complementary function will be a linear combination of the modes $\expkone$, $\expktwo$, $\expq$ and their complex conjugates.
Owing to the quadratic dependence on~$u_1$ in \cref{O(eps^2):1}, the particular solution will contain all quadratic combinations of the same six modes, 
so $u_2$ will take the form
\begin{equation}
\begin{split}
    u_2 ={}&  \zeta_{z_1} \expkone +\zeta_{z_2} \expktwo + \zeta_{w_1} \expq + c.c.\\
    \addlinespace
    &+ \delta_z z_1^2 e^{2i\boldsymbol{k_1} \cdot \boldsymbol{x}} + \delta_z z_2^2 e^{2i\boldsymbol{k_2} \cdot \boldsymbol{x}} + \delta_w w_1^2 e^{2i\boldsymbol{q_1} \cdot \boldsymbol{x}} + c.c.\\
    \addlinespace
    &+ \beta_{zz} z_1 \conj{z}_2 e^{i(\boldsymbol{k_1}-\boldsymbol{k_2})\cdot \boldsymbol{x}}+ \beta_{zw} z_1 w_1 e^{i(\boldsymbol{k_1}+\boldsymbol{q_1})\cdot \boldsymbol{x}} +\beta_{zw} z_2 w_1 e^{i(\boldsymbol{k_2}+\boldsymbol{q_1})\cdot \boldsymbol{x}} +c.c.\\
    \addlinespace
    &+\gamma_z |z_1|^2 +\gamma_z |z_2|^2 + \gamma_w |w_1|^2, \label{u2:general}
\end{split}
\end{equation}
where the $\delta$, $\beta$ and $\gamma$ coefficients are determined from \cref{O(eps^2):1}. 
The $\zeta_i$ are arbitrary for now, but are determined using the solvability conditions from higher order terms.
The $\kappa$-symmetry \cref{ksym} allows the $z_1$ and $z_2$ amplitudes to be interchanged, resulting in the coefficients for $e^{2i\boldsymbol{k_1} \cdot \boldsymbol{x}}$ and  $e^{2i\boldsymbol{k_2} \cdot \boldsymbol{x}}$ both being $\delta_z$. 
The same argument holds for $\beta_{zw}$ and $\gamma_z$.

Substituting \cref{u2:general} into \cref{O(eps^2):1} and comparing coefficients for each mode in $u_2$ yields
\begin{subequations}
\begin{align}
    \delta_z & = \frac{(Q_2 +Q_3-Q_1)q^4}{9\sigma_0\left(q^2-4\right)^2}, \label{deltaz} \\
    \addlinespace
    \delta_w &= \frac{q^2(Q_2+Q_3)-Q_1}{9\sigma_0\left(4q^2-1\right)^2}, \label{deltaw} \\
    \addlinespace
    \beta_{zz}  &= \frac{\left[2Q_2-2Q_1+Q_3\left(2-q^2\right)\right]q^4}{4\sigma_0\left(3-q^2\right)^2\left(2-q^2\right)^2}, \label{betazz} \\
    \addlinespace
    \beta_{zw} & = \frac{\left[-2Q_1 + Q_2\left(1+q^2\right)+q^2Q_3\right]}{4\sigma_0\left(q^2+1\right)^2}, \label{betaw} \\
    \gamma_z & = \frac{2}{\sigma_0} (Q_2-Q_3-Q_1), \label{gammaz} \\
    \gamma_w & = \frac{2}{\sigma_0}\left(q^2Q_2-q^2Q_3-Q_1\right). \label{gammaw}
\end{align}
\end{subequations}

The $\mathcal{O}(\varepsilon^3)$ terms of \cref{PDE:app} are
\begin{equation}
\begin{split}
     \mathcal{L}_0 u_3 = \pdev{u_1}{T_2}+\pdev{u_2}{T_1} - \mathcal{L}_2 u_1 &-2Q_1 u_1 u_2 -Q_2 u_1\nabla^2 u_2 -Q_2 u_2 \nabla^2 u_1 -2Q_3 \boldsymbol{\nabla}  u_1 \cdot \boldsymbol{\nabla} u_2 \\
     &- C_1u_1^3 -C_2 u_1^2 \nabla^2 u_1 - C_3 u_1|\boldsymbol{\nabla} u_1|^2. 
     \end{split} \label{O(eps^3):1} 
\end{equation}
As we did for the quadratic terms, we derive solvability constraints for each mode in $u_1$.
For $e^{i\boldsymbol{k_1}\cdot \boldsymbol{x}}$ we obtain
\begin{equation}
    \begin{split}
    \pdev{z_1}{T_2} +\pdev{\zeta_1}{T_1}  = {} & \mu z_1 + \Big[ 2Q_1 -\left(1+q^2\right)Q_2+q^2Q_3\Big] \Big[\conj{z}_2 \zeta_{w_1} +w_1 \conj{\zeta}_{z_2}\Big] \\
    &+ z_1 |z_1|^2 \Big[ 2Q_1(\delta_z +\gamma_z) - Q_2(5\delta_z + \gamma_z) + 4Q_3 \delta_z + 3C_1 -3C_2 +C_3  \Big] \\
    &+ z_1 |z_2|^2\Big[ 2Q_1(\beta_{zz} + \gamma_z) -Q_2\left(\beta_{zz} \left(5-q^2\right) + \gamma_z \right) + Q_3 \beta_{zz} \left(4-q^2\right) \\
    & \hspace{1.7cm}{} + 6C_1 - 6C_2 + 2C_3 \Big]\\
    &+ z_1 |w_1|^2\Big[2Q_1(\beta_{zw}+\gamma_w)-Q_2\left(\beta_{zw}\left(1+3q^2\right)+\gamma_w\right)+3q^2Q_3\beta_{zw}  \\
    &   \hspace{1.7cm}{}+ 6C_1 - \left(2+4q^2\right)C_2 + 2q^2C_3 \Big]. \label{z1:cubic}
\end{split} 
\end{equation}
The equations for the other modes of $u_1$ follow similarly. 

We combine \cref{z1:quad} and \cref{z1:cubic} into a single equation using the reconstitution method introduced by~\cite{Nayfeh1973}.
This involves combining $z_1$ and $\zeta_{z_1}$ into a single variable and restoring the original time~$t$.
It is important to note that this approach is only valid in the limit $Q_i=\mathcal{O}(\varepsilon)$ for $i=1,2,3$. 
This ensures the terms in the quadratic equations are small, hence the largest terms in each of the quadratic and cubic equations are of the same order. 
More information on the reconstitution method can be found in~\cite{Luongo1999,Rucklidge2009}.

To do the reconstitution, we define the new variables:
\begin{subequations}
\begin{align}
    Z_1 & = \varepsilon z_1 + \varepsilon^2 \zeta_{z_1} , \\
    Z_2 & = \varepsilon z_2 + \varepsilon^2 \zeta_{z_2} , \\
    W_1 & = \varepsilon w_1 + \varepsilon^2 \zeta_{w_1}. 
\end{align}
\end{subequations}
Then, we multiply \cref{z1:cubic} by $\varepsilon^3$ and \cref{z1:quad} by $\varepsilon^2$ and add them. 
We can write the leading order terms of this result in terms of the new variables $Z_1$, $Z_2$ and $W_1$:
\begin{equation}
\begin{split}
    \dev{Z_1}{t} ={} & \mu Z_1  +\Big[2Q_1 -\left(1+q^2\right)Q_2 +q^2Q_3\Big] \conj{Z}_2W_1 \\
    &+ Z_1 |Z_1|^2  \Big[2Q_1( \alpha_z+\gamma_z)-Q_2(5\alpha_z+\gamma_z)+4Q_3\alpha_z+3C_1 -3C_2 +C_3\Big] \\
    &+ Z_1 |Z_2|^2 \Big[2Q_1(\beta_{zz} +\gamma_z) -Q_2\left(\beta_{zz}\left(5-q^2\right)+\gamma_z\right) -2Q_3\beta_{zz}\left(4-q^2\right) \\
    & \hspace{1.8cm} {}+6C_1 - 6C_2 +2C_3\Big] \\
    &+ Z_1 |W_1|^2   \Big[2Q_1(\beta_{zw}+\gamma_w)-Q_2\left(\beta_{zw}\left(1+3q^2\right)+\gamma_w\right)+3q^2Q_3\beta_{zw}  \\
    & \hspace{1.8cm} {} +6C_1 - \left(2+4q^2\right) C_2 + 2q^2C_3\Big], \label{z_1:recon}
\end{split}
\end{equation}
where we have reversed the scalings introduced in \cref{scalings} and truncated to leading order. 
The equations for the evolution of $Z_2$ and $W_1$ are found similarly by considering the solvability constraints for the modes $\expktwo$ and $\expq$ respectively. 
The amplitude equation for $Z_1$ (and subsequent equations for $Z_2$ and $W_1$) are in the form of the original amplitude equations~\cref{z1dot:1,z2dot:1,wdot:1} where
\begin{subequations}
\begin{align}
    Q_{zw}  ={} & 2Q_1 -\left(1+q^2\right)Q_2 + q^2Q_3, \label{Q_{zw}} \\
    Q_{zz} = {} & 2Q_1 -2Q_2 +Q_3 \left(2-q^2\right), \label{Q_{zz}} \\
    A ={} & 2Q_1(\delta_z +\gamma_z) - Q_2(5\delta_z + \gamma_z) + 4Q_3 \delta_z + 3C_1 -3C_2 +C_3, \label{A:1}\\
    \begin{split}
    B_\alpha  ={} & 2Q_1( \beta_{zz} +\gamma_z) -Q_2\left(\beta_{zz}\left(5-q^2\right)+\gamma_z\right) +Q_3\beta_{zz}\left(4-q^2\right)\\
    &  + 6C_1 - 6C_2 + 2C_3,
    \end{split}\label{B_a:1} \\
    \begin{split}
    C_{90-\alpha/2} ={}& 2Q_1(\beta_{zw}+\gamma_w)-Q_2\left(\beta_{zw}\left(1+3q^2\right)+\gamma_w\right)+3q^2Q_3\beta_{zw} \\
    & + 6C_1 - \left(2+4q^2\right)C_2 + 2q^2C_3,
    \end{split}\label{C_90ma:1} \\
    D = {} &  2Q_1 (\alpha_w+\gamma_w)-q^2Q_2(5\alpha_w+\gamma_w)+4q^2Q_3\alpha_w+ 3C_1 - 3q^2C_2 +q^2 C_3, \label{D:1} \\ 
    \begin{split}
    E_{90-\alpha/2} ={} & 2Q_1(\beta_{zw}+\gamma_z)-Q_2\left(2\left(1+q^2\right)\beta_{zw}+q^2\gamma_z\right)+Q_3\left(2+q^2\right)\beta_{zw}\\
    &  + 6C_1 - \left(2q^2+4\right)C_2 + 2C_3,
    \end{split}\label{E_90ma:1} 
\end{align}
\end{subequations}
with the coefficients in $u_2$ ($\delta_z$, $\gamma_z$, etc.) given by \cref{deltaz}--\cref{gammaw}.
The expressions \cref{D:1} and \cref{E_90ma:1} are determined from the amplitude equation for \hbox{$W_1$}. 
The quadratic coefficients $Q_1$, $Q_2$ and $Q_3$ appear in the cubic terms because we made no assumptions on their magnitude.
If they had been scaled to $\mathcal{O}(\varepsilon)$, they would not have appeared.

The same approach can be applied for the eighteen mode ODE system, which is written in full below, reverting to lower-case letters for the variables:
{\allowdisplaybreaks
\begin{subequations}
\begin{align}
\begin{split}
    \dev{z_1}{t}  ={} & \mu z_1 + Q_{zw}\conj{z}_2w_1 +Q_{zhex}\conj{z}_3\conj{z}_5 \\
    \begin{split}
    &+ z_1 \left(A |z_1|^2 + B_{\alpha} |z_2|^2 + B_{60} |z_3|^2 + B_{60-\alpha} |z_4|^2 + B_{60}|z_5|^2 \right. \\
    & \hspace{1cm} \left.{} + B_{60+\alpha} |z_6|^2 + C_{90-\alpha/2} |w_1|^2 + C_{30+\alpha/2} |w_2|^2 + C_{30-\alpha/2}|w_3|^2 \right)
    \end{split} \\
    & + K_1 z_4 z_6 w_1 + K_2 \conj{z}_2 \conj{w}_2 \conj{w}_3 + K_3 z_4 \conj{z}_5 \conj{w}_2 + K_4 \conj{z}_3 z_6 \conj{w}_3  \\
    &\color{red} + K_{zww} z_4 \conj{w}_1 w_2 + K_{ww} w_2 \conj{w}_3^2,
\end{split} \label{z1hex:app}\\
\addlinespace
\begin{split}
    \dev{z_2}{t}  ={}& \mu z_2 + Q_{zw}\conj{z}_1w_1 +Q_{zhex}\conj{z}_4\conj{z}_6 \\
    \begin{split}
    &+ z_2 \left(B_{\alpha} |z_1|^2 + A |z_2|^2 + B_{60+\alpha} |z_3|^2 + B_{60} |z_4|^2 + B_{60-\alpha}|z_5|^2 \right. \\
    & \hspace{1cm} \left. {} + B_{60} |z_6|^2 + C_{90-\alpha/2} |w_1|^2 + C_{30-\alpha/2} |w_2|^2 + C_{30+\alpha/2}|w_3|^2 \right)
    \end{split} \\
    & + K_1 z_3 z_5 w_1 + K_2 \conj{z}_1 \conj{w}_2 \conj{w}_3 + K_3 \conj{z}_4 z_5 \conj{w}_3 + K_4 z_3 \conj{z}_6 \conj{w}_2 \\
    &\color{red} + K_{zww} z_5 \conj{w}_1 w_3 + K_{ww} \conj{w}_2^2 w_3,
\end{split} \label{z2hex:app}\\
\addlinespace
\begin{split}
    \dev{z_3}{t}  ={} & \mu z_3 + Q_{zw}\conj{z}_4w_2 +Q_{zhex}\conj{z}_1\conj{z}_5 \\
    \begin{split}
    &+ z_3 \left(B_{60} |z_1|^2 + B_{60+\alpha} |z_2|^2 + A |z_3|^2 + B_{\alpha} |z_4|^2 + B_{60}|z_5|^2 \right. \\
    & \hspace{1cm} \left.{}+ B_{60-\alpha} |z_6|^2 + C_{30-\alpha/2} |w_1|^2 + C_{90-\alpha/2} |w_2|^2 + C_{30+\alpha/2}|w_3|^2 \right)
    \end{split} \\
    & + K_1 z_2 z_6 w_2 + K_2 \conj{z}_4 \conj{w}_1 \conj{w}_3 + K_3 \conj{z}_1 z_6 \conj{w}_3 + K_4 z_2 \conj{z}_5 \conj{w}_1 \\ &\color{red} + K_{zww} z_6 \conj{w}_2 w_3 + K_{ww} \conj{w}_1^2 w_3,
\end{split} \label{z3hex:app}\\
\addlinespace
\begin{split}
    \dev{z_4}{t}  ={} & \mu z_4 + Q_{zw}\conj{z}_3w_2 +Q_{zhex}\conj{z}_2\conj{z}_6 \\
    \begin{split}
    &+ z_4 \left(B_{60-\alpha} |z_1|^2 + B_{60} |z_2|^2 + B_{\alpha} |z_3|^2 + A |z_4|^2 + B_{60+\alpha}|z_5|^2 \right. \\
    & \hspace{1cm} \left.{} + B_{60} |z_6|^2 + C_{30+\alpha/2} |w_1|^2 + C_{90-\alpha/2} |w_2|^2 + C_{30-\alpha/2}|w_3|^2 \right)
    \end{split} \\
    & + K_1 z_1 z_5 w_2 + K_2 \conj{z}_3 \conj{w}_1 \conj{w}_3 + K_3 z_1\conj{z}_6  \conj{w}_1 + K_4 \conj{z}_2 z_5 \conj{w}_3 \\
    &\color{red} + K_{zww} z_1 w_1 \conj{w}_2 + K_{ww} w_1 \conj{w}_3^2,
\end{split}\label{z4hex:app} \\
\addlinespace
\begin{split}
    \dev{z_5}{t}  = {} & \mu z_5 + Q_{zw}\conj{z}_6w_3 +Q_{zhex}\conj{z}_1\conj{z}_3 \\
    \begin{split}
    &+ z_5 \left(B_{60} |z_1|^2 + B_{60-\alpha} |z_2|^2 + B_{60} |z_3|^2 + B_{60+\alpha} |z_4|^2 + A|z_5|^2 \right. \\
    & \hspace{1cm} \left.{} + B_{\alpha} |z_6|^2 + C_{30+\alpha/2} |w_1|^2 + C_{30-\alpha/2} |w_2|^2 + C_{90-\alpha/2}|w_3|^2 \right)
    \end{split} \\
    & + K_1 z_2 z_4 w_3 + K_2 \conj{z}_6 \conj{w}_1 \conj{w}_2 + K_3 z_2 \conj{z}_3 \conj{w}_1 + K_4 \conj{z}_1 z_4 \conj{w}_2 \\ &\color{red} + K_{zww} z_2 w_1 \conj{w}_3 + K_{ww} w_1 \conj{w}_2^2 ,
\end{split} \label{z5hex:app}  \\
\addlinespace
\begin{split}
    \dev{z_6}{t}  = {}& \mu z_6 + Q_{zw}\conj{z}_5w_3 +Q_{zhex}\conj{z}_2\conj{z}_4 \\
    \begin{split}
    &+ z_6 \left(B_{60+\alpha} |z_1|^2 + B_{60} |z_2|^2 + B_{60-\alpha} |z_3|^2 + B_{60} |z_4|^2 + B_{\alpha}|z_5|^2\right. \\
    & \hspace{1cm} \left.{} + A|z_6|^2 + C_{30-\alpha/2} |w_1|^2 + C_{30+\alpha/2} |w_2|^2 + C_{90-\alpha/2}|w_3|^2 \right)
    \end{split} \\
    & + K_1 z_1 z_3 w_3 + K_2 \conj{z}_5 \conj{w}_1 \conj{w}_2 + K_3 \conj{z}_2 z_3  \conj{w}_2 + K_4 z_1\conj{z}_4 \conj{w}_1 \\ &\color{red} + K_{zww} z_3 w_2 \conj{w}_3 + K_{ww} \conj{w}_1^2 w_2,
\end{split}\label{z6hex:app} \\
\addlinespace
\begin{split}
    \dev{w_1}{t}  = {} & \nu w_1 + Q_{zz}z_1 z_2 +Q_{whex}\conj{w}_2\conj{w}_3 \\
    \begin{split}
    &+ w_1 \left(E_{90-\alpha/2} |z_1|^2 + E_{90-\alpha/2} |z_2|^2 + E_{30-\alpha/2} |z_3|^2 + E_{30+\alpha/2} |z_4|^2  \right. \\
    & \hspace{1cm} \left.{} + E_{30+\alpha/2}|z_5|^2 + E_{30-\alpha/2} |z_6|^2 + D |w_1|^2 + F_{60} |w_2|^2 + F_{60}|w_3|^2 \right)
    \end{split} \\
    & + L_{hex} z_2 \conj{z}_3 \conj{z}_5 + L_{hex} z_1 \conj{z}_4 \conj{z}_6  + L_1 \conj{z}_3 \conj{z}_4  \conj{w}_3 + L_1 \conj{z}_5 \conj{z}_6 \conj{w}_2 \\
    & \color{red} + L_{wzz} \conj{z}_1 z_4 w_2 + L_{wzz} \conj{z}_2 z_5 w_3 + L_{wwz} \conj{z}_3 \conj{w}_1 w_3 + L_{wwz} \conj{z}_6 \conj{w}_1 w_2 \\
    &\color{red}+ L_{wz} z_4 w_3^2 + L_{wz} z_5 w_2^2,
\end{split} \label{w1hex:app} \\
\addlinespace
\begin{split}
    \dev{w_2}{t}  = {} & \nu w_2 + Q_{zz}z_3 z_4 +Q_{whex}\conj{w}_1\conj{w}_3 \\
    \begin{split}
    &+ w_2 \left(E_{30+\alpha/2} |z_1|^2 + E_{30-\alpha/2} |z_2|^2 + E_{90-\alpha/2} |z_3|^2 + E_{90-\alpha/2} |z_4|^2 \right. \\
    & \hspace{1cm} \left.{} + E_{30-\alpha/2}|z_5|^2  + E_{30+\alpha/2} |z_6|^2+ F_{60} |w_1|^2 + D |w_2|^2 + F_{60}|w_3|^2 \right)
    \end{split} \\
    & + L_{hex}  \conj{z}_1 z_4 \conj{z}_5 + L_{hex} \conj{z}_2 z_3 \conj{z}_6  + L_1 \conj{z}_5 \conj{z}_6  \conj{w}_1 + L_1 \conj{z}_1 \conj{z}_2 \conj{w}_3 \\
    & \color{red} + L_{wzz} \conj{z}_3 z_6 w_3 + L_{wzz} z_1 \conj{z}_4 w_1 + L_{wwz} \conj{z}_5 w_1 \conj{w}_2 + L_{wwz} \conj{z}_2 \conj{w}_2 w_3 \\
    &\color{red} + L_{wz} z_6 w_1^2 + L_{wz} z_1 w_3^2,
\end{split} \label{w2hex:app} \\
\addlinespace
\begin{split}
    \dev{w_3}{t}  = {} & \nu w_3 + Q_{zz}z_5 z_6 +Q_{whex}\conj{w}_1\conj{w}_2 \\
    \begin{split}
    &+ w_3 \left(E_{30-\alpha/2} |z_1|^2 + E_{30+\alpha/2} |z_2|^2 + E_{30+\alpha/2} |z_3|^2 + E_{30-\alpha/2} |z_4|^2  \right. \\
    & \hspace{1cm} \left.{} + E_{90-\alpha/2}|z_5|^2 + E_{90-\alpha/2} |z_6|^2 + F_{60} |w_1|^2 + F_{60} |w_2|^2 + D|w_3|^2 \right)
    \end{split} \\
    & + L_{hex}  \conj{z}_1 \conj{z}_3 z_6 + L_{hex} \conj{z}_2 \conj{z}_4 z_5  + L_1 \conj{z}_1 \conj{z}_2  \conj{w}_2 + L_1 \conj{z}_3 \conj{z}_4 \conj{w}_1 \\
    & \color{red} + L_{wzz} z_2\conj{z}_5 w_1 + L_{wzz} z_3 \conj{z}_6 w_2 + L_{wwz} \conj{z}_1 w_2 \conj{w}_3 + L_{wwz} \conj{z}_4 w_1 \conj{w}_3 \\
    &\color{red}+ L_{wz} z_2 w_2^2 + L_{wz} z_3 w_1^2.
\end{split} \label{w3hex:app} 
\end{align}
\end{subequations}
}%
As before, the red terms (final line of \cref{z1hex:app}--\cref{z6hex:app} and final two lines of \cref{w1hex:app}--\cref{w3hex:app}) are only present when $q=1/\sqrt{7}$, as these terms arise from non-generic 4WIs for this wavenumber. 
The weakly nonlinear expressions of the ODE coefficients in the eighteen mode system for $q=1/\sqrt{7}$ are
\allowdisplaybreaks{
\begin{align}
Q_{zw} = {} & 2Q_1 - \frac{8}{7}Q_2 + \frac{1}{7} Q_3, \label{new:Qzw:1} \\
\addlinespace
Q_{zz}  = {} & 2Q_1 - 2Q_2 + \frac{13}{7} Q_3, \label{new:Qzz:1} \\
\addlinespace
Q_{zhex} ={} & 2Q_1 - 2Q_2 +Q_3, \label{new:Qzhex:1}\\
\addlinespace
Q_{whex} = {} &2Q_1 - \frac{2}{7}Q_2 +\frac{1}{7}Q_3, \label{new:Qwhex:1} \\
\addlinespace
\begin{split}
    A = {}& \frac{1}{6561 \sigma_0}\Big(-26246 Q_1^2 + 39373 Q_1 Q_2 - 26246 Q_1 Q_3 - 13127 Q_2^2 \\
    &{} + 13121 Q_2 Q_3 + 4 Q_3^2\Big) + 3 C_1 - 3 C_2 + C_3,
\end{split} \label{new:A:1} \\
\addlinespace
\begin{split}
    B_\alpha  = {}&\frac{1}{270400\sigma_0}\Big(-1081796 Q_1^2 + 1623072 Q_1 Q_2 - 1081796 Q_1 Q_3 - 541276 Q_2^2 \\
    &{}+ 540736 Q_2 Q_3 + 351 Q_3^2 \Big) + 6 C_1 - 6 C_2 + 2C_3,
\end{split} \label{new:B_a:1}\\
\addlinespace
\begin{split}
    B_{60}  ={} & \frac{1}{1600 \sigma_0} \Big(-6404 Q_1^2 + 9612 Q_1 Q_2 - 6404 Q_1 Q_3 - 3208 Q_2^2 \\
    & + 3202 Q_2 Q_3 + 3 Q_3^2 \Big)  + 6 C_1 - 6 C_2 + 2C_3,
\end{split} \label{new:B_60:1} \\
\addlinespace
\begin{split}
    B_{60-\alpha} ={} & \frac{1}{186624 \sigma_0} \Big( -1318228 Q_1^2 + 2100164 Q_1 Q_2 - 1318228 Q_1 Q_3 - 781936 Q_2^2\\
    & + 816478 Q_2 Q_3 - 95953 Q_3^2 \Big) + 6 C_1 - 6 C_2 + 2C_3,
\end{split} \label{new:B_60ma:1} \\
\addlinespace
\begin{split}
    B_{60+\alpha} = {}&  \frac{2}{2205225 \sigma_0} \Big(-4493750 Q_1^2 + 6815413 Q_1 Q_2 - 4493750 Q_1 Q_3 - 2321663 Q_2^2 \\
    & + 2291081 Q_2 Q_3 - 6812 Q_3^2 \Big) + 6 C_1 - 6 C_2 + 2C_3,
\end{split} \label{new:B_60pa:1} \\
\addlinespace
\begin{split}
    C_{90-\alpha/2} = {}& \frac{1}{1792 \sigma_0} \Big(-8540 Q_1^2 + 6372 Q_1 Q_2 - 1220 Q_1 Q_3 - 1072 Q_2^2 \\
    & + 610 Q_2 Q_3 + 21 Q_3^2 \Big) + 6 C_1 - \frac{18}{7} C_2 + \frac{2}{7}C_3,
\end{split} \label{new:C_90ma:1} \\
\addlinespace
\begin{split}
    C_{30+\alpha/2} ={} & \frac{2}{1715175 \sigma_0}  \Big(-5561066 Q_1^2 + 4215499 Q_1 Q_2 - 794438 Q_1 Q_3 - 698009 Q_2^2\\
    & + 286241 Q_2 Q_3 + 91504 Q_3^2 \Big) + 6 C_1 - \frac{18}{7} C_2 + \frac{2}{7}C_3,
\end{split} \label{new:C_30pa:1} \\
\addlinespace
\begin{split}
    C_{30-\alpha/2} ={} & \frac{1}{18144 \sigma_0} \Big(-128828 Q_1^2 + 95362 Q_1 Q_2 - 18404 Q_1 Q_3 - 14648 Q_2^2 \\
    &+ 4001 Q_2 Q_3 + 4375 Q_3^2 \Big) +  6 C_1 - \frac{18}{7} C_2 + \frac{2}{7}C_3,
\end{split} \label{new:C_30ma:1} \\
\addlinespace
\begin{split}
    D  ={}& \frac{1}{3969 \sigma_0} \Big(-20678 Q_1^2 + 5803 Q_1 Q_2 - 2954 Q_1 Q_3 - 407 Q_2^2 \\
    &+ 113 Q_2 Q_3 + 196 Q_3^2 \Big)  + 3 C_1 - \frac{3}{7} C_2 + \frac{1}{7}C_3, 
\end{split} \label{new:D:1} \\
\addlinespace
\begin{split}
    E_{90-\alpha/2} ={} & \frac{1}{1792 \sigma_0} \Big(-8540 Q_1^2 + 10032 Q_1 Q_2 - 8540 Q_1 Q_3 - 1408 Q_2^2\\
    & + 1240 Q_2 Q_3 + 105 Q_3^2 \Big) + 6 C_1 - \frac{30}{7} C_2 + 2C_3, 
\end{split} \label{new:E_90ma:1} \\
\addlinespace
\begin{split}
    E_{30+\alpha/2} = {}& \frac{2}{1715175 \sigma_0} \Big(-5561066 Q_1^2 + 6598813 Q_1 Q_2 - 5561066 Q_1 Q_3 - 1219817 Q_2^2 \\
    & + 1577153 Q_2 Q_3 - 403088 Q_3^2\Big)  + 6C_1 - \frac{30}{7} C_2 + 2C_3, 
\end{split} \label{new:E_30pa:1} \\
\addlinespace
\begin{split}
    E_{30-\alpha/2} = {}& \frac{1}{18144 \sigma_0} \Big(-128828 Q_1^2 + 150574 Q_1 Q_2 - 128828 Q_1 Q_3 - 28424 Q_2^2 \\
    &+ 39953 Q_2 Q_3 - 12425 Q_3^2 \Big) + 6 C_1 - \frac{30}{7} C_2 + 2C_3,
\end{split} \label{new:E_30ma:1} \\
\addlinespace
\begin{split}
    F_{60} ={} & \frac{1}{3136 \sigma_0} \Big(-22148 Q_1^2 + 6804 Q_1 Q_2 - 3164 Q_1 Q_3 - 520 Q_2^2 \\
    & + 226 Q_2 Q_3 + 147 Q_3^2 \Big) + 6 C_1 - \frac{6}{7}C_2 + \frac{2}{7}C_3,
\end{split} \label{new:F_60:1} \\
\addlinespace
\begin{split}
    K_1  = {}&\frac{1}{193600 \sigma_0} \Big(-605444 Q_1^2 + 786492 Q_1 Q_2 - 345968 Q_1 Q_3 - 251728 Q_2^2 \\
    &+ 225130 Q_2 Q_3 - 42303 Q_3^2 \Big)  + 6C_1 - \frac{30}{7}C_2 + \frac{8}{7}C_3,
\end{split} \label{new:K1:1} \\
\addlinespace
\begin{split}
    K_2 ={} & \frac{1}{193600 \sigma_0} \Big(-605444 Q_1^2 + 527016 Q_1 Q_2 - 86492 Q_1 Q_3 - 103456 Q_2^2 \\
    & - 12356 Q_2 Q_3 + 46911 Q_3^2 \Big) +  6C_1 - \frac{18}{7}C_2 + \frac{2}{7}C_3,
\end{split} \label{new:K2:1} \\
\addlinespace
\begin{split}
    K_3 = {}& \frac{1}{193600 \sigma_0} \Big( -605444 Q_1^2 + 786492 Q_1 Q_2 - 345968 Q_1 Q_3 - 253648 Q_2^2 \\
    & + 230890 Q_2 Q_3 - 46143 Q_3^2 \Big) + 6C_1 - \frac{30}{7}C_2 +\frac{8}{7}C_3,
\end{split} \label{new:K3:1} \\
\addlinespace
\begin{split}
    K_4  ={} & \frac{1}{193600 \sigma_0}  \Big(-605444 Q_1^2 + 786492 Q_1 Q_2 - 345968 Q_1 Q_3 - 179128 Q_2^2 \\
    & + 7330 Q_2 Q_3 + 102897 Q_3^2 \Big)  + 6C_1 - \frac{30}{7}C_2 +\frac{8}{7}C_3,
\end{split} \label{new:K4:1} \\
\addlinespace
\begin{split}
    L_1 = {}& \frac{1}{193600 \sigma_0} \Big(-605444 Q_1^2 + 786492 Q_1 Q_2 - 605444 Q_1 Q_3 - 251728 Q_2^2 \\
    & + 373402 Q_2 Q_3 - 131517 Q_3^2 \Big)  + 6C_1 -\frac{30}{7}C_2 +2C_3,
\end{split} \label{new:L1:1} \\
\addlinespace
\begin{split}
    L_{hex} ={} & \frac{1}{193600 \sigma_0} \Big(-605444 Q_1^2 + 1045968 Q_1 Q_2 - 864920 Q_1 Q_3 - 440524 Q_2^2 \\
   & + 732460 Q_2 Q_3 - 301779 Q_3^2 \Big) + 6C_1 -6C_2 +\frac{20}{7}C_3,
\end{split} \label{new:Lhex:1} \\
\addlinespace
\begin{split}
    \color{red} K_{zww} ={} & \color{red} \frac{1}{20736 \sigma_0}  \Big(-129556 Q_1^2 + 121436 Q_1 Q_2 - 18508 Q_1 Q_3 - 23200 Q_2^2 \\
    & \color{red}+ 862 Q_2 Q_3 + 3263 Q_3^2 \Big) + 6 C_1 - \frac{18}{7}C_2 +\frac{13}{14}C_3,
    \end{split} \label{new:K_zww:1} \\
\addlinespace
\begin{split}
    \color{red} K_{ww} = {} & \ \color{red} \frac{1}{5184 \sigma_0} \Big( -22148 Q_1^2 + 9940 Q_1 Q_2 + 6328 Q_1 Q_3 - 968 Q_2^2 \\
    & \color{red} - 1258 Q_2 Q_3 - 371 Q_3^2 \Big) + 3C_1 -\frac{3}{7}C_2 -\frac{2}{7}C_3, 
\end{split}\label{new:K_ww:1} \\
\addlinespace
\begin{split}
    \color{red} L_{wzz} = {}& \color{red} \frac{1}{20736 \sigma_0} \Big(-129556 Q_1^2 + 176960 Q_1 Q_2 - 129556 Q_1 Q_3 - 51040 Q_2^2\\
    & \color{red} + 68032 Q_2 Q_3 - 19717 Q_3^2 \Big) + 6C_1 - \frac{30}{7} C_2 + \frac{4}{7}C_3,
\end{split} \label{new:Lwzz:1}\\
\addlinespace
\begin{split}
    \color{red} L_{wwz} ={} & \color{red} \frac{1}{2592 \sigma_0} \Big(-22148 Q_1^2 + 19432 Q_1 Q_2 - 12656 Q_1 Q_3 - 3386 Q_2^2 \\
    & \color{red} + 3770 Q_2 Q_3 - 755 Q_3^2 \Big) + 6C_1 - \frac{18}{7} C_2 + \frac{8}{7}C_3,
\end{split} \label{new:Lwwz:1} \\
\addlinespace
 \begin{split}
     \color{red} L_{wz} ={} & \color{red} \frac{1}{5184\sigma_0} \Big(-22148 Q_1^2 + 19432 Q_1Q_2 - 12656 Q_1 Q_3 - 3296 Q_2^2 \\
     & \color{red} + 3500 Q_2 Q_3 - 575 Q_3^2 \Big) + 3C_1 - \frac{9}{7} C_2 + \frac{4}{7}C_3.
 \end{split}\label{new:Lwz:1} 
\end{align}
}%
The coefficients for other values of~$q$ can be calculated in a similar manner. 

\section{Pattern Classification} \label{app:pattern classification}
In \cref{patternclassification} we introduced a pattern classification method to categorize the patterns in the PDE solutions. Here we present some of the finer details of the method including how TC and STC are differentiated, the classification of patterns with defects, and the thresholds for the metrics categorizing the time dependence of solutions.

\subsection{Fuzziness and Defects} \label{app:pattern_classification:defects}
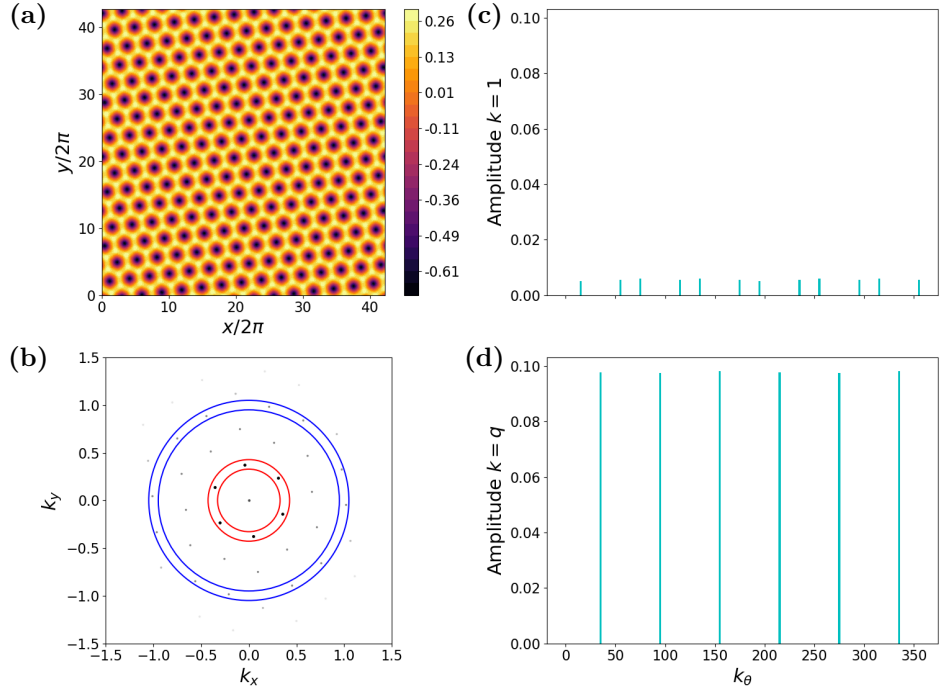
\begin{figure}
\centering
\begin{tikzpicture}
\node[anchor=south west,inner sep=0] (image) at (0,0) 
{\includegraphics[width=123mm]{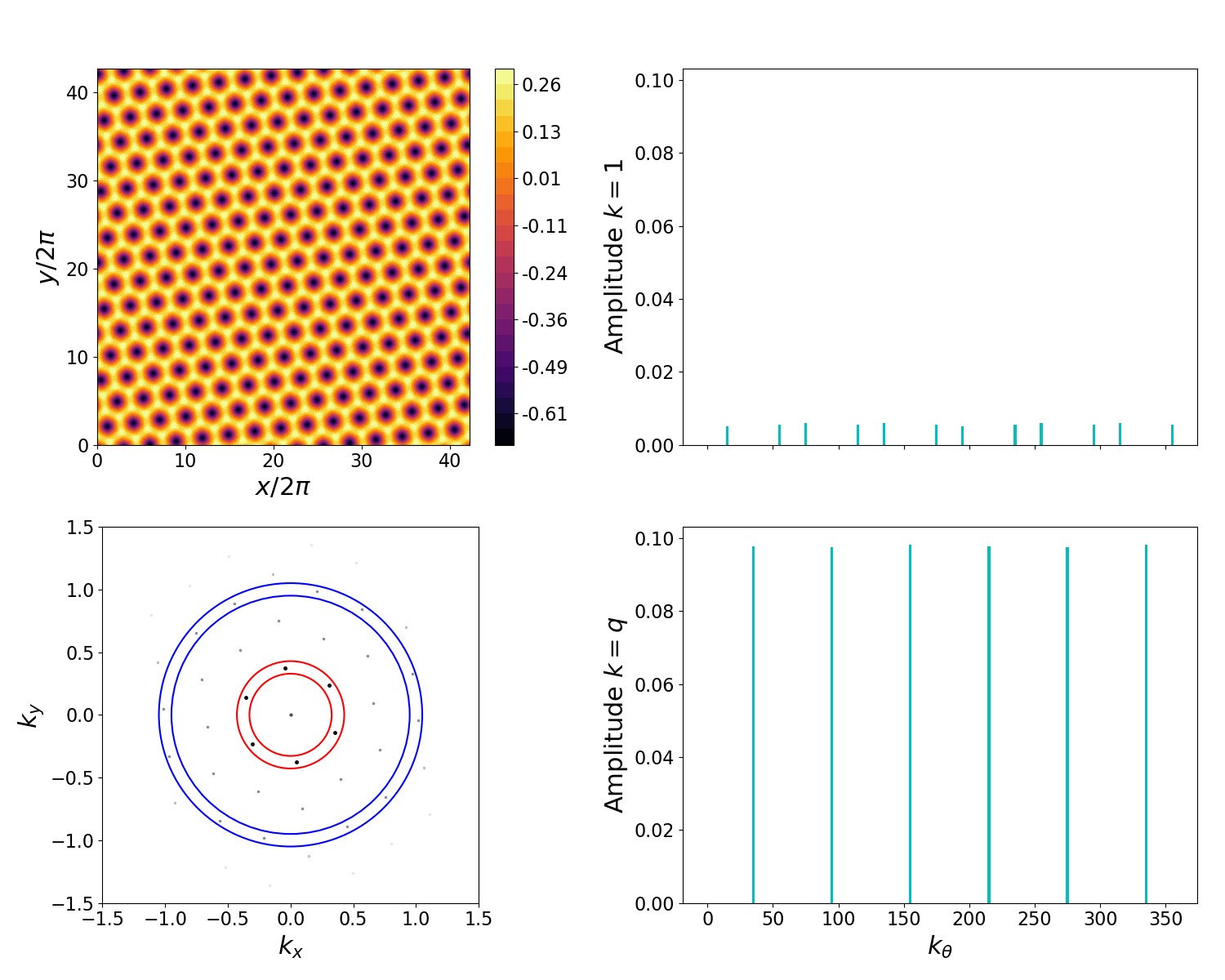}};
\begin{scope}[x={(image.south east)},y={(image.north west)}]
\node[fill=none] at (0,0.92) {\small \textbf{(a)}};
\node[fill=none] at (0,0.46) {\small \textbf{(b)}};
\node[fill=none] at (0.49,0.92) {\small \textbf{(c)}};
\node[fill=none] at (0.49,0.46) {\small \textbf{(d)}};
\end{scope}
\end{tikzpicture}
\caption{Solution of $w$-hexagons with $r=0.5$ and $\chi=175^\circ$ where \hbox{$(\mu,\nu)=(r\cos \chi, r \sin \chi)$}, and with $Q_1 = -1.06$, $Q_2 = -2$, $Q_3 = -0.8$, $C_1 = -1.3$, $C_2 = -5$, $C_3 = -15$ and $\sigma_0=-2$.
Panel~(a) shows the solution at the final computed time,
and (b)~the corresponding Fourier spectrum.
Panels~(c) and~(d) show the largest amplitudes of modes in each segment of the two annuli as a function of an angle $k_\theta$. 
We define $k_\theta$ as the angular lower bound (in degrees) of each segment of the annulus, with the angle measured from the positive $k_x$ axis.
The small contributions in the $k=1$ annulus arise from 4WIs and are not large enough to be classed as a peak or fuzziness.} 
\label{fuzziness}
\end{figure}

\begin{figure}
\centering
\begin{tikzpicture}
\node[anchor=south west,inner sep=0] (image) at (0,0) 
{\includegraphics[width=110mm]{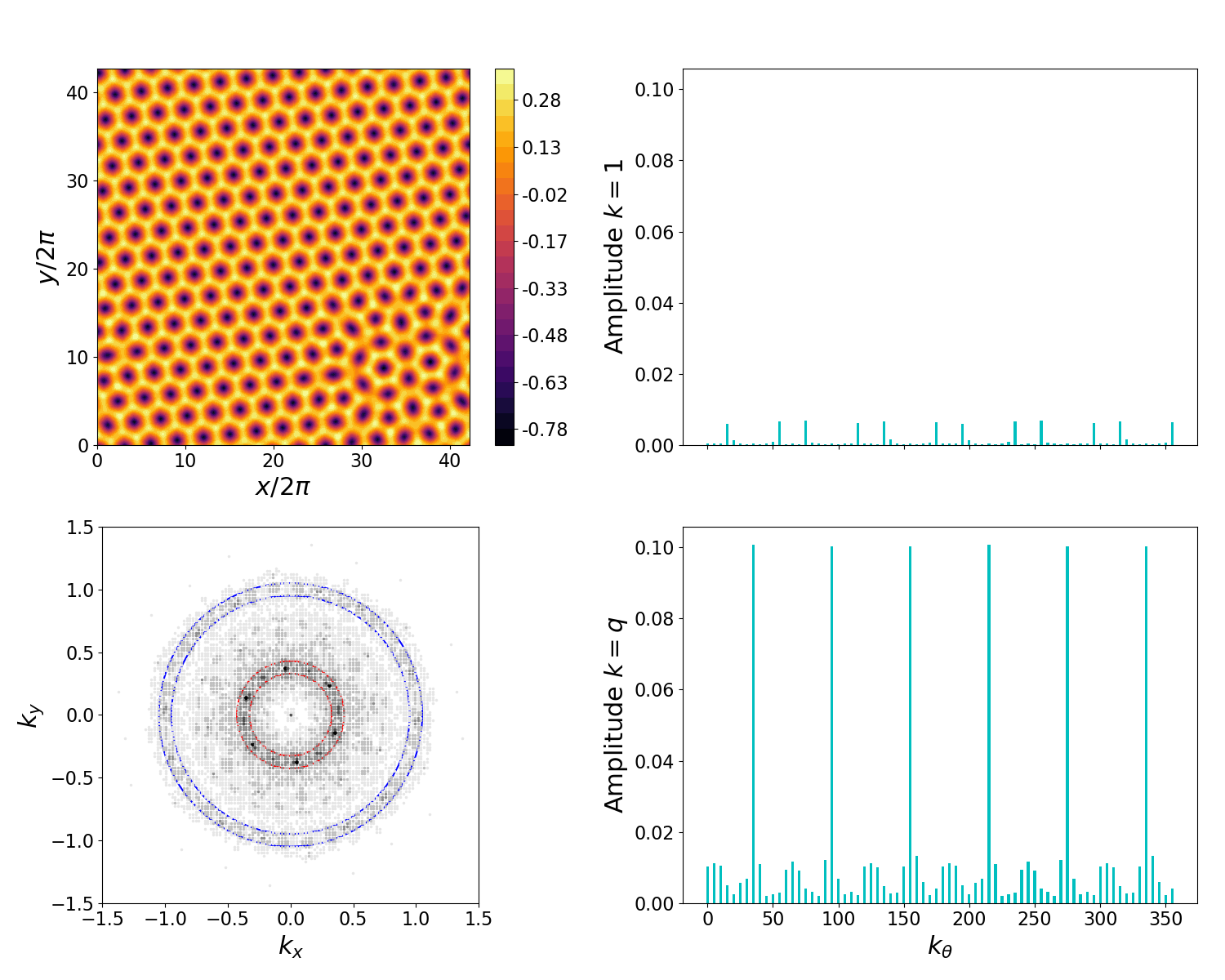}};
\begin{scope}[x={(image.south east)},y={(image.north west)}]
\node[fill=none] at (0,0.92) {\small \textbf{(a)}};
\end{scope}
\end{tikzpicture}
\begin{tikzpicture}
\node[anchor=south west,inner sep=0] (image) at (0,0) 
{\includegraphics[width=110mm]{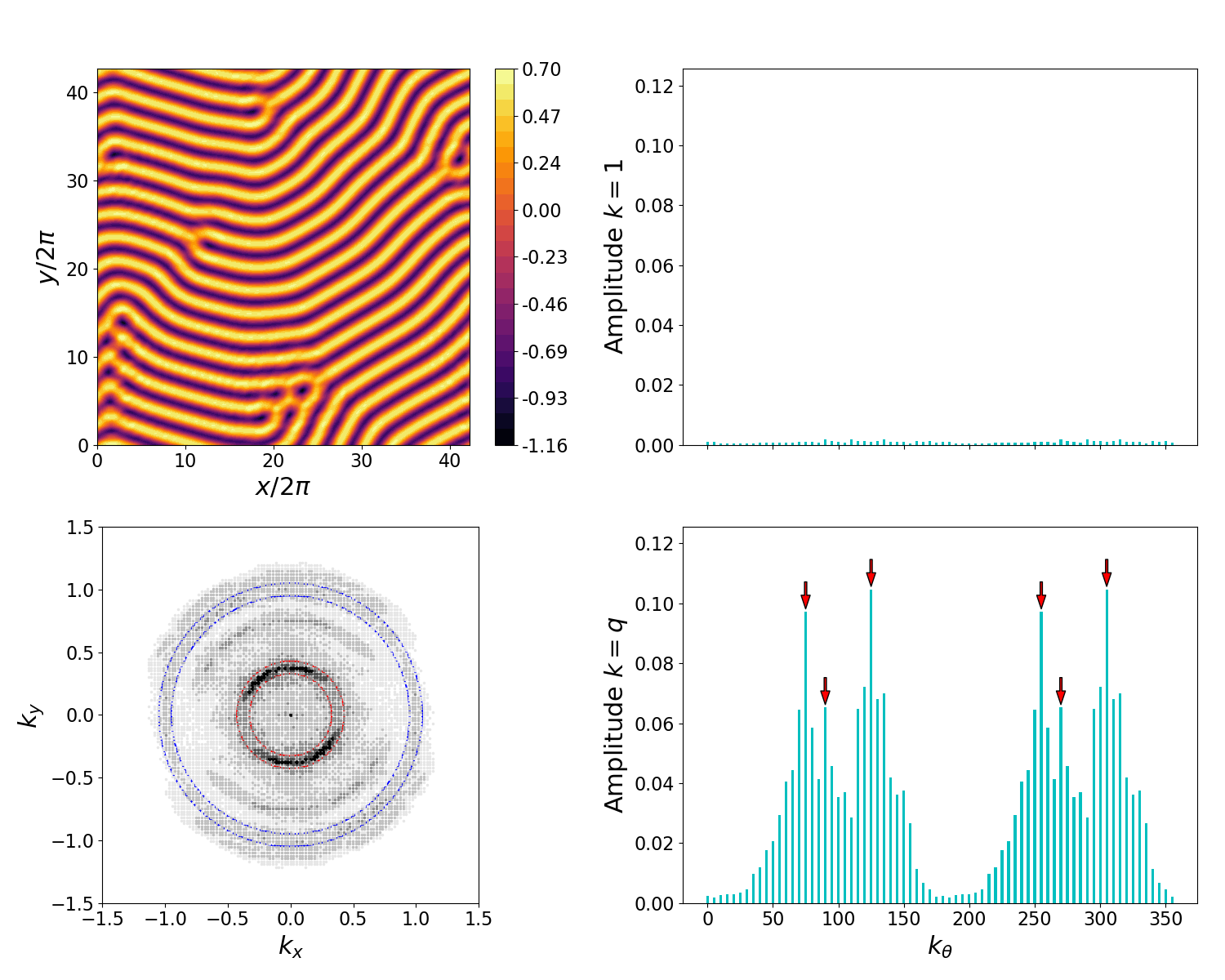}};
\begin{scope}[x={(image.south east)},y={(image.north west)}]
\node[fill=none] at (0,0.92) {\small \textbf{(b)}};
\end{scope}
\end{tikzpicture}
\caption{Two different patterns each with six peaks in the $k=q$ annulus for $Q_1 = -1.06$, $Q_2 = -2$, $Q_3 = -0.8$, $C_2 = -5$, $C_3 = -15$ and $\sigma_0=-2$.
The values of $C_1$, $r$ and $\chi$ are:
(a)~$C_1=-1.3$, $r=0.5$ and $\chi =170^\circ$ ($w$-hexagons); 
(b)~$C_1=-1.4$, $r=0.45$ and $\chi =105^\circ$ ($w$-stripes).
The red arrows in (b) mark the locations of the six peaks.
The layout of the panels is the same as in \cref{fuzziness}.}
\label{multipleorientations}
\end{figure}

One of the classification criteria in our method is based on the number of peaks in the Fourier spectrum of the pattern close to the two critical circles.
When a pattern has defects or modulation, the values for $P_1$ and $P_q$---the number of peaks in the $k=1$ and $k=q$ annuli respectively---may not be the same as those in \cref{tab:equilibriacriteria} for the given pattern.
Here we introduce the concept of \emph{fuzzy peaks} in the Fourier spectrum to help identify patterns with defects and modulation.
\Cref{fuzziness} shows a PDE solution ($w$-hexagons) with no fuzziness, and \cref{multipleorientations} shows two examples with fuzziness ($w$-hexagons and $w$-stripes with defects).
In \cref{patternclassification} we described how we discretized the annuli in the Fourier spectrum into segments, such that a segment contained a \emph{peak} if a mode in that segment had an amplitude that is larger than one third of the largest amplitude across both circles and larger than its two neighbors on either side.
We now define a segment to contain \emph{fuzziness} if the largest amplitude of that segment is greater than two-fifteenths of the largest amplitude across both circles and that segment does not contain a peak. 
If there is at least one fuzzy segment in either annulus, then we say the pattern has fuzziness, and a fuzzy peak is a peak with adjoining fuzzy segments.

We differentiate between perfect patterns and patterns with modulation or defects by considering the level of fuzziness in the Fourier spectrum.
Counting the number of segments containing fuzziness is also used to differentiate between TC and STC.
If a TC solution has more than half of the segments in each annulus containing either fuzziness or a peak, then we classify the solution as~\hbox{STC}.

Both of the solutions in \cref{multipleorientations} have fuzziness and six peaks in the $k=q$ annulus, but only one of these corresponds to $w$-hexagons. 
The other solution consists of multiple orientations of $w$-stripes (with defects), and so would be incorrectly classified by just counting peaks.
To ensure examples such as this are categorized correctly, we compute a local Fourier transform on a small region of the domain for any solution that has six or more fuzzy peaks in the $k=q$ annulus.
We do not find any examples of solutions that were incorrectly classified from the number of peaks on the $k=1$ circle (such as multiple orientations of $z$-stripes). 
Therefore, we do not repeat this analysis for the case with six or more peaks on the $k=1$ circle.

\begin{figure} 
\centering
\begin{tikzpicture}
\node[anchor=south west,inner sep=0] (image) at (0,0) 
{\includegraphics[width=110mm]{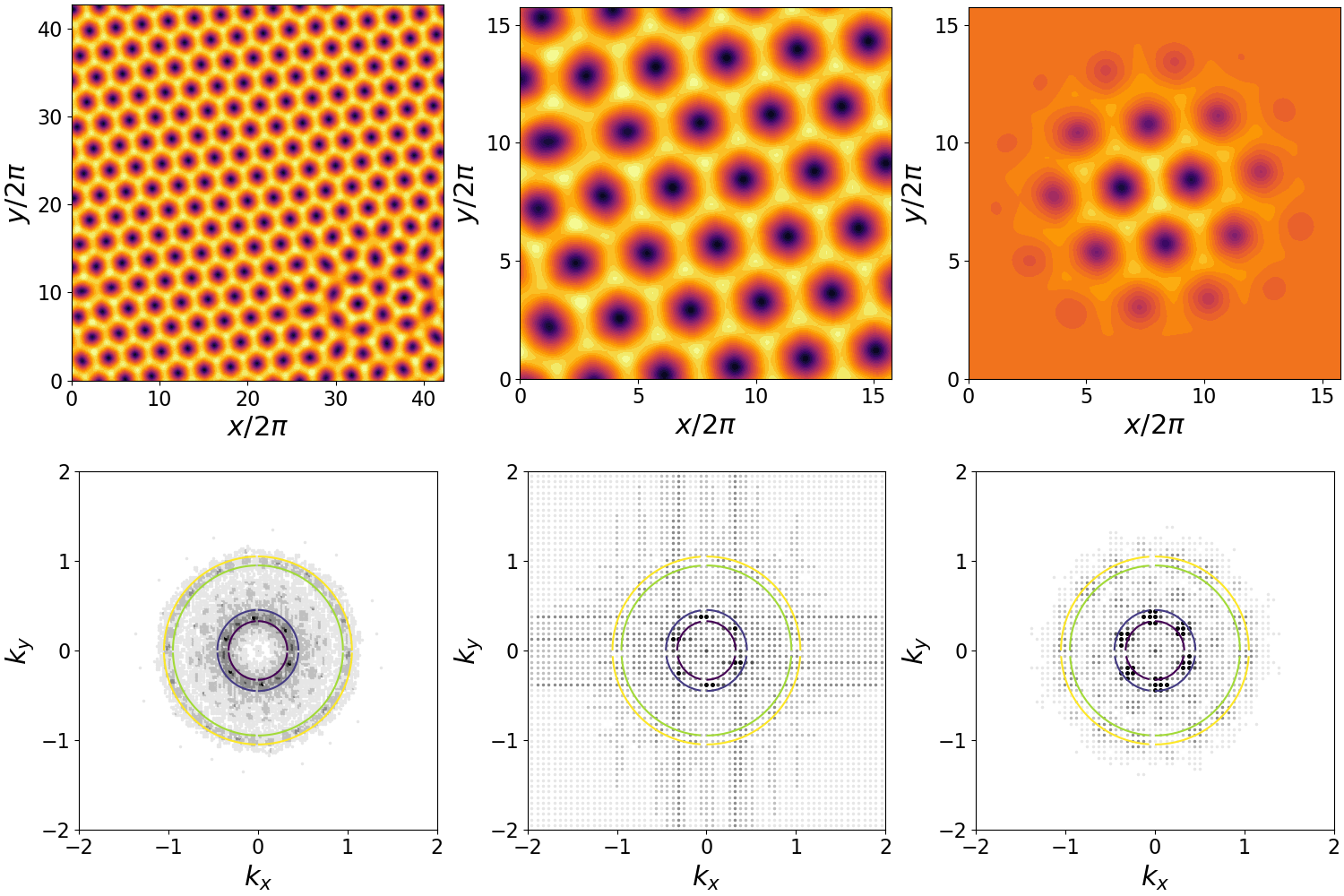}};
\begin{scope}[x={(image.south east)},y={(image.north west)}]
\node[fill=none] at (0,0.97) {\small \textbf{(a)}};
\end{scope}
\end{tikzpicture}
\begin{tikzpicture}
\node[anchor=south west,inner sep=0] (image) at (0,0) 
{\includegraphics[width=110mm]{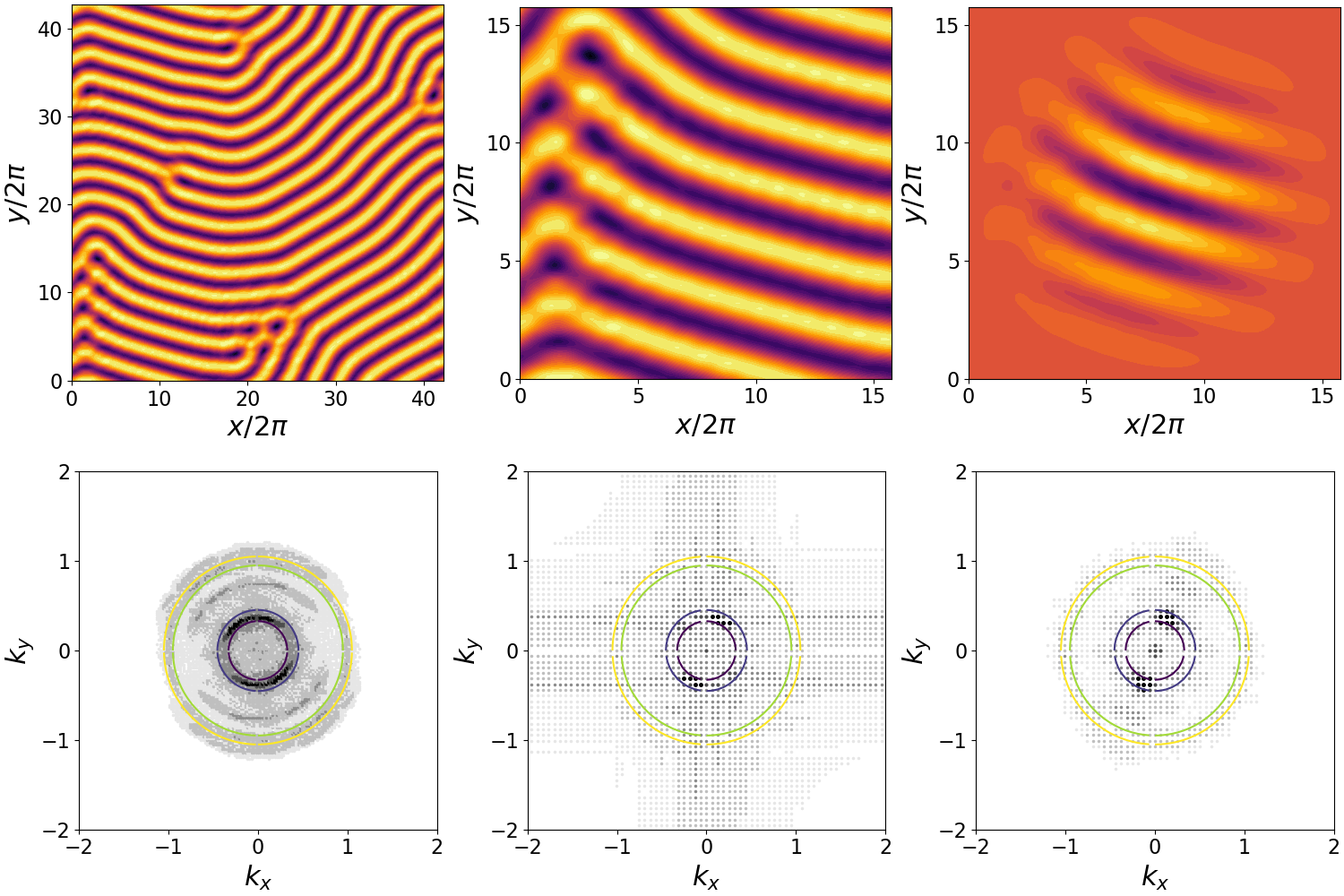}};
\begin{scope}[x={(image.south east)},y={(image.north west)}]
\node[fill=none] at (0,0.97) {\small \textbf{(b)}};
\end{scope}
\end{tikzpicture}
\caption{Examples of windowing analysis on the solutions in \cref{multipleorientations}. 
The $w$-stripe pattern (bottom two rows) was initially incorrectly classified, but after applying the Hann window, the solution is classified correctly.}
\label{Hann}
\end{figure}

We perform the local Fourier analysis on three equally sized square patches of the domain to try and capture a section of the pattern with no defects.
The patches are $6 \times 6$ repetitions of the longer wavelength ($k=q$) pattern, which is approximately $15.87\times 15.87$ repetitions of the shorter wavelength ($k=1$) pattern.
The width of the patch is three-eights of that in the full domain, so three-eights of the number of Fourier modes in the full domain is used when computing the local Fourier transform.
Following the approach introduced by~\cite{JANY2020}, we apply a 2D Hann window to each patch. 
This is a smoothing function that sets the boundaries of the pattern to zero.
Without the smoothing function, the patch would not satisfy periodic boundary conditions, which would result in high-frequency contributions in its Fourier spectrum.
A Fourier transform is then applied to the patch, and we proceed with counting the peaks in each annulus as outlined in \cref{patternclassification}, but taking segments of $10^\circ$ instead of $5^\circ$ to account for the less dense Fourier mesh.

The local Fourier analysis for both patterns of \cref{multipleorientations} is in \cref{Hann}.
The left column shows the full simulated domain of the solution and its Fourier spectrum.
The middle column shows one of the patches taken from the solution and its Fourier spectrum, both before the Hann window has been applied. 
There are large contributions in the Fourier spectrum for the untreated patch close to the $k_x$ and $k_y$ axes owing to the patch not satisfying periodic boundary conditions.
The final column shows the patch and Fourier spectrum with the Hann window applied. 
There are now no large wavenumber contributions to the spectrum and both solutions are classified correctly: six fuzzy peaks for the $w$-hexagons and two fuzzy peaks for the $w$-stripes.

A significant portion of the pattern is lost when the Hann window is applied, therefore the size of the patch needs to be large enough such that there are enough Fourier modes to correctly classify the pattern, whilst not being too large to reduce the likelihood that there are defects within the patch.
We repeat this analysis for three patches, in case some of the patches contain defects.
We have found three patches to be enough to classify all of our simulations correctly.

\subsection{Time Dependence Thresholds}
In \cref{patternclassification} we defined three metrics: $\tilde{\mathcal{C}}_1$, $\bar{\mathcal{C}}_2$ and $\bar{\mathcal{C}}_3$ to classify the time dependence of each solution.
The first of these is $\tilde{\mathcal{C}}_1$ and has two thresholds that categorize the solutions as one of equilibrium, slow time-dependent or fast time-dependent.
The second is $\bar{\mathcal{C}}_2$ and has one threshold that differentiates between small time variations and large time variations. 
This is determined by computing the rate of change of $\mathcal{C}_1(t)$.
The final metric is $\bar{\mathcal{C}}_3$, which categorizes patterns with no spatial change, patterns with small spatial change and patterns with large spatial change.
The thresholds for each metric is given in \cref{tab:criteria1}.
\begin{table}[h] 
    \centering
    \resizebox{\columnwidth}{!}{%
    \begin{tabular}{|c|c|}
    \hline
    Value & Time Dependence \\
    \hline \hline
    \rule{0pt}{12pt} 
       $\tilde{\mathcal{C}}_1<5\times 10^{-8}$  & Equilibrium \\
       $5\times10^{-8}\leq \tilde{\mathcal{C}}_1<1\times 10^{-3}$  & Slow \\
       $1\times10^{-3}\leq \tilde{\mathcal{C}}_1$ & Fast \\
       \hline
    \end{tabular}
    {\renewcommand{\arraystretch}{1}
    \begin{tabular}{|c|c|}
    \hline
    Value & $SV$ \\
    \hline \hline 
\rule{0pt}{16pt} 
       $\bar{\mathcal{C}}_2<1\times10^{-4}$  &  True \\
       \rule{0pt}{18pt} 
       $\bar{\mathcal{C}}_2\geq1\times10^{-4}$  & False 
       \rule{0pt}{19pt} 
       \\
       \hline
    \end{tabular}}
    \begin{tabular}{|c|c|}
    \hline
    Value  & Spatial Change \\
    \hline \hline
    \rule{0pt}{12pt} 
       $\bar{\mathcal{C}}_3\leq5\times10^{-4}$  & Both False \\
       $5\times10^{-4}<\bar{\mathcal{C}}_3\leq5\times10^{-3}$  & $SSC=$ True \\
       $5\times10^{-3}<\bar{\mathcal{C}}_3$ & $LSC=$ True \\
       \hline
    \end{tabular}%
    }
    \caption{
    Classification values for each of quantities $\tilde{\mathcal{C}}_1$, $\bar{\mathcal{C}}_2$ and $\bar{\mathcal{C}}_3$. 
    We use $SV$ to refer to small time variations in the solution and $SSC$ ($LSC$) for small (large) spatial change.
    These intervals were determined by directly observing correlations between the values of these three quantities and the patterned solution for multiple sets of simulations.}
    \label{tab:criteria1}
\end{table}
\section*{Acknowledgments}
The authors thank Ron Lifshitz, Priya Subramanian, Dan Hill and David Lloyd for stimulating conversations. LP is grateful for a Leeds Doctoral Scholarship from the University of Leeds. The data associated with this paper are openly available from the University of Leeds Data Repository (https://doi.org/10.5518/1819)~\cite{Pinkney_data}. This work was undertaken on ARC4 and Aire, part of the High Performance Computing facilities at the University of Leeds,~\hbox{UK}.
For the purpose of open access, the authors have applied a Creative Commons Attribution (CC BY) license to any Author Accepted Manuscript version arising from this submission.


\bibliographystyle{siamplain_unsorted}
\bibliography{references}

\end{document}